\documentclass[%
 amsmath,amssymb,
 twocolumn,aps
]{revtex4-2}
\usepackage{bm}

\usepackage[normalem]{ulem}
\newcommand\redsout{\bgroup\markoverwith{\textcolor{red}{\rule[0.5ex]{2pt}{0.4pt}}}\ULon}

\renewcommand{\vec}[1]{{\boldsymbol #1}}
\usepackage{graphicx}
\usepackage{dcolumn}
\usepackage{bm}
\usepackage{xcolor}
\usepackage[caption=false]{subfig}
\usepackage{tikz}
\usepackage{xfrac}
\usepackage{blindtext}
\usepackage{times}
\usepackage{tabularx}
\usepackage{dsfont}

\definecolor{darkblue}{HTML}{004D6B}
\definecolor{darkred}{HTML}{8c1515}

\usepackage{hyperref}
\hypersetup{
	colorlinks=true,
	urlcolor=darkred,
	citecolor=darkblue,
	linkcolor=darkred,
	breaklinks
}

 \newcommand{\moire}{moir\'e\ }

\begin{document}

\title{Network of chiral one-dimensional channels and localized states emerging in a moiré system}

\author{Jeyong Park}
\affiliation{Institute for Theoretical Physics, University of Cologne, 50937 Cologne, Germany}
\author{Lasse Gresista}
\affiliation{Institute for Theoretical Physics, University of Cologne, 50937 Cologne, Germany}
\author{Simon Trebst}
\affiliation{Institute for Theoretical Physics, University of Cologne, 50937 Cologne, Germany}
\author{Achim Rosch}
\affiliation{Institute for Theoretical Physics, University of Cologne, 50937 Cologne, Germany}
\author{Jinhong Park}
\affiliation{Institute for Theoretical Physics, University of Cologne, 50937 Cologne, Germany}

\date{\today}

\begin{abstract}
    Moiré systems provide a highly tunable platform for engineering band structures and exotic correlated phases. Here, we theoretically study a model for a single layer of graphene subject to a smooth moiré electrostatic potential, induced by an insulating substrate layer. For sufficiently large moiré unit cells, we find that
   ultra-flat bands coexist with a triangular network of chiral one-dimensional (1D) channels. These  channels mediate 
   an effective interaction between localized modes with spin-, orbital- and valley degrees of freedom emerging 
   from the flat bands. The form of the interaction reflects the chiralilty and 1D nature of the network.
    We study this interacting model within an $SU(4)$ mean-field theory, semi-classical Monte-Carlo simulations, and an $SU(4)$ spin-wave theory, focusing on commensurate order stabilized by 
   local two-site and chiral three-site interactions. 
    By tuning a gate voltage, one can trigger a non-coplanar phase characterized by a peculiar coexistence of three different types of order:  ferromagnetic spin order in one valley, non-coplanar chiral spin order in the other valley, and 120$^\circ$ order in the remaining spin and valley-mixed degrees of freedom. Quantum and classical fluctuations have qualitatively different effects on the observed phases and can, for example, create a finite spin-chirality purely via fluctuation effects. 
 \end{abstract}

\maketitle


\section{Introduction}

Stacking a two-dimensional van der Waals material on top of other van der Waals materials (with or without a relative twist) defines a class of quantum material known as moiré materials~\cite{Geim2013,Andrei2021}.
Due to their highly tunable experimental knobs for engineering band structures,  thereby facilitating the emergence of correlated phases \cite{Chen2019,Xie2022, Jin2021, Xu2020, Li2021}, such  moiré materials have recently met with tremendous interest. 
A prototypical example is twisted bilayer graphene (TBG)~\cite{Cao2018a, Cao2018,Bistritzer2011}, where two sheets of graphene are stacked with a relative twist. At twist angles $\sim1.1 ^\circ$, the so-called 'magic angle', flat bands emerge near the charge neutrality point~\cite{Cao2018a, Cao2018, Bistritzer2011,tarnopolsky2019origin}, which amplifies the effect of interaction to exhibit various correlated phases~\cite{po2018origin, Kennes2018,Wu2018, Lian2019, Roy2019,Yankowitz2019,Isobe2018,Choi2019,Kerelsky2019,Xie2021, Ledwith2020,song2022magic, Thomson2018, Bultinck2020, Kwan2021, Hofmann2022,Chou2022}.
Besides TBG, a wealth of different types of exotic bands and interaction effects have been discovered in 
multilayer moiré systems~\cite{Geim2013,Andrei2021,Mora2019, Khalaf2019, Park2020, Chen2020,He2020, Ramires2021,Park2022}.

\begin{figure}[h!]
	\centering
	\includegraphics[width= \columnwidth]{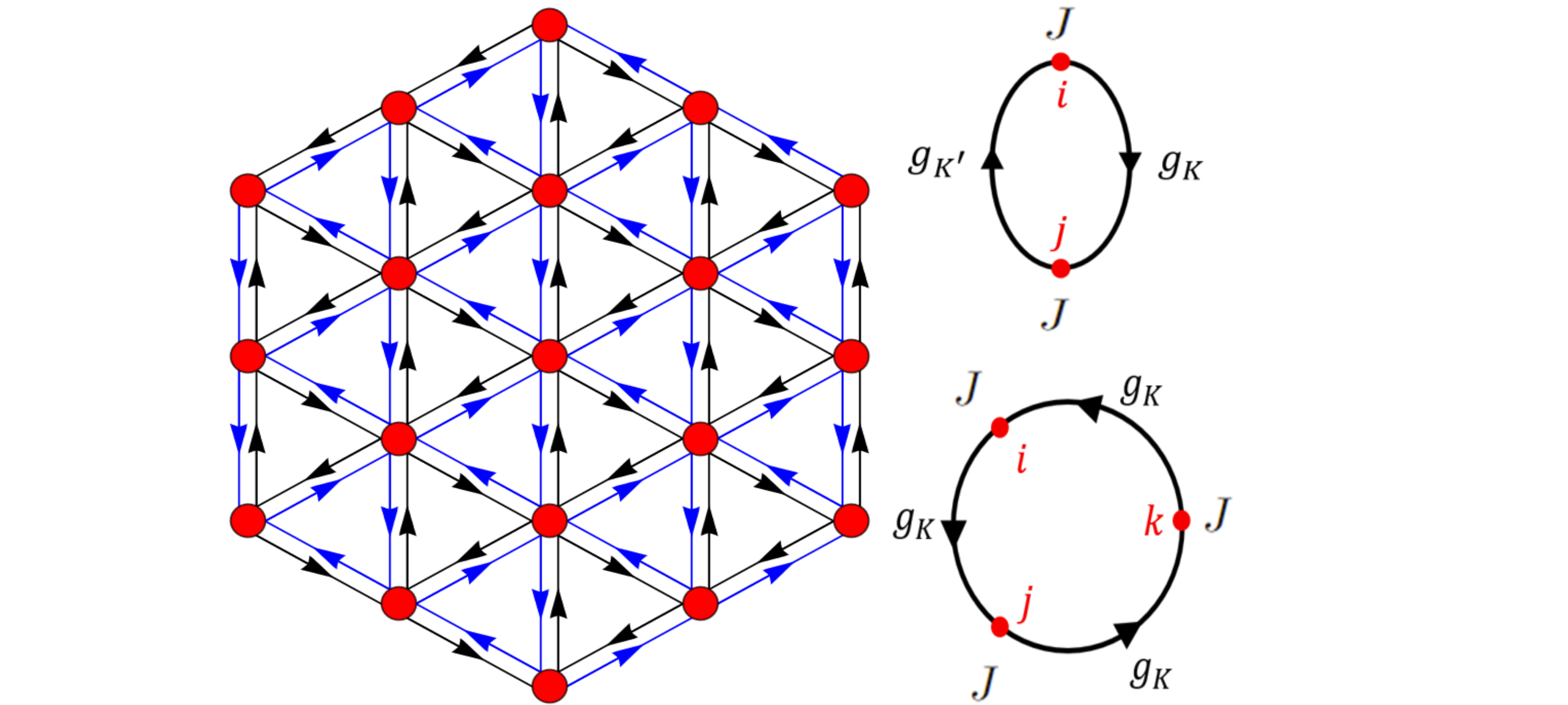}
	\caption{ {\bf Network model} consisting of one-dimensional chiral channels (blue arrows for the $K$ valley, black for the $K'$ valley) and localized states (red circles). 
	Upper (Lower) right: the diagram for the two (three)-spin interaction.
	}
	\label{fig3}
\end{figure}

 In this manuscript, we address one of the simplest models of a moiré system: a {\em single} layer of graphene subject to a moiré potential induced by a substrate layer. Despite its simplicity, it 
shows -- even without fine tuning --  remarkably rich physics. 
For sufficiently large moiré unit cells 
two kinds of moiré bands emerge: one-dimensional chiral channels (1DCCs) and ultra-flat bands.
Along lines where the gap arising from the \moire potential changes sign, a network of topologically protected 1DCCs is developed,
as depicted in Fig.~\ref{fig3}. 
At the same time,  an extra set of localized modes emerges at the junction where six 1DCCs join (red dots in  Fig.~\ref{fig3}). These modes only hybridize weakly with the 1DCCs and with the neighboring localized modes giving rise to ultra-flat bands. This coexistence of localized modes and propagating 1DCCs and the resulting peculiar interaction physics are the main results of this paper.

The emergence of a network of 1D chiral channels in moiré systems has been previously discussed~\cite{San-Jose2013, Efimkin2018, Ramires2018, Rickhaus2018, Huang2018, Yoo2019, Xu2019,Chou2020,DeBeule2020, DeBeule2021,Chou2021}. In an early study, San-Jose and Prada \cite{San-Jose2013} pointed out that  a network of topologically protected 1D helical channels forms in TBG subject to an out-of-plane electric field,
see also Refs.~\cite{Efimkin2018, DeBeule2021}. Experimentally, signatures of these 1D channels have been observed in transport~\cite{Rickhaus2018, Yoo2019, Xu2019} and scanning tunneling spectroscopy~\cite{Huang2018}.
In contrast to our model, such systems do not exhibit the coexistence of flat bands and 1D channels. 
Moreover, a coexistence of propagating {\em two-dimensional} Dirac dispersing bands and flat bands has been reported in mirror symmetric twisted trilayer graphene~\cite{Khalaf2019, Park2020}. In this setting, 
Ramires and Lado discussed heavy fermion physics, emerging  from the interaction of localized and propagating modes~\cite{Ramires2021}.
From a more general point of view,  the emergence of localized and propagating bands in moiré systems has been investigated in Ref.~\cite{Attig2021} using concepts of quantum chaos. Generic bands tend  {\em not} to be flat due to localization in {\em momentum space}, but these arguments cannot be applied to the bands discussed in our paper arising from the specific real-space structure of the \moire potential.

\begin{figure*}[t]
	\centering
	\includegraphics[width=2 \columnwidth]{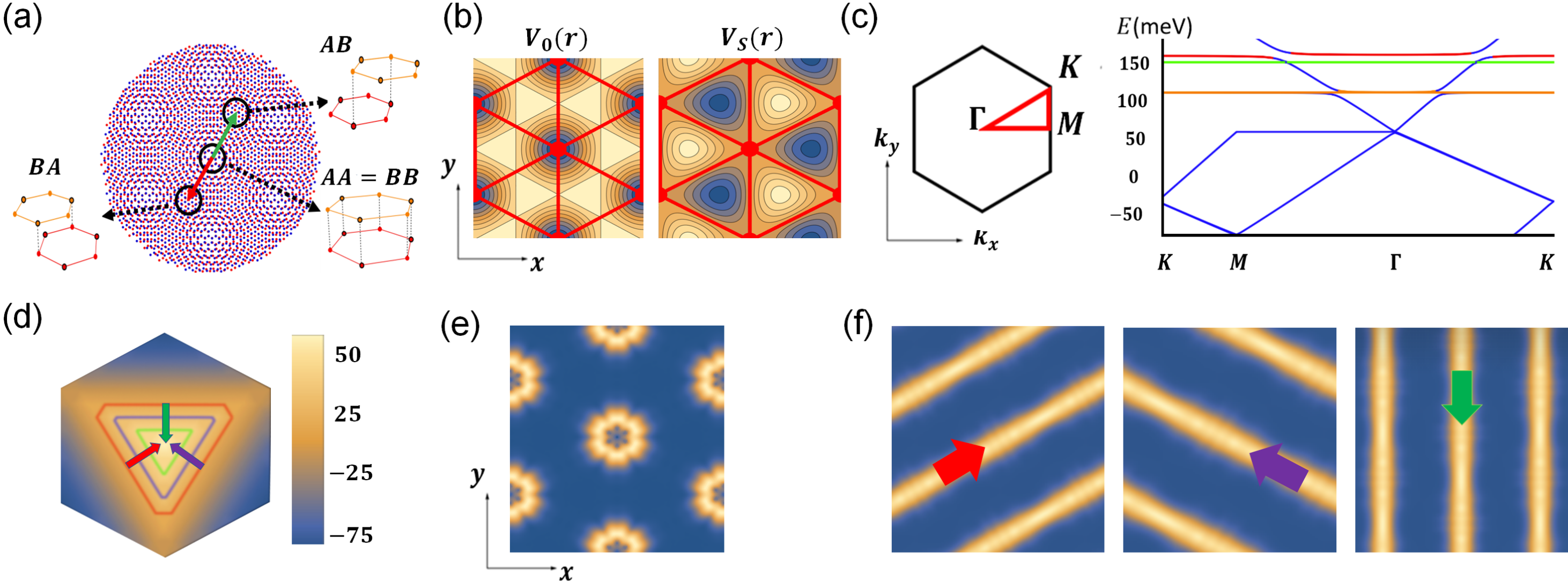}
	\caption{{\bf Moir\'e potential model.} {\bf (a)} Moir\'e pattern from a layer of the hexagonal lattice (blue dots) stacked on  top of another layer of an hexagonal lattice (red). Due to a relative rotation of the layers, different local stacking patterns (AA, AB, BA) occur (circles). The vectors $\vec{r}_{AB} = - \vec{r}_{BA}=(\frac{\sqrt{3}L}{6},\frac{L}{2}) $ connect regions of AA and AB stacking. {\bf (b)} The uniform and staggered potential, $V_0 (\vec{r})$ and $V_s (\vec{r})$, in real space.  
			{\bf (c)} Band structure for the effective Hamiltonian Eq.~\eqref{eq:Heff}  in the moir\'e Brillouin zone. 
			Flat bands are classified by different representations (red: $\rho_3$, green: $\rho_4$, orange: $\rho_6$) of the symmetry group $\textrm{Dic}_3$. 
			{\bf (d)} Density plot of the energy for one of the 1D dispersing bands and the Fermi surfaces at three different chemical potentials (red 0meV, blue 20meV, green 40meV). The arrows indicate the three different propagating directions. 
		{\bf (e)} Bloch wave functions $|\Psi_{n,\vec k}(\vec r)|^2$ of the green flat band at the $\Gamma$ point. {\bf (f)}  Bloch wave functions of 1D propagating bands at momenta indicated by the arrows in (d).
		Parameters: $u_{AA} = u_{BB} = 0$ and $u_{AB} = u_{BA} = \frac{8 \pi v}{\sqrt{3} L }$. 
	 }
	\label{fig2}
\end{figure*}

\section{Model}

We consider a single layer of graphene on top of some {\em insulating} substrate which shares the hexagonal structure with graphene but has either a slightly different lattice constant or is rotated by a small twist angle. 
As the substrate is gapped, it mainly affects graphene via electrostatic potential terms. Thus, at low-energies, the spinless single-particle Hamiltonian is approximated by
\begin{align}
	H_{\textrm{eff}} =  -iv (\partial_x s^x \tau^z + \partial_y s^y \tau^0) + V_0(\vec r) \mathds{1} +  V_s(\vec r) s^z 
	\label{eq:Heff}.
\end{align}
Here the Pauli matrices $\tau^i$ and $s^i$  act on the valley and sublattice space, respectively, and $v$ is the graphene Fermi velocity.
The staggered term $V_s =(V_A-V_B)/2$ 
describes the potential difference between the $A$ and $B$ sublattice, $V_A$ and $V_B$, and a constant $V_s$ opens a mass gap in the Dirac spectrum. 
The magnitude of $V_s$ has a maximum in regions of the moir\'e lattice where the atoms of different sublattices stack on the top of each other, i.e., AB or BA stacking as shown in Fig.~\ref{fig2}(a).
The uniform potential $V_0$ is given by $V_0=(V_A+V_B)/2$.
Due to the smoothness of moir\'e structures, we can focus on the lowest Fourier components of the potentials. Denoting the six smallest reciprocal lattice vectors of the moir\'e structure by $\vec G_i$, $i=1,\dots,6$, with $|\vec G_i| \equiv G =   \frac{4 \pi}{\sqrt{3} L}$, we obtain
\begin{align} \label{eq:potentials}
	V_{\beta}(\vec{r}) &=\sum\limits_{\beta' = A,B}{\sum\limits_{i=1}^{6}{u_{\beta \beta'}e^{i 
	\vec{G}_i \cdot (\vec{r}-\vec{r}_{\beta \beta'})}}}\,\,
\end{align}
with sublattice $\beta=A, B$, the size of \moire unit cell $L$, and $\vec{r}_{AA}= \vec{r}_{BB}=0$, $\vec{r}_{AB} = - \vec{r}_{BA}$, see Fig.~\ref{fig2}(a). We consider a hexagonal substrate with equivalent $A$ and $B$ sublattices such that $u_{AA}=u_{BB}$ and $u_{AB}=u_{BA}$. In this case, the amplitudes of $V_s$ and $V_0$ are given by $u_s=u_{AB}$ and $u_0=-2 u_{AA}+u_{AB}$. 

As shown in Fig.~\ref{fig2}(b), $V_s$ vanishes along straight lines and thus changes its sign across those lines. At the same time, $V_0$ has minima at  high-symmetry points (red dots) in the center of the moir\'e unit cell where the lines cross.
These two regions lead to two very different types of bands: 1D dispersing bands and ultra-flat bands, see
Fig.~\ref{fig2}(c), computed by diagonalizing Eq.~\eqref{eq:Heff} in momentum space. 

\subsection{One-dimensional chiral channels}

The sign change of the mass term $V_s (\vec{r})$ induces a 1DCC, propagating along the straight lines in Fig.~\ref{fig2}(b) with the full speed of the graphene Fermi velocity. 1DCCs emerging from the $K$ and $K'$ valley move in opposite directions, as depicted by black and blue arrows in Fig.~\ref{fig3}. 
Fig.~\ref{fig2}(f) shows the Bloch wave function of the propagating bands which perfectly tracks the 
straight lines in Fig.~\ref{fig2}(b). Surprisingly, the wave functions show almost no modulation at their crossing points. 
 
\subsection{Localized states}

The flat bands in Fig.~\ref{fig2}(c) have their origins in  states localized close to the red dots in Fig.~\ref{fig2}(b), where $V_0$ has a minimum whereas $V_s$ is highly suppressed. 
A sufficiently strong \moire potential, $u_0 \gg v/L$, renders the states localized in real space as shown in Fig.~\ref{fig2}(e). These localized modes hybridize only weakly with the 1DCCs and  neighboring localized modes, leading to ultra-flat bands. The localized states with fixed valley index can be classified by the dicyclic symmetry group $\text{Dic}_3$. From $\text{Dic}_3$, one obtains three different types of localized states, labeled by two one-dimensional representations $\rho_3$, $\rho_4$ and a two-dimensional  irreducible representation $\rho_6$, see supplement~\cite{Supple}. 

\subsection{Network model} 

Combining localized and propagating states, we obtain the network model depicted in Fig.~\ref{fig3}. 
The kinetic Hamiltonian for the 1DCCs is given by \begin{align} \label{eq:kineticenergy}
    H_{\rm kin} = -i v \sum_{\alpha, n, i,\sigma} \alpha \int d\rho \Psi^\dagger_{n,i,\alpha,\sigma}(\rho)
   \partial_{\rho} \Psi_{n,i,\alpha,\sigma}(\rho). 
\end{align}
The operator $\Psi^\dagger_{n,i,\alpha,\sigma}(\rho)$ creates an electron with spin $\sigma=\uparrow/\downarrow$ and valley  $\alpha=\pm = K/K'$ in a 1DCC propagating along the lattice vector $\vec a_i$ with $i=1,2,3$; the center of the corresponding wave packet is located at $\vec R_{n,i,\rho}= n \vec a_{i+1} + \rho \vec{a}_i$ with integer $n$. We denote the location of crossing points of 1DCCs by $\vec R_m$ and define $\rho_{n,i,m}$ as the solution of $\vec R_m=\vec R_{n,i,\rho_{n,i,m}}$. In these notations, the inter-channel tunneling $H_{w}$ and the coupling of 1DCCs to localized states, $H_{\lambda}$, are given by 
\begin{align} 
H_{w}=&\sum_{\text{crossing at } \vec{R}_m}  \hat{w}_{i i'}\Psi^\dagger_{n,i,\alpha,\sigma}(\rho_{n,i,m})\Psi_{n',i',\alpha,\sigma}(\rho_{n',i',m}), \nonumber \\ 
H_{\lambda}=& \sum_{\text{crossing at } \vec{R}_m} 
\hat{\lambda}_{ij}\, d^\dagger_{m,\alpha,\sigma,j} \Psi_{n,i,\alpha,\sigma}(\rho_{n,i,m}) +h.c.\ . \label{eq:wireloctunneling}
\end{align}
We sum over all channels which cross at $\vec{R}_m$. $d^\dagger_{m,\alpha,\sigma,j}$ creates localized electronic states where $j$ denotes an extra orbital index if the localized states belong to the $\rho_6$ representation. 
The form of the matrices $\hat{w}_{i i'}$ and $\hat{\lambda}_{ij}$ is entirely determined by the symmetries of the system and the representation of  Dic$_3$ of the localized states.
Eqs.~\eqref{eq:kineticenergy} and \eqref{eq:wireloctunneling} describe the bandstructure with high precision after fitting the amplitude of $\hat w$ and $\hat \lambda$ and the energy of the localized states, see supplement~\cite{Supple}.

\subsection{Local interaction}

Since the flat bands are highly localized, there will be a Coulomb blockade for adding electrons to the localized sites, described by
\begin{align} \label{eq:onsiteinteraction}
    H_{U} = U \sum_{\vec{R}_m} \sum_{\xi \neq \xi'} n_{m, \xi} n_{m, \xi'},
\end{align}
where $\xi = \{\alpha, \sigma\}$ includes all local quantum numbers, i.e., valley $\alpha$ and spin $\sigma$ (and an extra orbital quantum number for the $\rho_6$ representation).  Since for large moir\'e unit cells the dominant contribution comes from the long-ranged part of the Coulomb interaction which is only sensitive to charge, $H_U$ is approximately $SU(4)$ (or $SU(8)$ for $\rho_6$) invariant. 
Using $U \sim \frac{e^2}{4 \pi \epsilon_0 d_{\rm loc}}$ with $d_{\rm loc}\approx 6.3$nm, we estimate $U \approx 230$meV for the parameters of Fig.~\ref{fig2}(e) which is more than an order of magnitude larger than the hybridization of impurity levels, $\lambda$. 

As $U \gg \lambda$, 
the system maps to a (generalized) Kondo lattice model, where local degrees of freedom couple only via the network of 1DCCs. 
For a localized state in the $\rho_4$ representation, one obtains an effective SU(4) symmetric coupling 
\begin{align} \label{eq:Kondocoupling}
   H_{J}\approx J L \!\!\! \sum_{\textrm{crossing at } \vec{R}_m}  \!\!\! \Gamma^\ell (\vec{R}_m) \cdot \tilde{\Psi}^\dagger_{\xi}(\vec{R}_m )\gamma^{\ell}_{\xi \xi'} \tilde{\Psi}_{\xi'}(\vec{R}_m ). 
\end{align}
Here $\gamma^\ell$, $\ell=1,\dots,15$, are the  4$\times$4 generators of $SU(4)$ acting on a linear combination of the three 1DCCs resulting from the hybridization matrix $\hat\lambda$~\cite{Supple},  $\tilde{\Psi}_{\xi} (\vec{R}_m) = \frac{1}{\sqrt{3}} \sum_{i=1,2,3} (-1)^i \Psi_{n_i, i, \xi} (\rho_{n_i, i, m})$. $ \Gamma^{\ell} (\vec{R}_m) =\sum_{\xi,\xi'} d^{\dagger}_{m, \xi} \gamma^{\ell}_{\xi \xi'} d_{m, \xi'} $ describes the local $SU(4)$ degree of freedom and  $J \sim \lambda^2/U$ is the Kondo coupling. 

\begin{figure*}
	\centering
	\includegraphics[width= 1.9\columnwidth]{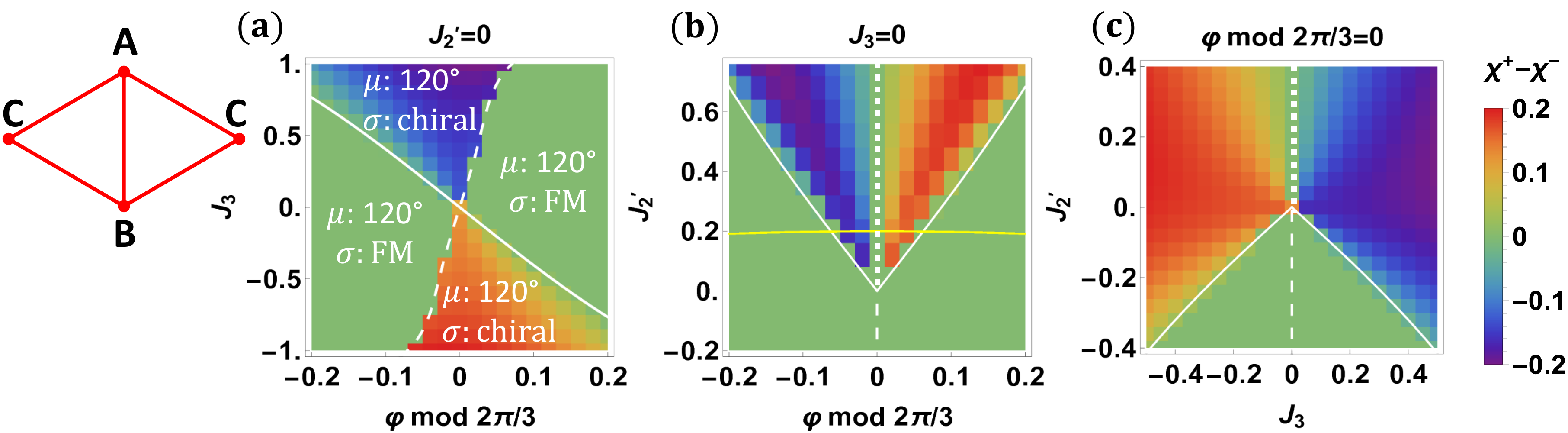}
	\caption{{\bf Mean-field phase diagrams} with the order parameter, $\chi^{+} - \chi^{-}$, for small $J_3, J_2'$, and $\varphi$, defined modulo $2 \pi/3$ (cf. Eq.~\eqref{eq:simspinvalleyHam}). All phases shown in the figure show a $120^\circ$ order in the $\vec \mu$ valley degrees of freedom. They differ by their spin order which are distinguished by 
$\chi^{\pm} \equiv \langle\vec{\sigma} P^\pm\rangle_A \cdot \langle\vec{\sigma} P^\pm\rangle_B \vec{\times}  \langle\vec{\sigma} P^\pm\rangle_C$, the valley-projected chirality, defined on a triangle (cf. the inset). Three different types of spin-order occur: spin-ferromagnic order (green regions), a chiral spin-order (red or blue regions) in one of the two valleys, and an in-plane 120$^\circ$ spin order in one of the two valleys (white dotted line). $J_2$ is set to 1. 
 	Panel \textbf{(a)}: $(\varphi, J_3)$ space with $J_2' = 0$, panel \textbf{(b)}: $(\varphi, J_2')$ with $J_3=0$, panel \textbf{(c)}: $(J_3, J_2')$ with $\varphi=0$. The yellow line in panel (b) shows schematically how parameters change as a function of a gate voltage for $k_F L$ close to $-\frac{\pi}{2}$. While the solid white lines represent second order transitions, the dashed white line a first order transition where the type of $120^\circ$ order changes from left- to right circulating on a triangle $\triangleright$ (opening angle $\theta=0, \pi$, respectively, see \cite{Supple}).
}
	\label{fig4:phasediagram}
\end{figure*}

The interaction between the localized states is mediated by the network of 1DCCs. The resulting RKKY interaction is obtained from a perturbation theory both in  $J$ and the inter-channel tunneling $w$. 
From a standard RKKY diagram (see Fig.~\ref{fig3}) to order $J^2 w^0$, we obtain the two-spin interaction term
\begin{align} \label{eq:twospininteraction}
 H_{\textrm{2s}} &= -\frac{J^2 L^2}{12\pi v}\sum_{(m_1 \to m_2)_c} \bigg (\frac{e^{2 i \bm{k_F}^{i} \cdot \vec{\rho}_{m_1 m_2}}}{|\vec{\rho}_{m_1 m_2}|} 
 \nonumber \\ &\times  \left(1+ \vec{\sigma}_{m_1}\cdot \vec{\sigma}_{m_2}\right)\tau^{-}_{m_1}\tau^{+}_{m_2}  
  + h.c. \bigg ). 
\end{align}
$\vec{\sigma}_m \equiv \vec{\sigma} (\vec{R}_m)$ and $\vec{\tau}_m  \equiv \vec{\tau} (\vec{R}_m)$ are Pauli matrices acting on spin and valley at $\vec{R}_m$, and $\tau^{\pm} \equiv \tau^{x} \pm i \tau^{y}$. 
The summation $(m_1 \to m_2)_c$ runs only over localized states connected by the same 1DCC, separated by $\vec \rho_{m_1 m_2} \equiv \vec{R}_{m_2} -  \vec{R}_{m_1}$
parallel to the Fermi velocity of the $+$ valley channels. 
Importantly, the RKKY term 
necessarily requires two valley flip processes, $\tau^{-}_{m_1}\tau^{+}_{m_2}$: both a valley $+$ and $-$ channel running in opposite directions are needed to form a closed loop connecting two sites, see Fig.~\ref{fig3}. This process breaks the $SU(4)$ symmetry.

Closed loops can also be formed by triangles in Fig.~\ref{fig3}, remarkably, inducing a chiral interaction to order $J^3$. From the diagram depicted in Fig.~\ref{fig3}, we obtain 
\begin{align} \label{eq:threespininteraction}
H_{3s} &= \frac{8 J^3 L^3}{27 \sqrt{3} \pi v^2}\sum_{p = \triangleright/\triangleleft,(m_1, m_2, m_3)_p} 
\frac{\textrm{cos}(3 k_F |\vec{\rho}_{m_1, m_2}|)}{ |\vec{\rho}_{m_1, m_2}|} 
\nonumber \\ & \times p \ 
 \vec{\sigma}_{m_1}\cdot (\vec{\sigma}_{m_2} \vec{\times} \vec{\sigma}_{m_3}) \Big(\prod_{i=1}^3 P_{m_i}^{+} - \prod_{i=1}^3 P_{m_i}^{-} \Big). 
\end{align}
The summation runs over the right- and left-oriented triangles, $p = \triangleright/\triangleleft=\pm 1$, in Fig.~\ref{fig3}, where $(m_1, m_2, m_3)_p$ denotes the three sites of each triangle (in anti-clockwise order). $k_F$ is the Fermi momentum of 1DCCs. 
We defined the projector $P_{m}^{\pm} \equiv \frac{(\tau^0_{m} \pm \tau^{z}_{m})}{2}$ on valley $\pm$ at $\vec{R}_m$.
The chiral spin-interaction,  $\vec{\sigma}_{m_1}\cdot (\vec{\sigma}_{m_2} \vec{\times} \vec{\sigma}_{m_3})$, is induced by the chiral motion of the 1DCCs within each triangle (even in the absence of spin-orbit interaction).
The direction of the chiral currents determines the sign of the chiral interaction which changes when moving from $\triangleright$ to $\triangleleft$ or from valley $+$ to $-$. 

There is also a non-chiral contribution from the same diagram and from a similar diagram to order $J^2w$,
\begin{align} \label{eq:twospintermJprime}
    &H'_{2s} = \frac{ J^2 L^3}{27 \sqrt{3} \pi v^2} \sum_{p = \triangleright/\triangleleft,(m_1, m_2, m_3)_p} 
    \frac{\sin(3 k_F |\vec{\rho}_{m_1, m_2}|)}{|\vec{\rho}_{m_1, m_2}|}
    \nonumber \\
    &\times   (1+ \vec{\sigma}_{m_1} \cdot \vec{\sigma}_{m_2}) \Big ( J \tau^z_{m_3} (\tau^z_{m_1}+ \tau^z_{m_2}) + w (1+ \tau^z_{m_1}\tau^z_{m_2} ) \Big) \nonumber \\
    & + \text{permutations}.
\end{align}
We sum over the 6  permutations for renaming $m_1$, $m_2$ and $m_3$.

\section{Mean-field Phase diagram}
\label{sec:meanfieldphasediagram}

To study the interplay of  Eqs.~\eqref{eq:twospininteraction}-\eqref{eq:twospintermJprime}, we consider a simplified Hamiltonian which contains only nearest neighbor interactions,  $H_{\rm{sv}}=H_2+H_3+H_{2'}$, with 
\begin{align} \label{eq:simspinvalleyHam}
H_2=&J_2 \sum_{\langle m_1 \to m_2 \rangle_c}(1 + \vec{\sigma}_{m_1} \cdot \vec{\sigma}_{m_2} )  (e^{i \varphi} \tau^{+}_{m_1} \tau^{-}_{m_2} + h.c.) \nonumber \\
H_3=&J_3 \!\!\! \sum_{p = \triangleright/\triangleleft,(m_1, m_2, m_3)_p} \!\!\!\! p \Big(\prod_{i=1}^3 P_{m_i}^{+} - \prod_{i=1}^3 P_{m_i}^{-} \Big)\nonumber \\
& \hspace*{4cm} \times  \vec{\sigma}_{m_1}\cdot(\vec{\sigma}_{m_2} \vec{\times} \vec{\sigma}_{m_3}) \nonumber \\
H_{2'}=&J_2' \sum_{\langle m_1 \to m_2 \rangle_c}(1 + \vec{\sigma}_{m_1} \cdot \vec{\sigma}_{m_2} )  (1 + \tau^{z}_{m_1} \tau^{z}_{m_2})\,\,.
\end{align}
Here $J_2>0$ is the largest coupling constant with $\varphi=2 k_F L+\pi$, while $J_2' \sim \sin(3 k_F L)$ and $J_3 \sim \cos(3 k_F L)$. From $H_{2s}'$, Eq.~\eqref{eq:twospintermJprime}, we take, for simplicity, only the term $\sim J^2 w$ into account (assuming $w>J$) but we checked that the $J^3$ contribution to $H_{2s}'$ does not lead to qualitative changes.
The continuous symmetries of $H_{\rm{sv}}$ are $U(1) \times SU(2) \times SU(2)$ generated by $\tau^z$, $P^+ \vec \sigma$ and $P^- \vec \sigma$. Remarkably, one can rotate the spin-orientation of the two valleys independently.

Assuming that the localized states are filled with one electron, the states on the $SU(4)$ space are spanned by a 4-component complex vector. In this basis, we solve the self-consistent mean-field equations at $T=0$ iteratively.
We find that either a one- or a three-sublattice solution has the lowest energy. 
As $J_2$ is the largest term, we first analyze the case $J_3 = J'_2=0$. 
The parameter $\varphi$ in Eq.~\eqref{eq:simspinvalleyHam} can be viewed as an Aharonov-Bohm phase arising from a staggered magnetic flux. As  $3 \varphi$ is the total phase along a triangular loop, one can always `gauge away' changes of $\varphi$ by $\frac{2 \pi}{3}$ using $\tau_z$ rotations by $0$, $2 \pi/3$, $4 \pi/3$ on the A, B, C sublattices. For $\varphi=0$, we obtain a variant of the Kugel-Khomskii model \cite{Kugel1982}
\begin{align} \label{eq:H0}
 H_0 &= 2 J_2 \sum_{\langle m_1 \to m_2 \rangle_c}(1 + \vec{\sigma}_{m_1} \cdot \vec{\sigma}_{m_2} )  (
     \tau^{x}_{m_1} \tau^{x}_{m_2} +  \tau^{y}_{m_1} \tau^{y}_{m_2})\nonumber \\
    &= 2 J_2\sum_{\langle m_1 \to m_2 \rangle_c} \vec{\mu}^1_{m_1}\cdot \vec{\mu}^1_{m_2}+ \vec{\mu}^2_{m_1} \cdot \vec{\mu}^2_{m_2}
  \end{align} with four component vectors given by $\vec \mu^1_m=(\tau^x_m, \tau^y_m \sigma^x_m, \tau^y_m \sigma^y_m, \tau^y_m \sigma^z_m)$ and  $\vec \mu^2_m=(\tau^y_m, \tau^x_m \sigma^x_m, \tau^x_m \sigma^y_m, \tau^x_m \sigma^z_m)$. 
  The ground states have a three-site unit-cell where the vectors $\langle \vec \mu^n_m\rangle$, $n=1,2$, have the norm $1$, and show 120$^\circ$ order such that $\langle \vec \mu^n_{m_1}\rangle\cdot  \langle \vec \mu^n_{m_2}\rangle=\cos(2 \pi/3)=-\frac{1}{2}$ for neighboring sites. Note that this specific type of 120$^\circ$ order is realized with 4-component vectors. 
  Surprisingly, the above described 120$^\circ$ order has an extra degree of freedom that is revealed by the magnetization vectors $\langle P^{\pm} \vec{\sigma}\rangle$ in the two valleys. These vectors have length $1/2$ in the ground-state manifold. 
  In one of the two valleys, the magnetization is always ferromagnetic, but  in the other valley a non-coplanar spin configuration is possible, leading to a finite staggered chirality $\chi^{\pm}$  with $\chi^{\pm} = \langle \vec{\sigma}_{m_1} P^\pm_{m_1} \cdot ( \vec{\sigma}_{m_2} P^\pm_{m_2} \vec{\times} \vec{\sigma}_{m_3} P^\pm_{m_3})\rangle$. In the supplementary material \cite{Supple}, we describe how the mean-field solution can be parameterized by a continuous angle $\theta$ and a discrete variable $\pm$, describing the opening angle of non-coplanar valley-projected spins on the three sublattices and also which of the valley sector exhibits ferromagnetic order.
  
States with an arbitrary chirality, $-\frac{1}{8} \le  \chi^{\pm} \le \frac{1}{8}$, are degenerate (within mean-field theory) if only $H_0$, Eq.~\eqref{eq:H0}, is considered, see  supplement \cite{Supple}. Thus, $H_0$ defines a highly singular point in the phase diagram and even small perturbations can select one of the states in the ground-state manifold of $H_0$. For example, for an infinitesimal $J_3>0$ perturbation, states are selected which have either the minimal value $\chi^+=-\frac{1}{8}$ with $\chi^-=0$ or the maximal value of  $\chi^-=\frac{1}{8}$ with $\chi^+=0$.
Such a staggered (or uniform) chiral order has, e.g., been extensively studied in the spin-$1/2$~\cite{Wietek2017,Gong2017} or the half-filled Hubbard model~\cite{Szasz2020, Chen2022,Sur2022,Kuhlenkamp2022} on the triangular lattice. 

In contrast, the perturbation by a finite $\varphi$ 
stabilizes a phase where $\langle P^{\pm} \sigma\rangle$ orders ferromagnetically for both valleys, $\chi^+=\chi^-=0$, while $\langle\vec  \mu^n\rangle$ displays a coplanar 120$^\circ$ ordered phase. The presence of both $\varphi$ and $J_3$ leads to the phase diagram of Fig.~\ref{fig4:phasediagram}(a). A finite $J_2'>0$, however, suppresses such ferromagnetic configuration, selecting a state where $\langle P^{\pm} \sigma\rangle$ is non-collinear but coplanar, forming a 120$^\circ$ order in either $\langle P^{+} \sigma\rangle$ or $\langle P^{-} \sigma\rangle$ on top of the 120$^\circ$ order in $\langle \vec \mu^n\rangle$. The resulting phase diagrams are shown in Fig.~\ref{fig4:phasediagram}(b) and (c).

\section{Classical fluctuations}

The mean-field theory discussed above, ignores the effect of both quantum and classical fluctuations. To capture fluctuation effects, we have (i) performed an $SU(4)$ spin-wave calculation (or, more precisely, spin-valley-wave calculation) both in the classical and quantum regime. Details of the $SU(4)$ spin-wave theory are given in the supplementary material~\cite{Supple}. Furthermore,  we have (ii) calculated finite temperature properties of the semi-classical version of our $SU(4)$ model using Monte Carlo calculations.

 A semi-classical variant of our $SU(4)$ model can formally be obtained by making a product ansatz for the wavefunction, $|\Psi \rangle=\prod_{m} |\Psi_{m}\rangle$, where $|\Psi_{m}\rangle$ is a single-site 4-component normalized wave function with an arbitrary phase per site. 
 A semi-classical state for a system of size $N \times N$ is thus parameterized by $(4\cdot 2-2) N^2$ real numbers. 
 At $T=0$, this semi-classical model reproduces the mean-field results discussed in Sec.~\ref{sec:meanfieldphasediagram}.
  Thermal expectation values at a finite temperature $T = 1/\beta$ can be approximately calculated by sampling the space of product-state wavefunctions according to the Boltzmann distribution $\sim\exp(-\beta \langle \psi|H|\psi\rangle)$ \cite{stoudenmire2009, hickey2014} using a standard Markov chain Monte Carlo algorithm \cite{LandauBinder}. Employing local Metropolis updates, a typical Monte Carlo run consists of $N_m = 1 \cdot 10^6$ thermalization sweeps followed by $N_m = 4 \cdot 10^6$ measurement sweeps, or up to $N_m = 10^7$ sweeps close to the transition temperature. We use linear lattice sizes of up to $N=72$ with periodic boundary conditions. Additional details on the simulations are provided in the supplementary material \cite{Supple}.

\begin{figure}[b]
	\centering
	\includegraphics{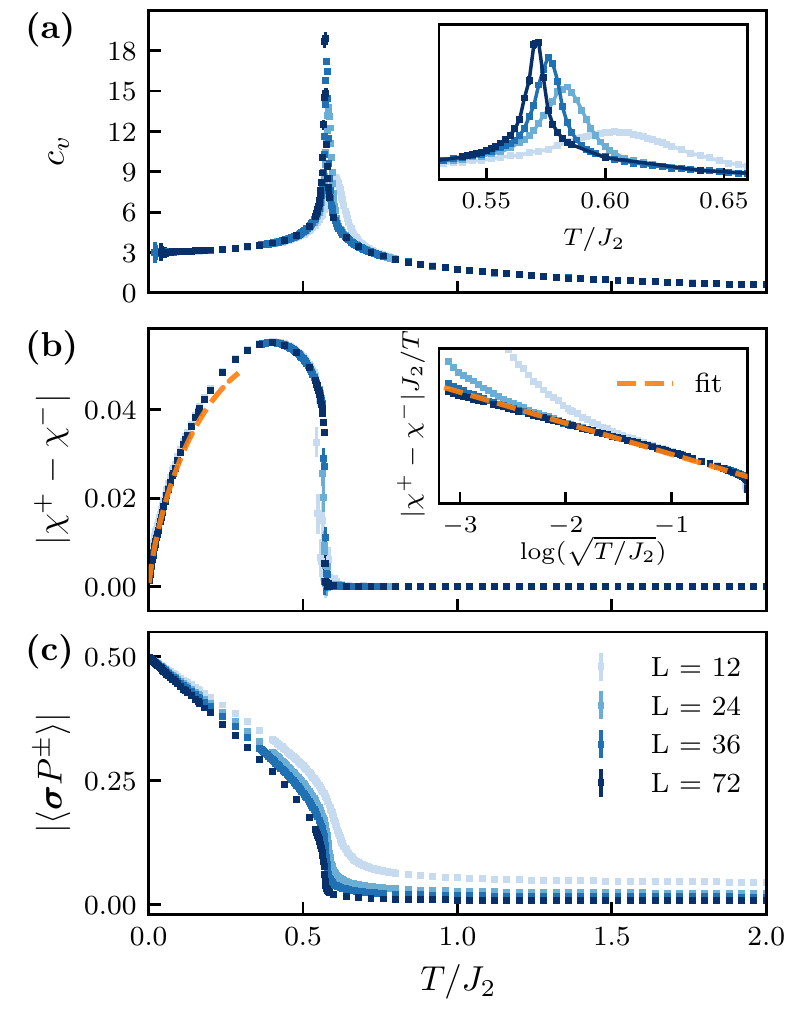}
	\caption{{\bf Thermodyamics and thermal order-by-disorder transition.} 
		Shown are Monte Carlo results for the $J_2$ model, Eq.~\eqref{eq:H0}, in the semi-classical approximation
		for different linear system sizes $N=12, 24, 36$ and $72$. 
		Numerical errors are smaller than the size of the symbols. 
		Panel \textbf{(a)}: The specific heat shows a pronounced peak indicating a thermal phase transition 
		which gets sharper upon increasing $N$. 
		Panel \textbf{(b)}: At finite temperature $T$ and inside the ordered phase a finite spin-chirality develops. 
		The $T$-dependence at low temperature is singular and approximately proportional to $T \log 1/\sqrt{T}$, see inset. 
		Inset: Fit to the analytical result \eqref{eq:chipmClassicalMainText} at low $T$ (dashed orange line).
		Panel \textbf{(c)}: By an order-by-disorder mechanism, the system selects a state with ferromagnetic spin order at low $T$. 
		At finite $T$ the order parameter is suppressed by thermal fluctuations linear in $T$. 
		The prefactor of the linear correction increases with system size, reflecting the suppression of long-ranged order by thermal 
		fluctuations, consistent with the Mermin-Wagner theorem \cite{Mermin-1966}. 
	}
	\label{Fig5:NumericsH0}
\end{figure}

  \begin{figure*}
	\centering
	\includegraphics[width=\linewidth]{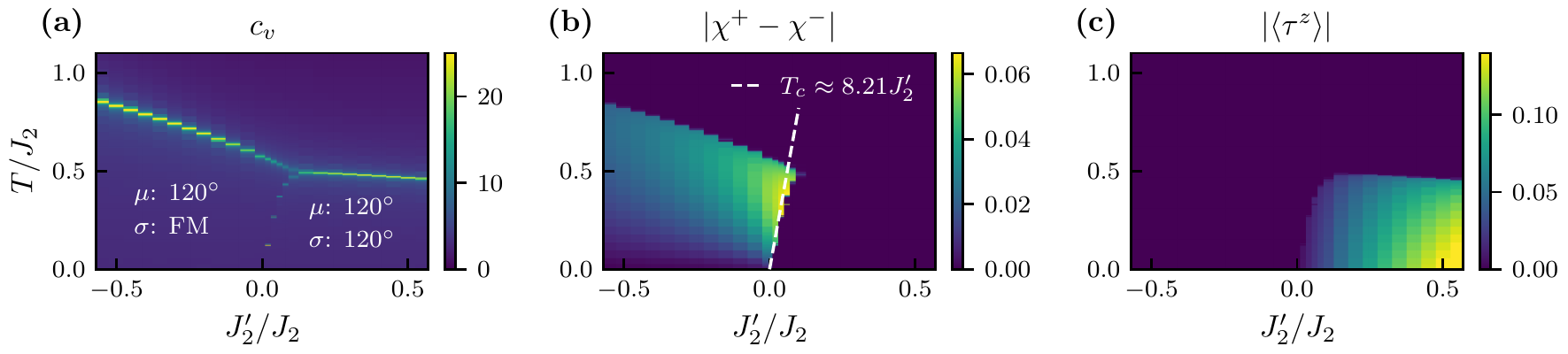}
	\caption{{\bf Phase diagram from semi-classical Monte Carlo calculations}. 
	Shown is Monte Carlo data for the specific heat (a), spin-chirality (b), and valley-magnetization (c),
	obtained from simulations of a model with $\varphi = J_3=0$ as function of temperature and $J_2'$ 
	for a fixed linear system size $N=36$.
	Panel (a): Specific heat. 
	At high temperatures one obtains a paramagnetic phase while at low $T$ one finds  two ordered phases. 
	All phase transitions appear to be of first-order type. For $J_2'<0$ one obtains 120$^\circ$ valley order coexisting with spin-ferromagnetic order, while for $J_2'>0$ a state with 120$^\circ$ valley order and coplanar 120$^\circ$ spin order in one of the two valleys is realized. 
	The transition temperature between the two phases follows the analytically estimated slope $T_c \approx 8.21 J_2^\prime$ (white dashed line in panel (b)). Panel (b): The spin-ferromagnetic order supports a finite spin-chirality $|\chi^+ - \chi^-|$ at finite $T$. 
	Panel (c): As the spins order coplanar in one valley but remain ferromagnetic in the other, $\tau_z$ develops a finite expectation value. 
}
	\label{Fig6:2dplotsphasediagram}
\end{figure*}

As the mean-field ground state of the $J_2$-only model $H_0$, Eq.~\eqref{eq:H0}, is degenerate, we focus our discussions on fluctuation effects around this state. The specific heat of the semi-classical model, Fig.~\ref{Fig5:NumericsH0}(a), shows a sharp peak 
 indicating a finite-temperature phase transition. The numerical data is both consistent with a weak first-order or a second-order transition, see supplementary material~\cite{Supple} which also discusses energy distributions at criticality.
We analyze two  types of order parameters, the  spin-chirality, Fig.~\ref{Fig5:NumericsH0}(b), and the valley-projected ferromagnetic order,  Fig.~\ref{Fig5:NumericsH0}(c), which show very different finite-size and temperature behavior as discussed below.

\begin{figure}[t]
	\centering
	\includegraphics[width= .9\linewidth]{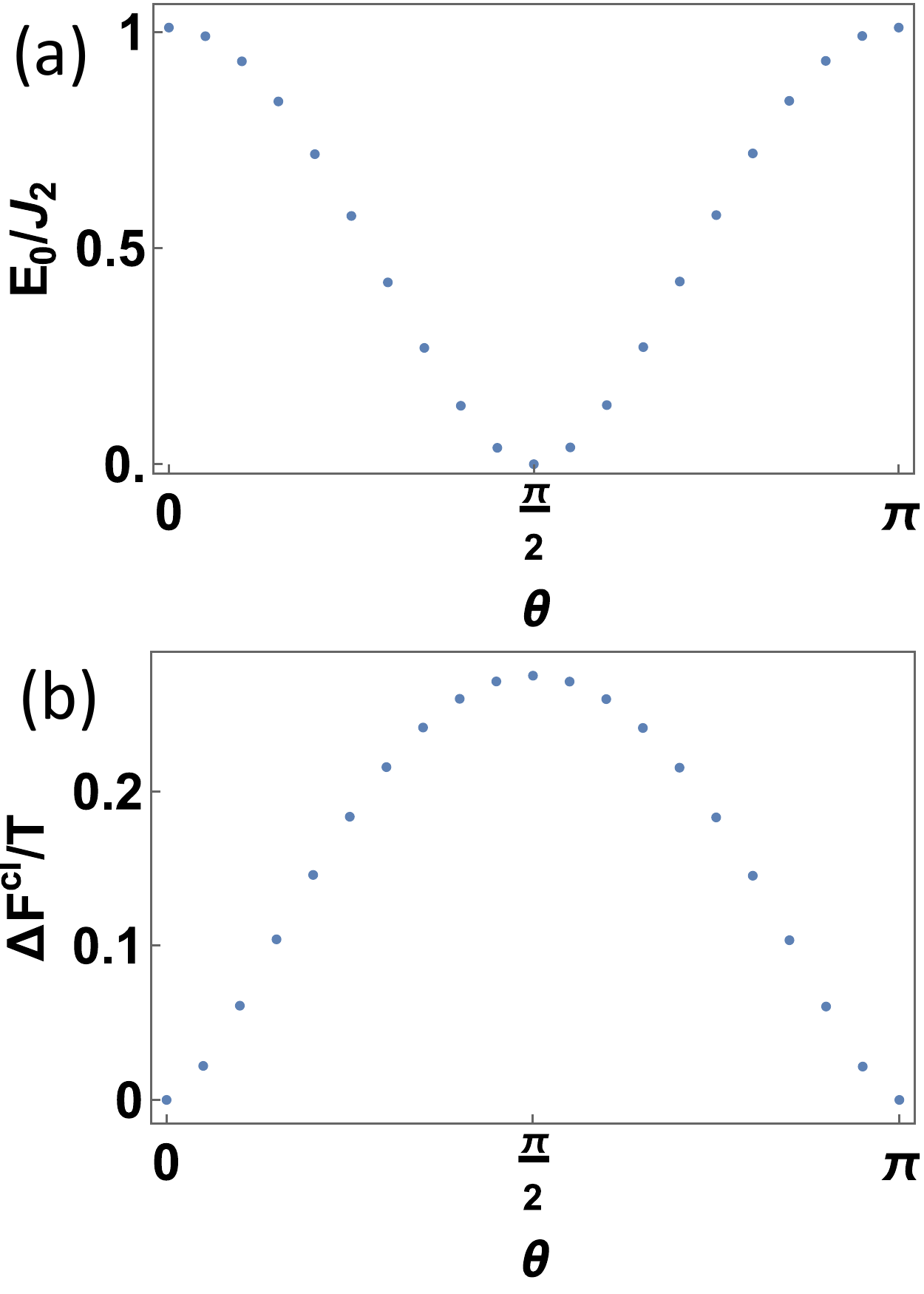}
	\caption{{\bf Order-by-disorder mechanisms.} 
 (a) Ground-state energy per site computed from $SU(4)$ spin-wave theory \cite{Supple} as function of $\theta$, parameterizing the opening angle of valley-projected spins in the three sublattices. 
 By the quantum order-by-disorder mechanism, a state with the an opening angle $\theta = \pi/2$ (i.e., the spin-coplanar 120 degree order in one of the two valleys) is selected from the mean-field ground-state manifold. (b) In the classical model, the free energy obtains at low $T$ a correction linear in $T$ from thermal fluctuations. In contrast with the quantum order-by-disorder mechanism, the thermal order-by-disorder leads to the selection of the spin ferromagnetic order in both of the valley sectors ($\theta = 0$ or $\pi$).}
	\label{fig:Orderbydisordermech}
\end{figure}

 For $T \to 0$, the spin-chirality, Fig.~\ref{Fig5:NumericsH0}(b), vanishes while the ferromagnetic magnetization in both valley sectors takes the value $1/2$, Fig.~\ref{Fig5:NumericsH0}(c). This shows that thermal fluctuations select the spin-ferromagnetic states, $\theta=0, \pi$ from the ground-state manifold. At the same time, the valley, more precisely $\vec \mu^{1,2}$, 
 exhibits 120$^\circ$ order (not shown). This `order-by-disorder'   selection \cite{Villain1980} of the classical ground state in the limit $T\to 0$ is also  found within our $SU(4)$ spin wave calculation, see Fig.~\ref{fig:Orderbydisordermech}(b) below and supplement \cite{Supple}: a fluctuation correction to the free energy linear in $T$ selects the ferromagnetic state.
 
 At finite $T$, the ferromagnetic order parameter shown in Fig.~\ref{Fig5:NumericsH0}(c) is suppressed linearly in $T$. The prefactor of this suppression increases with system size $N$. This is explained by an order-parameter suppression $\propto T \ln N$, well known from the Mermin-Wagner theorem \cite{Mermin-1966} in two spatial dimensions. Thus, there is nominally no long-ranged spin-order in the thermodynamic limit. 
 
 A remarkable result is that in the spin-ferromagnetic state the spin-chirality becomes finite at finite $T$, Fig.~\ref{Fig5:NumericsH0}(b), showing a highly singular $T$ dependence which is almost independent on system size $N$. 
 In the supplementary material \cite{Supple}, we use an $SU(4)$ spin-wave calculation to compute $\langle \hat{\chi}^+-\hat{\chi}^- \rangle$. The $SU(4)$ spin wave theory is formally derived using a $1/M$  expansion, where $M$ are the number of local bosons, $\sum_{\xi=1}^4 b^\dagger_{m,\xi}b_{m,\xi}=M$, used to describe the local  $SU(4)$ degree of freedom, see \cite{Supple}. For $M\to \infty$ one recovers mean-field and spin-waves are computed to leading order in $1/M$, where $M$ is set to its physical value, $M=1$ at the end of the calculation, corresponding to one localized electron per site. In the classical limit, at low-$T$ deviations from mean-field are small, which allows to make quantitative predictions based on spin wave theory.

 The naive spin-wave calculation in the classical limit predicts a divergent result reflecting the ground-state degeneracy of the $T=0$ state. This degeneracy is lifted by the order-by-disorder mechanism discussed above  which provides  a mass linear in $T$ to the chirality-mode. Taking this higher-order (in $1/M$ \cite{Supple}) effect into account we obtain
\begin{align} 
    \langle \hat{\chi}^+-\hat{\chi}^- \rangle \approx  \pm 0.22\, \frac{T}{J_2} \ln\left[\sqrt{T_0/T}\right] \,,
     \label{eq:chipmClassicalMainText}
\end{align}
in perfect agreement with the numerical data, see inset of Fig.~\ref{Fig5:NumericsH0}(b). The prefactor is fixed by our analytical results, see supplement \cite{Supple}, and the only fitting parameter is $T_0$.
We expect that such  a non-analytic $T$ dependence is generic for classical systems with a degenerate ground-state manifold where a ground state of the manifold is selected by thermal fluctuations.  Thus some `pseudo Goldstone modes' obtain masses linear in $T$, leading to non-analytic $T \log 1/T$ corrections in spatial dimension $d=2$ or a $c_1 T+ c_2 T^{3/2}$ correction in spatial dimension $d=3$ for observables coupling to the mode, see supplementary material \cite{Supple}.

The sign in Eq.~\eqref{eq:chipmClassicalMainText} is related to the spontaneous breaking of the $\mathbb Z_2$ symmetry, $e^{i \pi \tau_x/2}=i \tau_x$, which maps $\chi^+$ to $\chi^-$. Therefore,  $\langle \hat{\chi}^+-\hat{\chi}^- \rangle$ can be used, at $T>0$, as an Ising order parameter of this symmetry. 
The extremely sharp rise of $\langle \hat{\chi}^+-\hat{\chi}^- \rangle$ at the phase transition, see Fig.~\ref{Fig5:NumericsH0}(b), is both consistent with an Ising phase transition, $\langle \hat{\chi}^+-\hat{\chi}^- \rangle \sim (T_c-T)^{1/8}$, or a first-order transition, see supplement \cite{Supple}.
    
In Fig.~\ref{Fig6:2dplotsphasediagram} we show the phase diagram of the $J_2-J_2'$ model as a function of temperature $T$ and coupling $J_2'$. 
    At $T=0$, this simply reproduces the mean-field result. While for $J_2'<0$ a ferromagnetic spin-order coexists with a 120$^\circ$ valley order, one obtains a coplanar spin order in one of the two valley sectors for $J_2'>0$. Thus the valley symmetry is spontaneously broken in this phase, leading to a finite expectation value for $\tau_z$, see Fig.~\ref{Fig6:2dplotsphasediagram}(c).
Numerically, we find that the phase transition into the spin-coplanar phase at $J_2'>0$ both as a function of $T$ or $J_2'$ is always of first order; an analysis of the energy distribution is given in the supplement \cite{Supple}. At low $T$, the first-order phase transition separating the two ordered phases has a linear slope, $T_c \propto J_2'$. This arises because at $J_2'=0$ the spin-ferromagnetic state gains energy linear in $T$ due to the order-by-disorder mechanism described above. This linear-in-$T$ energy gain competes with a linear-in-$J_2'$ energy gain of the spin-coplanar phase, Fig.~\ref{fig:Orderbydisordermech}(b), which arises because $J_2'$  selects at $T=0$ one of the states from the ground-state manifold of $H_0$. Analytically, we obtain from this argument $T_c \approx 8.21 J_2'$, which quantitatively explains the numerically observed slope, as shown in Fig.~\ref{Fig6:2dplotsphasediagram}(b). As discussed above, the finite-$T$ transition
from the paramagnetic into spin ferromagnetic phase is accompanied by a $\mathbb Z_2$ symmetry breaking.

For all considered values of $J_2'$ the specific heat shows a low temperature saturation of $c_v(T\to 0) = 3$, indicating that both ordered states in Fig.~\ref{Fig6:2dplotsphasediagram} feature six harmonic modes \cite{Chalker-1992} per site as expected for an $SU(4)$ model locally described by 6 parameters as discussed above.

\section{Quantum fluctuations}

Above, we discussed the effect of thermal fluctuations and showed that at low $T$ an $SU(4)$ spin-wave calculation in the classical regime reproduces the main numerical findings qualitatively and quantitatively including the order-by-disorder mechanism and non-analytic $T$ dependences arising from pseudo Goldstone modes. 
While the $SU(4)$ spin-wave theory becomes exact in the classical case for $T\to 0$, this is not the case in the quantum model, where quantum fluctuations in the ground state can be large. $SU(4)$ spin wave theory only becomes exact in a large $M$ limit, see supplement \cite{Supple} but we expect that qualitative features of ordered phases (in contrast to spin-liquid phases) are well captured by this approach. 

In Fig.~\ref{fig:Orderbydisordermech}(a), we show the corrections due to quantum fluctuations to the ground-state energy of $H_0$, Eq.~\eqref{eq:H0}, as a function of the spin-opening angle $\theta$. The state with $\theta=\pi/2$, i.e., a coplanar 120$^\circ$ order of the spins in one of the valley sectors, is selected by quantum fluctuations. 
In contrast, as discussed above, thermal fluctuations select spin-ferromagnetic order ($\theta = 0$ or $\pi$). Thus, our system is one of the rare cases where quantum and classical fluctuations select very different types of ground states. As we show in the supplement, this arises, technically, because classically a state is selected where the {\em geometric} average of the excitation energies $E_{\vec{k}, n}$ is lowest, while quantum fluctuations select the state with the lowest {\em arithmetic} average of all $E_{\vec{k}, n}$. While in most systems the two averages show the same qualitative behavior, this is not the case in our system.

How will the quantum fluctuations modify the ground-state phase diagrams shown in Fig.~\ref{fig4:phasediagram}? The main effect of quantum fluctuations is that they break the degeneracy of mean-field ground state of $H_0$. As shown in Fig.~\ref{fig:Orderbydisordermech}(a), the ground-state energy obtains a $\theta$ dependence. An almost identical $\theta$ dependence can be obtained in the purely classical model by adding a $J_2'$ to the Hamiltonian with $J_2' \approx 0.45\,J_2$.
Thus, we speculate that the quantum fluctuations have a similar effect on the phase diagram as increasing $J_2'$ within mean-field theory. This procedure is well-controlled in an $1/M$ expansion, see supplement \cite{Supple}:
for large $M$, quantum corrections of order $1/M$ can be fully compensated by a shift of $J_2'$ by $-0.45\,J_2/M$ (up to corrections of order $1/M^2$).

Thus, we expect that the main effect of quantum fluctuations will be that in Fig.~\ref{fig4:phasediagram}(b) and (c) the phase boundaries are shifted along the $y$ direction, most likely accompanied by a rounding of the sharp kink where the phases meet. This extra rounding  would be a $1/M^2$ effect, which is more difficult to calculate. 


\section{Conclusions}

Our study reveals that one of the  simplest feasible moir\'e  systems, a single layer of graphene on a substrate, can exhibit surprisingly rich physics. 
Ultra-flat bands generating localized modes coexist with a network of chiral one-dimensional channels where electrons move very fast with a speed set by the Fermi velocity of graphene. 
A main advantage of such large-unit-cell system is that one can tune the electron density by external gates. 

Different types of localized modes with spin-, valley- and orbital degrees of freedom can be realized depending on how many electrons are loaded into the local level and the quantum numbers of the localized states, fixed by the representation of the relevant dicyclic group. 
The chiral nature of the channels connecting the localized modes gives rise to characteristic chiral- and non-chiral interactions.
We expect that a wealth of different phases 
with commensurate and incommensurate spin-, valley- and orbital order can be realized.

As an example, we studied one such model, focusing on commensurate order stabilized by two- and three spin interactions. We use mean-field theory, an $SU(4)$ spin-wave theory both in the classical and quantum regime, and Monte Carlo simulations of a semi-classical model. By tuning gate voltages one can control $k_F$ and thus the effective interactions. One can, for example, tune parameters along the yellow line shown in Fig.~\ref{fig4:phasediagram}b. 
This triggers a transition from a coplanar phase with ferromagnetic spin and 120$^\circ$ valley order into a non-coplanar phase characterized by a peculiar coexistence of three different types of order: ferromagnetic spin order in one valley, non-coplanar chiral spin order in the other valley, and 120$^\circ$ order in remaining spin and valley-mixed degrees of freedom. 

The peculiar form of the mean-field phase diagrams, where tiny perturbations can profoundly change the ground state, is governed by the proximity to a variant of the Kugel-Khomskii model, $H_0$, where the mean-field ground state is highly degenerate. For example, the tiniest chiral interactions arising from 3-spin interactions mediated by the chiral electronic channels, induce a state with a huge spin-chirality in one of the valleys.
The degeneracy of the mean-field ground state of $H_0$ is, however, lifted by quantum and thermal fluctuations. While in most systems, quantum and thermal fluctuations stabilize the same type of order by such an order-by-disorder mechanism, this is not the case in our model where quantum fluctuations prefer coplanar spin-order, while classical fluctuations favor ferromagnetic spin order on top of a 120$^\circ$ valley order. The ferromagnetic spin order is, however, highly unconventional.
Due to the coupling of spin- and valley degrees of freedom, quantum or thermal fluctuations around the spin-ferromagnetic state are \emph{always}  chiral with a finite spin-chirality. 
In the classical limit, this fluctuation effect
is enhanced, Eq~\eqref{eq:chipmClassicalMainText}, due to the coupling to a pseudo Goldstone mode characteristic for the classical order-by-disorder mechanism.

Our results on quantum fluctuations are based on a spin-wave calculation, which formally becomes exact in a large $M$ limit.  As $M=1$, this result remains speculative. An alternative scenario is that for $M=1$ quantum fluctuations around the highly degenerate mean-field state induce a spin-valley entangled quantum liquid. It would be interesting to test these very different scenarios in future numerical studies.
Furthermore, we expect that the system will host many more commensurate and incommensurate phases and, potentially, quantum liquids, when other localized modes and different electronic fillings are considered.

\begin{acknowledgments}

We thank Martin Zirnbauer, Guo-Yi Zhu, Ciarán Hickey, Shahal Ilani, Nick Bultinck, Johannes Hofmann, Peter Cha and Hongki Min for useful discussions. Financial support of the Deutsche Forschungsgemeinschaft (DFG, German Research Foundation) within CRC1238 (project number 277146847, C02 and C04) 
and CRC183 (project number 277101999, A01 and A04) is acknowledged. Jeyong Park also acknowledges BCGS (Bonn-Cologne Graduate School) and ML4Q (Matter and Light for Quantum Computing) for support.
The numerical simulations were performed on the Noctua 2 cluster at the Paderborn Center for Parallel Computing (PC2) and the CHEOPS cluster at RRZK Cologne. 
\end{acknowledgments}

\vspace{5mm}
{\em Note added:} 
Recently, a preprint by Wittig {\it et al.} was published on ArXiv~\cite{wittig2023} that also considers localized states coupled to a network of chiral modes in the twisted bilayer graphene subject to an interlayer bias.

\bibliographystyle{apsrev4-1-titles}
\bibliography{MoirePotentialModel}

\begin{thebibliography}{68}%
\makeatletter
\providecommand \@ifxundefined [1]{%
 \@ifx{#1\undefined}
}%
\providecommand \@ifnum [1]{%
 \ifnum #1\expandafter \@firstoftwo
 \else \expandafter \@secondoftwo
 \fi
}%
\providecommand \@ifx [1]{%
 \ifx #1\expandafter \@firstoftwo
 \else \expandafter \@secondoftwo
 \fi
}%
\providecommand \natexlab [1]{#1}%
\providecommand \enquote  [1]{``#1''}%
\providecommand \bibnamefont  [1]{#1}%
\providecommand \bibfnamefont [1]{#1}%
\providecommand \citenamefont [1]{#1}%
\providecommand \href@noop [0]{\@secondoftwo}%
\providecommand \href [0]{\begingroup \@sanitize@url \@href}%
\providecommand \@href[1]{\@@startlink{#1}\@@href}%
\providecommand \@@href[1]{\endgroup#1\@@endlink}%
\providecommand \@sanitize@url [0]{\catcode `\\12\catcode `\$12\catcode
  `\&12\catcode `\#12\catcode `\^12\catcode `\_12\catcode `\%12\relax}%
\providecommand \@@startlink[1]{}%
\providecommand \@@endlink[0]{}%
\providecommand \url  [0]{\begingroup\@sanitize@url \@url }%
\providecommand \@url [1]{\endgroup\@href {#1}{\urlprefix }}%
\providecommand \urlprefix  [0]{URL }%
\providecommand \Eprint [0]{\href }%
\providecommand \doibase [0]{http://doi.org/}%
\providecommand \selectlanguage [0]{\@gobble}%
\providecommand \bibinfo  [0]{\@secondoftwo}%
\providecommand \bibfield  [0]{\@secondoftwo}%
\providecommand \translation [1]{[#1]}%
\providecommand \BibitemOpen [0]{}%
\providecommand \bibitemStop [0]{}%
\providecommand \bibitemNoStop [0]{.\EOS\space}%
\providecommand \EOS [0]{\spacefactor3000\relax}%
\providecommand \BibitemShut  [1]{\csname bibitem#1\endcsname}%
\let\auto@bib@innerbib\@empty
\bibitem [{\citenamefont {Geim}\ and\ \citenamefont
  {Grigorieva}(2013)}]{Geim2013}%
  \BibitemOpen
  \bibfield  {author} {\bibinfo {author} {\bibfnamefont {A.~K.}\ \bibnamefont
  {Geim}}\ and\ \bibinfo {author} {\bibfnamefont {I.~V.}\ \bibnamefont
  {Grigorieva}},\ }\bibfield  {title} {\emph {\bibinfo {title} {{Van der Waals
  heterostructures}}},\ }\href {https://www.nature.com/articles/nature12385}
  {\bibfield  {journal} {\bibinfo  {journal} {Nature}\ }\textbf {\bibinfo
  {volume} {499}},\ \bibinfo {pages} {419} (\bibinfo {year}
  {2013})}\BibitemShut {NoStop}%
\bibitem [{\citenamefont {Andrei}\ \emph {et~al.}(2021)\citenamefont {Andrei},
  \citenamefont {Efetov}, \citenamefont {Jarillo-Herrero}, \citenamefont
  {MacDonald}, \citenamefont {Mak}, \citenamefont {Senthil}, \citenamefont
  {Tutuc}, \citenamefont {Yazdani},\ and\ \citenamefont {Young}}]{Andrei2021}%
  \BibitemOpen
  \bibfield  {author} {\bibinfo {author} {\bibfnamefont {E.~Y.}\ \bibnamefont
  {Andrei}}, \bibinfo {author} {\bibfnamefont {D.~K.}\ \bibnamefont {Efetov}},
  \bibinfo {author} {\bibfnamefont {P.}~\bibnamefont {Jarillo-Herrero}},
  \bibinfo {author} {\bibfnamefont {A.~H.}\ \bibnamefont {MacDonald}}, \bibinfo
  {author} {\bibfnamefont {K.~F.}\ \bibnamefont {Mak}}, \bibinfo {author}
  {\bibfnamefont {T.}~\bibnamefont {Senthil}}, \bibinfo {author} {\bibfnamefont
  {E.}~\bibnamefont {Tutuc}}, \bibinfo {author} {\bibfnamefont
  {A.}~\bibnamefont {Yazdani}}, \ and\ \bibinfo {author} {\bibfnamefont
  {A.~F.}\ \bibnamefont {Young}},\ }\bibfield  {title} {\emph {\bibinfo {title}
  {{The marvels of moir{\'{e}} materials}}},\ }\href
  {https://www.nature.com/articles/s41578-021-00284-1} {\bibfield  {journal}
  {\bibinfo  {journal} {Nature Reviews Materials}\ }\textbf {\bibinfo {volume}
  {6}},\ \bibinfo {pages} {201} (\bibinfo {year} {2021})}\BibitemShut {NoStop}%
\bibitem [{\citenamefont {Chen}\ \emph {et~al.}(2019)\citenamefont {Chen},
  \citenamefont {Sharpe}, \citenamefont {Gallagher}, \citenamefont {Rosen},
  \citenamefont {Fox}, \citenamefont {Jiang}, \citenamefont {Lyu},
  \citenamefont {Li}, \citenamefont {Watanabe}, \citenamefont {Taniguchi},
  \citenamefont {Jung}, \citenamefont {Shi}, \citenamefont {Goldhaber-Gordon},
  \citenamefont {Zhang},\ and\ \citenamefont {Wang}}]{Chen2019}%
  \BibitemOpen
  \bibfield  {author} {\bibinfo {author} {\bibfnamefont {G.}~\bibnamefont
  {Chen}}, \bibinfo {author} {\bibfnamefont {A.~L.}\ \bibnamefont {Sharpe}},
  \bibinfo {author} {\bibfnamefont {P.}~\bibnamefont {Gallagher}}, \bibinfo
  {author} {\bibfnamefont {I.~T.}\ \bibnamefont {Rosen}}, \bibinfo {author}
  {\bibfnamefont {E.~J.}\ \bibnamefont {Fox}}, \bibinfo {author} {\bibfnamefont
  {L.}~\bibnamefont {Jiang}}, \bibinfo {author} {\bibfnamefont
  {B.}~\bibnamefont {Lyu}}, \bibinfo {author} {\bibfnamefont {H.}~\bibnamefont
  {Li}}, \bibinfo {author} {\bibfnamefont {K.}~\bibnamefont {Watanabe}},
  \bibinfo {author} {\bibfnamefont {T.}~\bibnamefont {Taniguchi}}, \bibinfo
  {author} {\bibfnamefont {J.}~\bibnamefont {Jung}}, \bibinfo {author}
  {\bibfnamefont {Z.}~\bibnamefont {Shi}}, \bibinfo {author} {\bibfnamefont
  {D.}~\bibnamefont {Goldhaber-Gordon}}, \bibinfo {author} {\bibfnamefont
  {Y.}~\bibnamefont {Zhang}}, \ and\ \bibinfo {author} {\bibfnamefont
  {F.}~\bibnamefont {Wang}},\ }\bibfield  {title} {\emph {\bibinfo {title}
  {{Signatures of tunable superconductivity in a trilayer graphene moir{\'{e}}
  superlattice}}},\ }\href {https://www.nature.com/articles/s41586-019-1393-y}
  {\bibfield  {journal} {\bibinfo  {journal} {Nature}\ }\textbf {\bibinfo
  {volume} {572}},\ \bibinfo {pages} {215} (\bibinfo {year}
  {2019})}\BibitemShut {NoStop}%
\bibitem [{\citenamefont {Xie}\ \emph {et~al.}(2022)\citenamefont {Xie},
  \citenamefont {Zhang}, \citenamefont {Hu}, \citenamefont {Mak},\ and\
  \citenamefont {Law}}]{Xie2022}%
  \BibitemOpen
  \bibfield  {author} {\bibinfo {author} {\bibfnamefont {Y.-M.}\ \bibnamefont
  {Xie}}, \bibinfo {author} {\bibfnamefont {C.-P.}\ \bibnamefont {Zhang}},
  \bibinfo {author} {\bibfnamefont {J.-X.}\ \bibnamefont {Hu}}, \bibinfo
  {author} {\bibfnamefont {K.~F.}\ \bibnamefont {Mak}}, \ and\ \bibinfo
  {author} {\bibfnamefont {K.~T.}\ \bibnamefont {Law}},\ }\bibfield  {title}
  {\emph {\bibinfo {title} {{Valley-Polarized Quantum Anomalous Hall State in
  Moir\'e ${\mathrm{MoTe}}_{2}/{\mathrm{WSe}}_{2}$ Heterobilayers}}},\ }\href
  {https://link.aps.org/doi/10.1103/PhysRevLett.128.026402} {\bibfield
  {journal} {\bibinfo  {journal} {Phys. Rev. Lett.}\ }\textbf {\bibinfo
  {volume} {128}},\ \bibinfo {pages} {026402} (\bibinfo {year}
  {2022})}\BibitemShut {NoStop}%
\bibitem [{\citenamefont {Jin}\ \emph {et~al.}(2021)\citenamefont {Jin},
  \citenamefont {Tao}, \citenamefont {Li}, \citenamefont {Xu}, \citenamefont
  {Tang}, \citenamefont {Zhu}, \citenamefont {Liu}, \citenamefont {Watanabe},
  \citenamefont {Taniguchi}, \citenamefont {Hone}, \citenamefont {Fu},
  \citenamefont {Shan},\ and\ \citenamefont {Mak}}]{Jin2021}%
  \BibitemOpen
  \bibfield  {author} {\bibinfo {author} {\bibfnamefont {C.}~\bibnamefont
  {Jin}}, \bibinfo {author} {\bibfnamefont {Z.}~\bibnamefont {Tao}}, \bibinfo
  {author} {\bibfnamefont {T.}~\bibnamefont {Li}}, \bibinfo {author}
  {\bibfnamefont {Y.}~\bibnamefont {Xu}}, \bibinfo {author} {\bibfnamefont
  {Y.}~\bibnamefont {Tang}}, \bibinfo {author} {\bibfnamefont {J.}~\bibnamefont
  {Zhu}}, \bibinfo {author} {\bibfnamefont {S.}~\bibnamefont {Liu}}, \bibinfo
  {author} {\bibfnamefont {K.}~\bibnamefont {Watanabe}}, \bibinfo {author}
  {\bibfnamefont {T.}~\bibnamefont {Taniguchi}}, \bibinfo {author}
  {\bibfnamefont {J.~C.}\ \bibnamefont {Hone}}, \bibinfo {author}
  {\bibfnamefont {L.}~\bibnamefont {Fu}}, \bibinfo {author} {\bibfnamefont
  {J.}~\bibnamefont {Shan}}, \ and\ \bibinfo {author} {\bibfnamefont {K.~F.}\
  \bibnamefont {Mak}},\ }\bibfield  {title} {\emph {\bibinfo {title} {{Stripe
  phases in WSe2/WS2 moiré superlattices}}},\ }\href
  {https://www.nature.com/articles/s41563-021-00959-8} {\bibfield  {journal}
  {\bibinfo  {journal} {Nature Materials}\ }\textbf {\bibinfo {volume} {20}},\
  \bibinfo {pages} {940} (\bibinfo {year} {2021})}\BibitemShut {NoStop}%
\bibitem [{\citenamefont {Xu}\ \emph {et~al.}(2020)\citenamefont {Xu},
  \citenamefont {Liu}, \citenamefont {Rhodes}, \citenamefont {Watanabe},
  \citenamefont {Taniguchi}, \citenamefont {Hone}, \citenamefont {Elser},
  \citenamefont {Mak},\ and\ \citenamefont {Shan}}]{Xu2020}%
  \BibitemOpen
  \bibfield  {author} {\bibinfo {author} {\bibfnamefont {Y.}~\bibnamefont
  {Xu}}, \bibinfo {author} {\bibfnamefont {S.}~\bibnamefont {Liu}}, \bibinfo
  {author} {\bibfnamefont {D.~A.}\ \bibnamefont {Rhodes}}, \bibinfo {author}
  {\bibfnamefont {K.}~\bibnamefont {Watanabe}}, \bibinfo {author}
  {\bibfnamefont {T.}~\bibnamefont {Taniguchi}}, \bibinfo {author}
  {\bibfnamefont {J.}~\bibnamefont {Hone}}, \bibinfo {author} {\bibfnamefont
  {V.}~\bibnamefont {Elser}}, \bibinfo {author} {\bibfnamefont {K.~F.}\
  \bibnamefont {Mak}}, \ and\ \bibinfo {author} {\bibfnamefont
  {J.}~\bibnamefont {Shan}},\ }\bibfield  {title} {\emph {\bibinfo {title}
  {{Correlated insulating states at fractional fillings of moiré
  superlattices}}},\ }\href {https://www.nature.com/articles/s41586-020-2868-6}
  {\bibfield  {journal} {\bibinfo  {journal} {Nature}\ }\textbf {\bibinfo
  {volume} {587}},\ \bibinfo {pages} {214} (\bibinfo {year}
  {2020})}\BibitemShut {NoStop}%
\bibitem [{\citenamefont {Li}\ \emph {et~al.}(2021)\citenamefont {Li},
  \citenamefont {Jiang}, \citenamefont {Shen}, \citenamefont {Zhang},
  \citenamefont {Li}, \citenamefont {Tao}, \citenamefont {Devakul},
  \citenamefont {Watanabe}, \citenamefont {Taniguchi}, \citenamefont {Fu},
  \citenamefont {Shan},\ and\ \citenamefont {Mak}}]{Li2021}%
  \BibitemOpen
  \bibfield  {author} {\bibinfo {author} {\bibfnamefont {T.}~\bibnamefont
  {Li}}, \bibinfo {author} {\bibfnamefont {S.}~\bibnamefont {Jiang}}, \bibinfo
  {author} {\bibfnamefont {B.}~\bibnamefont {Shen}}, \bibinfo {author}
  {\bibfnamefont {Y.}~\bibnamefont {Zhang}}, \bibinfo {author} {\bibfnamefont
  {L.}~\bibnamefont {Li}}, \bibinfo {author} {\bibfnamefont {Z.}~\bibnamefont
  {Tao}}, \bibinfo {author} {\bibfnamefont {T.}~\bibnamefont {Devakul}},
  \bibinfo {author} {\bibfnamefont {K.}~\bibnamefont {Watanabe}}, \bibinfo
  {author} {\bibfnamefont {T.}~\bibnamefont {Taniguchi}}, \bibinfo {author}
  {\bibfnamefont {L.}~\bibnamefont {Fu}}, \bibinfo {author} {\bibfnamefont
  {J.}~\bibnamefont {Shan}}, \ and\ \bibinfo {author} {\bibfnamefont {K.~F.}\
  \bibnamefont {Mak}},\ }\bibfield  {title} {\emph {\bibinfo {title} {{Quantum
  anomalous Hall effect from intertwined moiré bands}}},\ }\href
  {https://www.nature.com/articles/s41586-021-04171-1} {\bibfield  {journal}
  {\bibinfo  {journal} {Nature}\ }\textbf {\bibinfo {volume} {600}},\ \bibinfo
  {pages} {641} (\bibinfo {year} {2021})}\BibitemShut {NoStop}%
\bibitem [{\citenamefont {Cao}\ \emph {et~al.}(2018{\natexlab{a}})\citenamefont
  {Cao}, \citenamefont {Fatemi}, \citenamefont {Demir}, \citenamefont {Fang},
  \citenamefont {Tomarken}, \citenamefont {Luo}, \citenamefont
  {Sanchez-Yamagishi}, \citenamefont {Watanabe}, \citenamefont {Taniguchi},
  \citenamefont {Kaxiras}, \citenamefont {Ashoori},\ and\ \citenamefont
  {Jarillo-Herrero}}]{Cao2018a}%
  \BibitemOpen
  \bibfield  {author} {\bibinfo {author} {\bibfnamefont {Y.}~\bibnamefont
  {Cao}}, \bibinfo {author} {\bibfnamefont {V.}~\bibnamefont {Fatemi}},
  \bibinfo {author} {\bibfnamefont {A.}~\bibnamefont {Demir}}, \bibinfo
  {author} {\bibfnamefont {S.}~\bibnamefont {Fang}}, \bibinfo {author}
  {\bibfnamefont {S.~L.}\ \bibnamefont {Tomarken}}, \bibinfo {author}
  {\bibfnamefont {J.~Y.}\ \bibnamefont {Luo}}, \bibinfo {author} {\bibfnamefont
  {J.~D.}\ \bibnamefont {Sanchez-Yamagishi}}, \bibinfo {author} {\bibfnamefont
  {K.}~\bibnamefont {Watanabe}}, \bibinfo {author} {\bibfnamefont
  {T.}~\bibnamefont {Taniguchi}}, \bibinfo {author} {\bibfnamefont
  {E.}~\bibnamefont {Kaxiras}}, \bibinfo {author} {\bibfnamefont {R.~C.}\
  \bibnamefont {Ashoori}}, \ and\ \bibinfo {author} {\bibfnamefont
  {P.}~\bibnamefont {Jarillo-Herrero}},\ }\bibfield  {title} {\emph {\bibinfo
  {title} {{Correlated insulator behaviour at half-filling in magic-angle
  graphene superlattices}}},\ }\href
  {https://www.nature.com/articles/nature26154} {\bibfield  {journal} {\bibinfo
   {journal} {Nature}\ }\textbf {\bibinfo {volume} {556}},\ \bibinfo {pages}
  {80} (\bibinfo {year} {2018}{\natexlab{a}})}\BibitemShut {NoStop}%
\bibitem [{\citenamefont {Cao}\ \emph {et~al.}(2018{\natexlab{b}})\citenamefont
  {Cao}, \citenamefont {Fatemi}, \citenamefont {Fang}, \citenamefont
  {Watanabe}, \citenamefont {Taniguchi}, \citenamefont {Kaxiras},\ and\
  \citenamefont {Jarillo-Herrero}}]{Cao2018}%
  \BibitemOpen
  \bibfield  {author} {\bibinfo {author} {\bibfnamefont {Y.}~\bibnamefont
  {Cao}}, \bibinfo {author} {\bibfnamefont {V.}~\bibnamefont {Fatemi}},
  \bibinfo {author} {\bibfnamefont {S.}~\bibnamefont {Fang}}, \bibinfo {author}
  {\bibfnamefont {K.}~\bibnamefont {Watanabe}}, \bibinfo {author}
  {\bibfnamefont {T.}~\bibnamefont {Taniguchi}}, \bibinfo {author}
  {\bibfnamefont {E.}~\bibnamefont {Kaxiras}}, \ and\ \bibinfo {author}
  {\bibfnamefont {P.}~\bibnamefont {Jarillo-Herrero}},\ }\bibfield  {title}
  {\emph {\bibinfo {title} {{Unconventional superconductivity in magic-angle
  graphene superlattices}}},\ }\href
  {https://www.nature.com/articles/nature26160} {\bibfield  {journal} {\bibinfo
   {journal} {Nature}\ }\textbf {\bibinfo {volume} {556}},\ \bibinfo {pages}
  {43} (\bibinfo {year} {2018}{\natexlab{b}})}\BibitemShut {NoStop}%
\bibitem [{\citenamefont {Bistritzer}\ and\ \citenamefont
  {MacDonald}(2011)}]{Bistritzer2011}%
  \BibitemOpen
  \bibfield  {author} {\bibinfo {author} {\bibfnamefont {R.}~\bibnamefont
  {Bistritzer}}\ and\ \bibinfo {author} {\bibfnamefont {A.~H.}\ \bibnamefont
  {MacDonald}},\ }\bibfield  {title} {\emph {\bibinfo {title} {{Moir{\'{e}}
  bands in twisted double-layer graphene}}},\ }\href
  {https://www.pnas.org/doi/abs/10.1073/pnas.1108174108} {\bibfield  {journal}
  {\bibinfo  {journal} {Proceedings of the National Academy of Sciences of the
  United States of America}\ }\textbf {\bibinfo {volume} {108}},\ \bibinfo
  {pages} {12233} (\bibinfo {year} {2011})}\BibitemShut {NoStop}%
\bibitem [{\citenamefont {Tarnopolsky}\ \emph {et~al.}(2019)\citenamefont
  {Tarnopolsky}, \citenamefont {Kruchkov},\ and\ \citenamefont
  {Vishwanath}}]{tarnopolsky2019origin}%
  \BibitemOpen
  \bibfield  {author} {\bibinfo {author} {\bibfnamefont {G.}~\bibnamefont
  {Tarnopolsky}}, \bibinfo {author} {\bibfnamefont {A.~J.}\ \bibnamefont
  {Kruchkov}}, \ and\ \bibinfo {author} {\bibfnamefont {A.}~\bibnamefont
  {Vishwanath}},\ }\bibfield  {title} {\emph {\bibinfo {title} {{Origin of
  Magic Angles in Twisted Bilayer Graphene}}},\ }\href {\doibase
  10.1103/PhysRevLett.122.106405} {\bibfield  {journal} {\bibinfo  {journal}
  {Phys. Rev. Lett.}\ }\textbf {\bibinfo {volume} {122}},\ \bibinfo {pages}
  {106405} (\bibinfo {year} {2019})}\BibitemShut {NoStop}%
\bibitem [{\citenamefont {Po}\ \emph {et~al.}(2018)\citenamefont {Po},
  \citenamefont {Zou}, \citenamefont {Vishwanath},\ and\ \citenamefont
  {Senthil}}]{po2018origin}%
  \BibitemOpen
  \bibfield  {author} {\bibinfo {author} {\bibfnamefont {H.~C.}\ \bibnamefont
  {Po}}, \bibinfo {author} {\bibfnamefont {L.}~\bibnamefont {Zou}}, \bibinfo
  {author} {\bibfnamefont {A.}~\bibnamefont {Vishwanath}}, \ and\ \bibinfo
  {author} {\bibfnamefont {T.}~\bibnamefont {Senthil}},\ }\bibfield  {title}
  {\emph {\bibinfo {title} {{Origin of Mott Insulating Behavior and
  Superconductivity in Twisted Bilayer Graphene}}},\ }\href
  {https://link.aps.org/doi/10.1103/PhysRevX.8.031089} {\bibfield  {journal}
  {\bibinfo  {journal} {Phys. Rev. X}\ }\textbf {\bibinfo {volume} {8}},\
  \bibinfo {pages} {031089} (\bibinfo {year} {2018})}\BibitemShut {NoStop}%
\bibitem [{\citenamefont {Kennes}\ \emph {et~al.}(2018)\citenamefont {Kennes},
  \citenamefont {Lischner},\ and\ \citenamefont {Karrasch}}]{Kennes2018}%
  \BibitemOpen
  \bibfield  {author} {\bibinfo {author} {\bibfnamefont {D.~M.}\ \bibnamefont
  {Kennes}}, \bibinfo {author} {\bibfnamefont {J.}~\bibnamefont {Lischner}}, \
  and\ \bibinfo {author} {\bibfnamefont {C.}~\bibnamefont {Karrasch}},\
  }\bibfield  {title} {\emph {\bibinfo {title} {Strong correlations and
  $d+\mathit{id}$ superconductivity in twisted bilayer graphene}},\ }\href
  {https://link.aps.org/doi/10.1103/PhysRevB.98.241407} {\bibfield  {journal}
  {\bibinfo  {journal} {Phys. Rev. B}\ }\textbf {\bibinfo {volume} {98}},\
  \bibinfo {pages} {241407} (\bibinfo {year} {2018})}\BibitemShut {NoStop}%
\bibitem [{\citenamefont {Wu}\ \emph {et~al.}(2018)\citenamefont {Wu},
  \citenamefont {MacDonald},\ and\ \citenamefont {Martin}}]{Wu2018}%
  \BibitemOpen
  \bibfield  {author} {\bibinfo {author} {\bibfnamefont {F.}~\bibnamefont
  {Wu}}, \bibinfo {author} {\bibfnamefont {A.~H.}\ \bibnamefont {MacDonald}}, \
  and\ \bibinfo {author} {\bibfnamefont {I.}~\bibnamefont {Martin}},\
  }\bibfield  {title} {\emph {\bibinfo {title} {{Theory of Phonon-Mediated
  Superconductivity in Twisted Bilayer Graphene}}},\ }\href
  {https://link.aps.org/doi/10.1103/PhysRevLett.121.257001} {\bibfield
  {journal} {\bibinfo  {journal} {Phys. Rev. Lett.}\ }\textbf {\bibinfo
  {volume} {121}},\ \bibinfo {pages} {257001} (\bibinfo {year}
  {2018})}\BibitemShut {NoStop}%
\bibitem [{\citenamefont {Lian}\ \emph {et~al.}(2019)\citenamefont {Lian},
  \citenamefont {Wang},\ and\ \citenamefont {Bernevig}}]{Lian2019}%
  \BibitemOpen
  \bibfield  {author} {\bibinfo {author} {\bibfnamefont {B.}~\bibnamefont
  {Lian}}, \bibinfo {author} {\bibfnamefont {Z.}~\bibnamefont {Wang}}, \ and\
  \bibinfo {author} {\bibfnamefont {B.~A.}\ \bibnamefont {Bernevig}},\
  }\bibfield  {title} {\emph {\bibinfo {title} {{Twisted Bilayer Graphene: A
  Phonon-Driven Superconductor}}},\ }\href
  {https://link.aps.org/doi/10.1103/PhysRevLett.122.257002} {\bibfield
  {journal} {\bibinfo  {journal} {Phys. Rev. Lett.}\ }\textbf {\bibinfo
  {volume} {122}},\ \bibinfo {pages} {257002} (\bibinfo {year}
  {2019})}\BibitemShut {NoStop}%
\bibitem [{\citenamefont {Roy}\ and\ \citenamefont
  {Juri{\v{c}}i{\'{c}}}(2019)}]{Roy2019}%
  \BibitemOpen
  \bibfield  {author} {\bibinfo {author} {\bibfnamefont {B.}~\bibnamefont
  {Roy}}\ and\ \bibinfo {author} {\bibfnamefont {V.}~\bibnamefont
  {Juri{\v{c}}i{\'{c}}}},\ }\bibfield  {title} {\emph {\bibinfo {title}
  {{Unconventional superconductivity in nearly flat bands in twisted bilayer
  graphene}}},\ }\href
  {https://journals.aps.org/prb/abstract/10.1103/PhysRevB.99.121407} {\bibfield
   {journal} {\bibinfo  {journal} {Phys. Rev. B}\ }\textbf {\bibinfo {volume}
  {99}},\ \bibinfo {pages} {121407} (\bibinfo {year} {2019})}\BibitemShut
  {NoStop}%
\bibitem [{\citenamefont {Yankowitz}\ \emph {et~al.}(2019)\citenamefont
  {Yankowitz}, \citenamefont {Chen}, \citenamefont {Polshyn}, \citenamefont
  {Zhang}, \citenamefont {Watanabe}, \citenamefont {Taniguchi}, \citenamefont
  {Graf}, \citenamefont {Young},\ and\ \citenamefont {Dean}}]{Yankowitz2019}%
  \BibitemOpen
  \bibfield  {author} {\bibinfo {author} {\bibfnamefont {M.}~\bibnamefont
  {Yankowitz}}, \bibinfo {author} {\bibfnamefont {S.}~\bibnamefont {Chen}},
  \bibinfo {author} {\bibfnamefont {H.}~\bibnamefont {Polshyn}}, \bibinfo
  {author} {\bibfnamefont {Y.}~\bibnamefont {Zhang}}, \bibinfo {author}
  {\bibfnamefont {K.}~\bibnamefont {Watanabe}}, \bibinfo {author}
  {\bibfnamefont {T.}~\bibnamefont {Taniguchi}}, \bibinfo {author}
  {\bibfnamefont {D.}~\bibnamefont {Graf}}, \bibinfo {author} {\bibfnamefont
  {A.~F.}\ \bibnamefont {Young}}, \ and\ \bibinfo {author} {\bibfnamefont
  {C.~R.}\ \bibnamefont {Dean}},\ }\bibfield  {title} {\emph {\bibinfo {title}
  {{Tuning superconductivity in twisted bilayer graphene}}},\ }\href
  {https://www.science.org/doi/10.1126/science.aav1910} {\bibfield  {journal}
  {\bibinfo  {journal} {Science}\ }\textbf {\bibinfo {volume} {363}},\ \bibinfo
  {pages} {1059} (\bibinfo {year} {2019})}\BibitemShut {NoStop}%
\bibitem [{\citenamefont {Isobe}\ \emph {et~al.}(2018)\citenamefont {Isobe},
  \citenamefont {Yuan},\ and\ \citenamefont {Fu}}]{Isobe2018}%
  \BibitemOpen
  \bibfield  {author} {\bibinfo {author} {\bibfnamefont {H.}~\bibnamefont
  {Isobe}}, \bibinfo {author} {\bibfnamefont {N.~F.}\ \bibnamefont {Yuan}}, \
  and\ \bibinfo {author} {\bibfnamefont {L.}~\bibnamefont {Fu}},\ }\bibfield
  {title} {\emph {\bibinfo {title} {{Unconventional Superconductivity and
  Density Waves in Twisted Bilayer Graphene}}},\ }\href
  {https://journals.aps.org/prx/abstract/10.1103/PhysRevX.8.041041} {\bibfield
  {journal} {\bibinfo  {journal} {Phys. Rev. X}\ }\textbf {\bibinfo {volume}
  {8}},\ \bibinfo {pages} {041041} (\bibinfo {year} {2018})}\BibitemShut
  {NoStop}%
\bibitem [{\citenamefont {Choi}\ \emph {et~al.}(2019)\citenamefont {Choi},
  \citenamefont {Kemmer}, \citenamefont {Peng}, \citenamefont {Thomson},
  \citenamefont {Arora}, \citenamefont {Polski}, \citenamefont {Zhang},
  \citenamefont {Ren}, \citenamefont {Alicea}, \citenamefont {Refael},
  \citenamefont {von Oppen}, \citenamefont {Watanabe}, \citenamefont
  {Taniguchi},\ and\ \citenamefont {Nadj-Perge}}]{Choi2019}%
  \BibitemOpen
  \bibfield  {author} {\bibinfo {author} {\bibfnamefont {Y.}~\bibnamefont
  {Choi}}, \bibinfo {author} {\bibfnamefont {J.}~\bibnamefont {Kemmer}},
  \bibinfo {author} {\bibfnamefont {Y.}~\bibnamefont {Peng}}, \bibinfo {author}
  {\bibfnamefont {A.}~\bibnamefont {Thomson}}, \bibinfo {author} {\bibfnamefont
  {H.}~\bibnamefont {Arora}}, \bibinfo {author} {\bibfnamefont
  {R.}~\bibnamefont {Polski}}, \bibinfo {author} {\bibfnamefont
  {Y.}~\bibnamefont {Zhang}}, \bibinfo {author} {\bibfnamefont
  {H.}~\bibnamefont {Ren}}, \bibinfo {author} {\bibfnamefont {J.}~\bibnamefont
  {Alicea}}, \bibinfo {author} {\bibfnamefont {G.}~\bibnamefont {Refael}},
  \bibinfo {author} {\bibfnamefont {F.}~\bibnamefont {von Oppen}}, \bibinfo
  {author} {\bibfnamefont {K.}~\bibnamefont {Watanabe}}, \bibinfo {author}
  {\bibfnamefont {T.}~\bibnamefont {Taniguchi}}, \ and\ \bibinfo {author}
  {\bibfnamefont {S.}~\bibnamefont {Nadj-Perge}},\ }\bibfield  {title} {\emph
  {\bibinfo {title} {{Electronic correlations in twisted bilayer graphene near
  the magic angle}}},\ }\href
  {https://www.nature.com/articles/s41567-019-0606-5} {\bibfield  {journal}
  {\bibinfo  {journal} {Nature Physics}\ }\textbf {\bibinfo {volume} {15}},\
  \bibinfo {pages} {1174} (\bibinfo {year} {2019})}\BibitemShut {NoStop}%
\bibitem [{\citenamefont {Kerelsky}\ \emph {et~al.}(2019)\citenamefont
  {Kerelsky}, \citenamefont {McGilly}, \citenamefont {Kennes}, \citenamefont
  {Xian}, \citenamefont {Yankowitz}, \citenamefont {Chen}, \citenamefont
  {Watanabe}, \citenamefont {Taniguchi}, \citenamefont {Hone}, \citenamefont
  {Dean}, \citenamefont {Rubio},\ and\ \citenamefont
  {Pasupathy}}]{Kerelsky2019}%
  \BibitemOpen
  \bibfield  {author} {\bibinfo {author} {\bibfnamefont {A.}~\bibnamefont
  {Kerelsky}}, \bibinfo {author} {\bibfnamefont {L.~J.}\ \bibnamefont
  {McGilly}}, \bibinfo {author} {\bibfnamefont {D.~M.}\ \bibnamefont {Kennes}},
  \bibinfo {author} {\bibfnamefont {L.}~\bibnamefont {Xian}}, \bibinfo {author}
  {\bibfnamefont {M.}~\bibnamefont {Yankowitz}}, \bibinfo {author}
  {\bibfnamefont {S.}~\bibnamefont {Chen}}, \bibinfo {author} {\bibfnamefont
  {K.}~\bibnamefont {Watanabe}}, \bibinfo {author} {\bibfnamefont
  {T.}~\bibnamefont {Taniguchi}}, \bibinfo {author} {\bibfnamefont
  {J.}~\bibnamefont {Hone}}, \bibinfo {author} {\bibfnamefont {C.}~\bibnamefont
  {Dean}}, \bibinfo {author} {\bibfnamefont {A.}~\bibnamefont {Rubio}}, \ and\
  \bibinfo {author} {\bibfnamefont {A.~N.}\ \bibnamefont {Pasupathy}},\
  }\bibfield  {title} {\emph {\bibinfo {title} {{Maximized electron
  interactions at the magic angle in twisted bilayer graphene}}},\ }\href
  {https://www.nature.com/articles/s41586-019-1431-9} {\bibfield  {journal}
  {\bibinfo  {journal} {Nature}\ }\textbf {\bibinfo {volume} {572}},\ \bibinfo
  {pages} {95} (\bibinfo {year} {2019})}\BibitemShut {NoStop}%
\bibitem [{\citenamefont {Xie}\ \emph {et~al.}(2021)\citenamefont {Xie},
  \citenamefont {Pierce}, \citenamefont {Park}, \citenamefont {Parker},
  \citenamefont {Khalaf}, \citenamefont {Ledwith}, \citenamefont {Cao},
  \citenamefont {Lee}, \citenamefont {Chen}, \citenamefont {Forrester},
  \citenamefont {Watanabe}, \citenamefont {Taniguchi}, \citenamefont
  {Vishwanath}, \citenamefont {Jarillo-Herrero},\ and\ \citenamefont
  {Yacoby}}]{Xie2021}%
  \BibitemOpen
  \bibfield  {author} {\bibinfo {author} {\bibfnamefont {Y.}~\bibnamefont
  {Xie}}, \bibinfo {author} {\bibfnamefont {A.~T.}\ \bibnamefont {Pierce}},
  \bibinfo {author} {\bibfnamefont {J.~M.}\ \bibnamefont {Park}}, \bibinfo
  {author} {\bibfnamefont {D.~E.}\ \bibnamefont {Parker}}, \bibinfo {author}
  {\bibfnamefont {E.}~\bibnamefont {Khalaf}}, \bibinfo {author} {\bibfnamefont
  {P.}~\bibnamefont {Ledwith}}, \bibinfo {author} {\bibfnamefont
  {Y.}~\bibnamefont {Cao}}, \bibinfo {author} {\bibfnamefont {S.~H.}\
  \bibnamefont {Lee}}, \bibinfo {author} {\bibfnamefont {S.}~\bibnamefont
  {Chen}}, \bibinfo {author} {\bibfnamefont {P.~R.}\ \bibnamefont {Forrester}},
  \bibinfo {author} {\bibfnamefont {K.}~\bibnamefont {Watanabe}}, \bibinfo
  {author} {\bibfnamefont {T.}~\bibnamefont {Taniguchi}}, \bibinfo {author}
  {\bibfnamefont {A.}~\bibnamefont {Vishwanath}}, \bibinfo {author}
  {\bibfnamefont {P.}~\bibnamefont {Jarillo-Herrero}}, \ and\ \bibinfo {author}
  {\bibfnamefont {A.}~\bibnamefont {Yacoby}},\ }\bibfield  {title} {\emph
  {\bibinfo {title} {{Fractional Chern insulators in magic-angle twisted
  bilayer graphene}}},\ }\href
  {https://www.nature.com/articles/s41586-021-04002-3} {\bibfield  {journal}
  {\bibinfo  {journal} {Nature}\ }\textbf {\bibinfo {volume} {600}},\ \bibinfo
  {pages} {439} (\bibinfo {year} {2021})}\BibitemShut {NoStop}%
\bibitem [{\citenamefont {Ledwith}\ \emph {et~al.}(2020)\citenamefont
  {Ledwith}, \citenamefont {Tarnopolsky}, \citenamefont {Khalaf},\ and\
  \citenamefont {Vishwanath}}]{Ledwith2020}%
  \BibitemOpen
  \bibfield  {author} {\bibinfo {author} {\bibfnamefont {P.~J.}\ \bibnamefont
  {Ledwith}}, \bibinfo {author} {\bibfnamefont {G.}~\bibnamefont
  {Tarnopolsky}}, \bibinfo {author} {\bibfnamefont {E.}~\bibnamefont {Khalaf}},
  \ and\ \bibinfo {author} {\bibfnamefont {A.}~\bibnamefont {Vishwanath}},\
  }\bibfield  {title} {\emph {\bibinfo {title} {{Fractional Chern insulator
  states in twisted bilayer graphene: An analytical approach}}},\ }\href
  {https://link.aps.org/doi/10.1103/PhysRevResearch.2.023237} {\bibfield
  {journal} {\bibinfo  {journal} {Phys. Rev. Res.}\ }\textbf {\bibinfo {volume}
  {2}},\ \bibinfo {pages} {023237} (\bibinfo {year} {2020})}\BibitemShut
  {NoStop}%
\bibitem [{\citenamefont {Song}\ and\ \citenamefont
  {Bernevig}(2022)}]{song2022magic}%
  \BibitemOpen
  \bibfield  {author} {\bibinfo {author} {\bibfnamefont {Z.-D.}\ \bibnamefont
  {Song}}\ and\ \bibinfo {author} {\bibfnamefont {B.~A.}\ \bibnamefont
  {Bernevig}},\ }\bibfield  {title} {\emph {\bibinfo {title} {{Magic-Angle
  Twisted Bilayer Graphene as a Topological Heavy Fermion Problem}}},\ }\href
  {https://link.aps.org/doi/10.1103/PhysRevLett.129.047601} {\bibfield
  {journal} {\bibinfo  {journal} {Phys. Rev. Lett.}\ }\textbf {\bibinfo
  {volume} {129}},\ \bibinfo {pages} {047601} (\bibinfo {year}
  {2022})}\BibitemShut {NoStop}%
\bibitem [{\citenamefont {Thomson}\ \emph {et~al.}(2018)\citenamefont
  {Thomson}, \citenamefont {Chatterjee}, \citenamefont {Sachdev},\ and\
  \citenamefont {Scheurer}}]{Thomson2018}%
  \BibitemOpen
  \bibfield  {author} {\bibinfo {author} {\bibfnamefont {A.}~\bibnamefont
  {Thomson}}, \bibinfo {author} {\bibfnamefont {S.}~\bibnamefont {Chatterjee}},
  \bibinfo {author} {\bibfnamefont {S.}~\bibnamefont {Sachdev}}, \ and\
  \bibinfo {author} {\bibfnamefont {M.~S.}\ \bibnamefont {Scheurer}},\
  }\bibfield  {title} {\emph {\bibinfo {title} {{Triangular antiferromagnetism
  on the honeycomb lattice of twisted bilayer graphene}}},\ }\href
  {https://journals.aps.org/prb/abstract/10.1103/PhysRevB.98.075109} {\bibfield
   {journal} {\bibinfo  {journal} {Phys. Rev. B}\ }\textbf {\bibinfo {volume}
  {98}},\ \bibinfo {pages} {075109} (\bibinfo {year} {2018})}\BibitemShut
  {NoStop}%
\bibitem [{\citenamefont {Bultinck}\ \emph {et~al.}(2020)\citenamefont
  {Bultinck}, \citenamefont {Khalaf}, \citenamefont {Liu}, \citenamefont
  {Chatterjee}, \citenamefont {Vishwanath},\ and\ \citenamefont
  {Zaletel}}]{Bultinck2020}%
  \BibitemOpen
  \bibfield  {author} {\bibinfo {author} {\bibfnamefont {N.}~\bibnamefont
  {Bultinck}}, \bibinfo {author} {\bibfnamefont {E.}~\bibnamefont {Khalaf}},
  \bibinfo {author} {\bibfnamefont {S.}~\bibnamefont {Liu}}, \bibinfo {author}
  {\bibfnamefont {S.}~\bibnamefont {Chatterjee}}, \bibinfo {author}
  {\bibfnamefont {A.}~\bibnamefont {Vishwanath}}, \ and\ \bibinfo {author}
  {\bibfnamefont {M.~P.}\ \bibnamefont {Zaletel}},\ }\bibfield  {title} {\emph
  {\bibinfo {title} {{Ground State and Hidden Symmetry of Magic-Angle Graphene
  at even Integer Filling}}},\ }\href
  {https://journals.aps.org/prx/abstract/10.1103/PhysRevX.10.031034} {\bibfield
   {journal} {\bibinfo  {journal} {Phys. Rev. X}\ }\textbf {\bibinfo {volume}
  {10}},\ \bibinfo {pages} {031034} (\bibinfo {year} {2020})}\BibitemShut
  {NoStop}%
\bibitem [{\citenamefont {Kwan}\ \emph {et~al.}(2021)\citenamefont {Kwan},
  \citenamefont {Wagner}, \citenamefont {Soejima}, \citenamefont {Zaletel},
  \citenamefont {Simon}, \citenamefont {Parameswaran},\ and\ \citenamefont
  {Bultinck}}]{Kwan2021}%
  \BibitemOpen
  \bibfield  {author} {\bibinfo {author} {\bibfnamefont {Y.~H.}\ \bibnamefont
  {Kwan}}, \bibinfo {author} {\bibfnamefont {G.}~\bibnamefont {Wagner}},
  \bibinfo {author} {\bibfnamefont {T.}~\bibnamefont {Soejima}}, \bibinfo
  {author} {\bibfnamefont {M.~P.}\ \bibnamefont {Zaletel}}, \bibinfo {author}
  {\bibfnamefont {S.~H.}\ \bibnamefont {Simon}}, \bibinfo {author}
  {\bibfnamefont {S.~A.}\ \bibnamefont {Parameswaran}}, \ and\ \bibinfo
  {author} {\bibfnamefont {N.}~\bibnamefont {Bultinck}},\ }\bibfield  {title}
  {\emph {\bibinfo {title} {{Kekul{\'{e}} Spiral Order at All Nonzero Integer
  Fillings in Twisted Bilayer Graphene}}},\ }\href
  {https://journals.aps.org/prx/abstract/10.1103/PhysRevX.11.041063} {\bibfield
   {journal} {\bibinfo  {journal} {Phys. Rev. X}\ }\textbf {\bibinfo {volume}
  {11}},\ \bibinfo {pages} {041063} (\bibinfo {year} {2021})}\BibitemShut
  {NoStop}%
\bibitem [{\citenamefont {Hofmann}\ \emph {et~al.}(2022)\citenamefont
  {Hofmann}, \citenamefont {Khalaf}, \citenamefont {Vishwanath}, \citenamefont
  {Berg},\ and\ \citenamefont {Lee}}]{Hofmann2022}%
  \BibitemOpen
  \bibfield  {author} {\bibinfo {author} {\bibfnamefont {J.~S.}\ \bibnamefont
  {Hofmann}}, \bibinfo {author} {\bibfnamefont {E.}~\bibnamefont {Khalaf}},
  \bibinfo {author} {\bibfnamefont {A.}~\bibnamefont {Vishwanath}}, \bibinfo
  {author} {\bibfnamefont {E.}~\bibnamefont {Berg}}, \ and\ \bibinfo {author}
  {\bibfnamefont {J.~Y.}\ \bibnamefont {Lee}},\ }\bibfield  {title} {\emph
  {\bibinfo {title} {{Fermionic Monte Carlo Study of a Realistic Model of
  Twisted Bilayer Graphene}}},\ }\href
  {https://journals.aps.org/prx/abstract/10.1103/PhysRevX.12.011061} {\bibfield
   {journal} {\bibinfo  {journal} {Phys. Rev. X}\ }\textbf {\bibinfo {volume}
  {12}},\ \bibinfo {pages} {011061} (\bibinfo {year} {2022})}\BibitemShut
  {NoStop}%
\bibitem [{\citenamefont {Chou}\ and\ \citenamefont {Sarma}(2022)}]{Chou2022}%
  \BibitemOpen
  \bibfield  {author} {\bibinfo {author} {\bibfnamefont {Y.-Z.}\ \bibnamefont
  {Chou}}\ and\ \bibinfo {author} {\bibfnamefont {S.~D.}\ \bibnamefont
  {Sarma}},\ }\bibfield  {title} {\emph {\bibinfo {title} {Kondo lattice model
  in magic-angle twisted bilayer graphene}},\ }\href
  {https://arxiv.org/abs/2211.15682} {\  (\bibinfo {year} {2022})},\ \Eprint
  {http://arxiv.org/abs/2211.15682} {arXiv:2211.15682} \BibitemShut {NoStop}%
\bibitem [{\citenamefont {Mora}\ \emph {et~al.}(2019)\citenamefont {Mora},
  \citenamefont {Regnault},\ and\ \citenamefont {Bernevig}}]{Mora2019}%
  \BibitemOpen
  \bibfield  {author} {\bibinfo {author} {\bibfnamefont {C.}~\bibnamefont
  {Mora}}, \bibinfo {author} {\bibfnamefont {N.}~\bibnamefont {Regnault}}, \
  and\ \bibinfo {author} {\bibfnamefont {B.~A.}\ \bibnamefont {Bernevig}},\
  }\bibfield  {title} {\emph {\bibinfo {title} {{Flatbands and Perfect Metal in
  Trilayer Moir{\'{e}} Graphene}}},\ }\href
  {https://journals.aps.org/prl/abstract/10.1103/PhysRevLett.123.026402}
  {\bibfield  {journal} {\bibinfo  {journal} {Phys. Rev. Lett.}\ }\textbf
  {\bibinfo {volume} {123}},\ \bibinfo {pages} {026402} (\bibinfo {year}
  {2019})}\BibitemShut {NoStop}%
\bibitem [{\citenamefont {Khalaf}\ \emph {et~al.}(2019)\citenamefont {Khalaf},
  \citenamefont {Kruchkov}, \citenamefont {Tarnopolsky},\ and\ \citenamefont
  {Vishwanath}}]{Khalaf2019}%
  \BibitemOpen
  \bibfield  {author} {\bibinfo {author} {\bibfnamefont {E.}~\bibnamefont
  {Khalaf}}, \bibinfo {author} {\bibfnamefont {A.~J.}\ \bibnamefont
  {Kruchkov}}, \bibinfo {author} {\bibfnamefont {G.}~\bibnamefont
  {Tarnopolsky}}, \ and\ \bibinfo {author} {\bibfnamefont {A.}~\bibnamefont
  {Vishwanath}},\ }\bibfield  {title} {\emph {\bibinfo {title} {{Magic angle
  hierarchy in twisted graphene multilayers}}},\ }\href
  {https://journals.aps.org/prb/abstract/10.1103/PhysRevB.100.085109}
  {\bibfield  {journal} {\bibinfo  {journal} {Phys. Rev. B}\ }\textbf {\bibinfo
  {volume} {100}},\ \bibinfo {pages} {085109} (\bibinfo {year}
  {2019})}\BibitemShut {NoStop}%
\bibitem [{\citenamefont {Park}\ \emph {et~al.}()\citenamefont {Park},
  \citenamefont {Cao}, \citenamefont {Watanabe}, \citenamefont {Taniguchi},\
  and\ \citenamefont {Jarillo-Herrero}}]{Park2020}%
  \BibitemOpen
  \bibfield  {author} {\bibinfo {author} {\bibfnamefont {J.~M.}\ \bibnamefont
  {Park}}, \bibinfo {author} {\bibfnamefont {Y.}~\bibnamefont {Cao}}, \bibinfo
  {author} {\bibfnamefont {K.}~\bibnamefont {Watanabe}}, \bibinfo {author}
  {\bibfnamefont {T.}~\bibnamefont {Taniguchi}}, \ and\ \bibinfo {author}
  {\bibfnamefont {P.}~\bibnamefont {Jarillo-Herrero}},\ }\bibfield  {title}
  {\emph {\bibinfo {title} {{Tunable Phase Boundaries and Ultra-Strong Coupling
  Superconductivity in Mirror Symmetric Magic-Angle Trilayer Graphene}}},\
  }\href {http://arxiv.org/abs/2012.01434
  http://dx.doi.org/10.1038/s41586-021-03192-0} {\ }\Eprint
  {http://arxiv.org/abs/2012.01434} {arXiv:2012.01434} \BibitemShut {NoStop}%
\bibitem [{\citenamefont {Chen}\ \emph {et~al.}(2020)\citenamefont {Chen},
  \citenamefont {He}, \citenamefont {Zhang}, \citenamefont {Hsieh},
  \citenamefont {Fei}, \citenamefont {Watanabe}, \citenamefont {Taniguchi},
  \citenamefont {Cobden}, \citenamefont {Xu}, \citenamefont {Dean},\ and\
  \citenamefont {Yankowitz}}]{Chen2020}%
  \BibitemOpen
  \bibfield  {author} {\bibinfo {author} {\bibfnamefont {S.}~\bibnamefont
  {Chen}}, \bibinfo {author} {\bibfnamefont {M.}~\bibnamefont {He}}, \bibinfo
  {author} {\bibfnamefont {Y.~H.}\ \bibnamefont {Zhang}}, \bibinfo {author}
  {\bibfnamefont {V.}~\bibnamefont {Hsieh}}, \bibinfo {author} {\bibfnamefont
  {Z.}~\bibnamefont {Fei}}, \bibinfo {author} {\bibfnamefont {K.}~\bibnamefont
  {Watanabe}}, \bibinfo {author} {\bibfnamefont {T.}~\bibnamefont {Taniguchi}},
  \bibinfo {author} {\bibfnamefont {D.~H.}\ \bibnamefont {Cobden}}, \bibinfo
  {author} {\bibfnamefont {X.}~\bibnamefont {Xu}}, \bibinfo {author}
  {\bibfnamefont {C.~R.}\ \bibnamefont {Dean}}, \ and\ \bibinfo {author}
  {\bibfnamefont {M.}~\bibnamefont {Yankowitz}},\ }\bibfield  {title} {\emph
  {\bibinfo {title} {{Electrically tunable correlated and topological states in
  twisted monolayer–bilayer graphene}}},\ }\href
  {https://www.nature.com/articles/s41567-020-01062-6} {\bibfield  {journal}
  {\bibinfo  {journal} {Nature Physics}\ }\textbf {\bibinfo {volume} {17}},\
  \bibinfo {pages} {374} (\bibinfo {year} {2020})}\BibitemShut {NoStop}%
\bibitem [{\citenamefont {He}\ \emph {et~al.}(2020)\citenamefont {He},
  \citenamefont {Li}, \citenamefont {Cai}, \citenamefont {Liu}, \citenamefont
  {Watanabe}, \citenamefont {Taniguchi}, \citenamefont {Xu},\ and\
  \citenamefont {Yankowitz}}]{He2020}%
  \BibitemOpen
  \bibfield  {author} {\bibinfo {author} {\bibfnamefont {M.}~\bibnamefont
  {He}}, \bibinfo {author} {\bibfnamefont {Y.}~\bibnamefont {Li}}, \bibinfo
  {author} {\bibfnamefont {J.}~\bibnamefont {Cai}}, \bibinfo {author}
  {\bibfnamefont {Y.}~\bibnamefont {Liu}}, \bibinfo {author} {\bibfnamefont
  {K.}~\bibnamefont {Watanabe}}, \bibinfo {author} {\bibfnamefont
  {T.}~\bibnamefont {Taniguchi}}, \bibinfo {author} {\bibfnamefont
  {X.}~\bibnamefont {Xu}}, \ and\ \bibinfo {author} {\bibfnamefont
  {M.}~\bibnamefont {Yankowitz}},\ }\bibfield  {title} {\emph {\bibinfo {title}
  {{Symmetry breaking in twisted double bilayer graphene}}},\ }\href
  {https://www.nature.com/articles/s41567-020-1030-6} {\bibfield  {journal}
  {\bibinfo  {journal} {Nature Physics}\ }\textbf {\bibinfo {volume} {17}},\
  \bibinfo {pages} {26} (\bibinfo {year} {2020})}\BibitemShut {NoStop}%
\bibitem [{\citenamefont {Ramires}\ and\ \citenamefont
  {Lado}(2021)}]{Ramires2021}%
  \BibitemOpen
  \bibfield  {author} {\bibinfo {author} {\bibfnamefont {A.}~\bibnamefont
  {Ramires}}\ and\ \bibinfo {author} {\bibfnamefont {J.~L.}\ \bibnamefont
  {Lado}},\ }\bibfield  {title} {\emph {\bibinfo {title} {{Emulating Heavy
  Fermions in Twisted Trilayer Graphene}}},\ }\href
  {https://link.aps.org/doi/10.1103/PhysRevLett.127.026401} {\bibfield
  {journal} {\bibinfo  {journal} {Phys. Rev. Lett.}\ }\textbf {\bibinfo
  {volume} {127}},\ \bibinfo {pages} {26401} (\bibinfo {year}
  {2021})}\BibitemShut {NoStop}%
\bibitem [{\citenamefont {Park}\ \emph {et~al.}(2022)\citenamefont {Park},
  \citenamefont {Cao}, \citenamefont {Xia}, \citenamefont {Sun}, \citenamefont
  {Watanabe}, \citenamefont {Taniguchi},\ and\ \citenamefont
  {Jarillo-Herrero}}]{Park2022}%
  \BibitemOpen
  \bibfield  {author} {\bibinfo {author} {\bibfnamefont {J.~M.}\ \bibnamefont
  {Park}}, \bibinfo {author} {\bibfnamefont {Y.}~\bibnamefont {Cao}}, \bibinfo
  {author} {\bibfnamefont {L.-Q.}\ \bibnamefont {Xia}}, \bibinfo {author}
  {\bibfnamefont {S.}~\bibnamefont {Sun}}, \bibinfo {author} {\bibfnamefont
  {K.}~\bibnamefont {Watanabe}}, \bibinfo {author} {\bibfnamefont
  {T.}~\bibnamefont {Taniguchi}}, \ and\ \bibinfo {author} {\bibfnamefont
  {P.}~\bibnamefont {Jarillo-Herrero}},\ }\bibfield  {title} {\emph {\bibinfo
  {title} {{Robust superconductivity in magic-angle multilayer graphene
  family}}},\ }\href {https://www.nature.com/articles/s41563-022-01287-1}
  {\bibfield  {journal} {\bibinfo  {journal} {Nature Materials}\ }\textbf
  {\bibinfo {volume} {21}},\ \bibinfo {pages} {877} (\bibinfo {year}
  {2022})}\BibitemShut {NoStop}%
\bibitem [{\citenamefont {San-Jose}\ and\ \citenamefont
  {Prada}(2013)}]{San-Jose2013}%
  \BibitemOpen
  \bibfield  {author} {\bibinfo {author} {\bibfnamefont {P.}~\bibnamefont
  {San-Jose}}\ and\ \bibinfo {author} {\bibfnamefont {E.}~\bibnamefont
  {Prada}},\ }\bibfield  {title} {\emph {\bibinfo {title} {{Helical networks in
  twisted bilayer graphene under interlayer bias}}},\ }\href
  {https://link.aps.org/doi/10.1103/PhysRevB.88.121408} {\bibfield  {journal}
  {\bibinfo  {journal} {Phys. Rev. B}\ }\textbf {\bibinfo {volume} {88}},\
  \bibinfo {pages} {121408} (\bibinfo {year} {2013})}\BibitemShut {NoStop}%
\bibitem [{\citenamefont {Efimkin}\ and\ \citenamefont
  {MacDonald}(2018)}]{Efimkin2018}%
  \BibitemOpen
  \bibfield  {author} {\bibinfo {author} {\bibfnamefont {D.~K.}\ \bibnamefont
  {Efimkin}}\ and\ \bibinfo {author} {\bibfnamefont {A.~H.}\ \bibnamefont
  {MacDonald}},\ }\bibfield  {title} {\emph {\bibinfo {title} {{Helical network
  model for twisted bilayer graphene}}},\ }\href
  {https://link.aps.org/doi/10.1103/PhysRevB.98.035404} {\bibfield  {journal}
  {\bibinfo  {journal} {Phys. Rev. B}\ }\textbf {\bibinfo {volume} {98}},\
  \bibinfo {pages} {35404} (\bibinfo {year} {2018})}\BibitemShut {NoStop}%
\bibitem [{\citenamefont {Ramires}\ and\ \citenamefont
  {Lado}(2018)}]{Ramires2018}%
  \BibitemOpen
  \bibfield  {author} {\bibinfo {author} {\bibfnamefont {A.}~\bibnamefont
  {Ramires}}\ and\ \bibinfo {author} {\bibfnamefont {J.~L.}\ \bibnamefont
  {Lado}},\ }\bibfield  {title} {\emph {\bibinfo {title} {{Electrically Tunable
  Gauge Fields in Tiny-Angle Twisted Bilayer Graphene}}},\ }\href
  {https://link.aps.org/doi/10.1103/PhysRevLett.121.146801} {\bibfield
  {journal} {\bibinfo  {journal} {Phys. Rev. Lett.}\ }\textbf {\bibinfo
  {volume} {121}},\ \bibinfo {pages} {146801} (\bibinfo {year}
  {2018})}\BibitemShut {NoStop}%
\bibitem [{\citenamefont {Rickhaus}\ \emph {et~al.}(2018)\citenamefont
  {Rickhaus}, \citenamefont {Wallbank}, \citenamefont {Slizovskiy},
  \citenamefont {Pisoni}, \citenamefont {Overweg}, \citenamefont {Lee},
  \citenamefont {Eich}, \citenamefont {Liu}, \citenamefont {Watanabe},
  \citenamefont {Taniguchi}, \citenamefont {Ihn},\ and\ \citenamefont
  {Ensslin}}]{Rickhaus2018}%
  \BibitemOpen
  \bibfield  {author} {\bibinfo {author} {\bibfnamefont {P.}~\bibnamefont
  {Rickhaus}}, \bibinfo {author} {\bibfnamefont {J.}~\bibnamefont {Wallbank}},
  \bibinfo {author} {\bibfnamefont {S.}~\bibnamefont {Slizovskiy}}, \bibinfo
  {author} {\bibfnamefont {R.}~\bibnamefont {Pisoni}}, \bibinfo {author}
  {\bibfnamefont {H.}~\bibnamefont {Overweg}}, \bibinfo {author} {\bibfnamefont
  {Y.}~\bibnamefont {Lee}}, \bibinfo {author} {\bibfnamefont {M.}~\bibnamefont
  {Eich}}, \bibinfo {author} {\bibfnamefont {M.-H.}\ \bibnamefont {Liu}},
  \bibinfo {author} {\bibfnamefont {K.}~\bibnamefont {Watanabe}}, \bibinfo
  {author} {\bibfnamefont {T.}~\bibnamefont {Taniguchi}}, \bibinfo {author}
  {\bibfnamefont {T.}~\bibnamefont {Ihn}}, \ and\ \bibinfo {author}
  {\bibfnamefont {K.}~\bibnamefont {Ensslin}},\ }\bibfield  {title} {\emph
  {\bibinfo {title} {{Transport Through a Network of Topological Channels in
  Twisted Bilayer Graphene}}},\ }\href
  {https://doi.org/10.1021/acs.nanolett.8b02387} {\bibfield  {journal}
  {\bibinfo  {journal} {Nano Letters}\ }\textbf {\bibinfo {volume} {18}},\
  \bibinfo {pages} {6725} (\bibinfo {year} {2018})}\BibitemShut {NoStop}%
\bibitem [{\citenamefont {Huang}\ \emph {et~al.}(2018)\citenamefont {Huang},
  \citenamefont {Kim}, \citenamefont {Efimkin}, \citenamefont {Lovorn},
  \citenamefont {Taniguchi}, \citenamefont {Watanabe}, \citenamefont
  {MacDonald}, \citenamefont {Tutuc},\ and\ \citenamefont {LeRoy}}]{Huang2018}%
  \BibitemOpen
  \bibfield  {author} {\bibinfo {author} {\bibfnamefont {S.}~\bibnamefont
  {Huang}}, \bibinfo {author} {\bibfnamefont {K.}~\bibnamefont {Kim}}, \bibinfo
  {author} {\bibfnamefont {D.~K.}\ \bibnamefont {Efimkin}}, \bibinfo {author}
  {\bibfnamefont {T.}~\bibnamefont {Lovorn}}, \bibinfo {author} {\bibfnamefont
  {T.}~\bibnamefont {Taniguchi}}, \bibinfo {author} {\bibfnamefont
  {K.}~\bibnamefont {Watanabe}}, \bibinfo {author} {\bibfnamefont {A.~H.}\
  \bibnamefont {MacDonald}}, \bibinfo {author} {\bibfnamefont {E.}~\bibnamefont
  {Tutuc}}, \ and\ \bibinfo {author} {\bibfnamefont {B.~J.}\ \bibnamefont
  {LeRoy}},\ }\bibfield  {title} {\emph {\bibinfo {title} {{Topologically
  Protected Helical States in Minimally Twisted Bilayer Graphene}}},\ }\href
  {https://link.aps.org/doi/10.1103/PhysRevLett.121.037702} {\bibfield
  {journal} {\bibinfo  {journal} {Phys. Rev. Lett.}\ }\textbf {\bibinfo
  {volume} {121}},\ \bibinfo {pages} {37702} (\bibinfo {year}
  {2018})}\BibitemShut {NoStop}%
\bibitem [{\citenamefont {Yoo}\ \emph {et~al.}(2019)\citenamefont {Yoo},
  \citenamefont {Engelke}, \citenamefont {Carr}, \citenamefont {Fang},
  \citenamefont {Zhang}, \citenamefont {Cazeaux}, \citenamefont {Sung},
  \citenamefont {Hovden}, \citenamefont {Tsen}, \citenamefont {Taniguchi},
  \citenamefont {Watanabe}, \citenamefont {Yi}, \citenamefont {Kim},
  \citenamefont {Luskin}, \citenamefont {Tadmor}, \citenamefont {Kaxiras},\
  and\ \citenamefont {Kim}}]{Yoo2019}%
  \BibitemOpen
  \bibfield  {author} {\bibinfo {author} {\bibfnamefont {H.}~\bibnamefont
  {Yoo}}, \bibinfo {author} {\bibfnamefont {R.}~\bibnamefont {Engelke}},
  \bibinfo {author} {\bibfnamefont {S.}~\bibnamefont {Carr}}, \bibinfo {author}
  {\bibfnamefont {S.}~\bibnamefont {Fang}}, \bibinfo {author} {\bibfnamefont
  {K.}~\bibnamefont {Zhang}}, \bibinfo {author} {\bibfnamefont
  {P.}~\bibnamefont {Cazeaux}}, \bibinfo {author} {\bibfnamefont {S.~H.}\
  \bibnamefont {Sung}}, \bibinfo {author} {\bibfnamefont {R.}~\bibnamefont
  {Hovden}}, \bibinfo {author} {\bibfnamefont {A.~W.}\ \bibnamefont {Tsen}},
  \bibinfo {author} {\bibfnamefont {T.}~\bibnamefont {Taniguchi}}, \bibinfo
  {author} {\bibfnamefont {K.}~\bibnamefont {Watanabe}}, \bibinfo {author}
  {\bibfnamefont {G.-C.}\ \bibnamefont {Yi}}, \bibinfo {author} {\bibfnamefont
  {M.}~\bibnamefont {Kim}}, \bibinfo {author} {\bibfnamefont {M.}~\bibnamefont
  {Luskin}}, \bibinfo {author} {\bibfnamefont {E.~B.}\ \bibnamefont {Tadmor}},
  \bibinfo {author} {\bibfnamefont {E.}~\bibnamefont {Kaxiras}}, \ and\
  \bibinfo {author} {\bibfnamefont {P.}~\bibnamefont {Kim}},\ }\bibfield
  {title} {\emph {\bibinfo {title} {{Atomic and electronic reconstruction at
  the van der Waals interface in twisted bilayer graphene}}},\ }\href {\doibase
  10.1038/s41563-019-0346-z} {\bibfield  {journal} {\bibinfo  {journal} {Nature
  Materials}\ }\textbf {\bibinfo {volume} {18}},\ \bibinfo {pages} {448}
  (\bibinfo {year} {2019})}\BibitemShut {NoStop}%
\bibitem [{\citenamefont {Xu}\ \emph {et~al.}(2019)\citenamefont {Xu},
  \citenamefont {Berdyugin}, \citenamefont {Kumaravadivel}, \citenamefont
  {Guinea}, \citenamefont {{Krishna Kumar}}, \citenamefont {Bandurin},
  \citenamefont {Morozov}, \citenamefont {Kuang}, \citenamefont {Tsim},
  \citenamefont {Liu}, \citenamefont {Edgar}, \citenamefont {Grigorieva},
  \citenamefont {Fal'ko}, \citenamefont {Kim},\ and\ \citenamefont
  {Geim}}]{Xu2019}%
  \BibitemOpen
  \bibfield  {author} {\bibinfo {author} {\bibfnamefont {S.~G.}\ \bibnamefont
  {Xu}}, \bibinfo {author} {\bibfnamefont {A.~I.}\ \bibnamefont {Berdyugin}},
  \bibinfo {author} {\bibfnamefont {P.}~\bibnamefont {Kumaravadivel}}, \bibinfo
  {author} {\bibfnamefont {F.}~\bibnamefont {Guinea}}, \bibinfo {author}
  {\bibfnamefont {R.}~\bibnamefont {{Krishna Kumar}}}, \bibinfo {author}
  {\bibfnamefont {D.~A.}\ \bibnamefont {Bandurin}}, \bibinfo {author}
  {\bibfnamefont {S.~V.}\ \bibnamefont {Morozov}}, \bibinfo {author}
  {\bibfnamefont {W.}~\bibnamefont {Kuang}}, \bibinfo {author} {\bibfnamefont
  {B.}~\bibnamefont {Tsim}}, \bibinfo {author} {\bibfnamefont {S.}~\bibnamefont
  {Liu}}, \bibinfo {author} {\bibfnamefont {J.~H.}\ \bibnamefont {Edgar}},
  \bibinfo {author} {\bibfnamefont {I.~V.}\ \bibnamefont {Grigorieva}},
  \bibinfo {author} {\bibfnamefont {V.~I.}\ \bibnamefont {Fal'ko}}, \bibinfo
  {author} {\bibfnamefont {M.}~\bibnamefont {Kim}}, \ and\ \bibinfo {author}
  {\bibfnamefont {A.~K.}\ \bibnamefont {Geim}},\ }\bibfield  {title} {\emph
  {\bibinfo {title} {{Giant oscillations in a triangular network of
  one-dimensional states in marginally twisted graphene}}},\ }\href
  {https://doi.org/10.1038/s41467-019-11971-7} {\bibfield  {journal} {\bibinfo
  {journal} {Nature Communications}\ }\textbf {\bibinfo {volume} {10}},\
  \bibinfo {pages} {4008} (\bibinfo {year} {2019})}\BibitemShut {NoStop}%
\bibitem [{\citenamefont {Chou}\ \emph {et~al.}(2020)\citenamefont {Chou},
  \citenamefont {Wu},\ and\ \citenamefont {Das~Sarma}}]{Chou2020}%
  \BibitemOpen
  \bibfield  {author} {\bibinfo {author} {\bibfnamefont {Y.-Z.}\ \bibnamefont
  {Chou}}, \bibinfo {author} {\bibfnamefont {F.}~\bibnamefont {Wu}}, \ and\
  \bibinfo {author} {\bibfnamefont {S.}~\bibnamefont {Das~Sarma}},\ }\bibfield
  {title} {\emph {\bibinfo {title} {{Hofstadter butterfly and Floquet
  topological insulators in minimally twisted bilayer graphene}}},\ }\href
  {\doibase 10.1103/PhysRevResearch.2.033271} {\bibfield  {journal} {\bibinfo
  {journal} {Phys. Rev. Res.}\ }\textbf {\bibinfo {volume} {2}},\ \bibinfo
  {pages} {033271} (\bibinfo {year} {2020})}\BibitemShut {NoStop}%
\bibitem [{\citenamefont {De~Beule}\ \emph {et~al.}(2020)\citenamefont
  {De~Beule}, \citenamefont {Dominguez},\ and\ \citenamefont
  {Recher}}]{DeBeule2020}%
  \BibitemOpen
  \bibfield  {author} {\bibinfo {author} {\bibfnamefont {C.}~\bibnamefont
  {De~Beule}}, \bibinfo {author} {\bibfnamefont {F.}~\bibnamefont {Dominguez}},
  \ and\ \bibinfo {author} {\bibfnamefont {P.}~\bibnamefont {Recher}},\
  }\bibfield  {title} {\emph {\bibinfo {title} {Aharonov-bohm oscillations in
  minimally twisted bilayer graphene}},\ }\href {\doibase
  10.1103/PhysRevLett.125.096402} {\bibfield  {journal} {\bibinfo  {journal}
  {Phys. Rev. Lett.}\ }\textbf {\bibinfo {volume} {125}},\ \bibinfo {pages}
  {096402} (\bibinfo {year} {2020})}\BibitemShut {NoStop}%
\bibitem [{\citenamefont {{De Beule}}\ \emph {et~al.}(2021)\citenamefont {{De
  Beule}}, \citenamefont {Dominguez},\ and\ \citenamefont
  {Recher}}]{DeBeule2021}%
  \BibitemOpen
  \bibfield  {author} {\bibinfo {author} {\bibfnamefont {C.}~\bibnamefont {{De
  Beule}}}, \bibinfo {author} {\bibfnamefont {F.}~\bibnamefont {Dominguez}}, \
  and\ \bibinfo {author} {\bibfnamefont {P.}~\bibnamefont {Recher}},\
  }\bibfield  {title} {\emph {\bibinfo {title} {{Network model and
  four-terminal transport in minimally twisted bilayer graphene}}},\ }\href
  {https://journals.aps.org/prb/abstract/10.1103/PhysRevB.104.195410}
  {\bibfield  {journal} {\bibinfo  {journal} {Phys. Rev. B}\ }\textbf {\bibinfo
  {volume} {104}},\ \bibinfo {pages} {195410} (\bibinfo {year}
  {2021})}\BibitemShut {NoStop}%
\bibitem [{\citenamefont {Chou}\ \emph {et~al.}(2021)\citenamefont {Chou},
  \citenamefont {Wu},\ and\ \citenamefont {Sau}}]{Chou2021}%
  \BibitemOpen
  \bibfield  {author} {\bibinfo {author} {\bibfnamefont {Y.-Z.}\ \bibnamefont
  {Chou}}, \bibinfo {author} {\bibfnamefont {F.}~\bibnamefont {Wu}}, \ and\
  \bibinfo {author} {\bibfnamefont {J.~D.}\ \bibnamefont {Sau}},\ }\bibfield
  {title} {\emph {\bibinfo {title} {Charge density wave and finite-temperature
  transport in minimally twisted bilayer graphene}},\ }\href {\doibase
  10.1103/PhysRevB.104.045146} {\bibfield  {journal} {\bibinfo  {journal}
  {Phys. Rev. B}\ }\textbf {\bibinfo {volume} {104}},\ \bibinfo {pages}
  {045146} (\bibinfo {year} {2021})}\BibitemShut {NoStop}%
\bibitem [{\citenamefont {Attig}\ \emph {et~al.}(2021)\citenamefont {Attig},
  \citenamefont {Park}, \citenamefont {Scherer}, \citenamefont {Trebst},
  \citenamefont {Altland},\ and\ \citenamefont {Rosch}}]{Attig2021}%
  \BibitemOpen
  \bibfield  {author} {\bibinfo {author} {\bibfnamefont {J.}~\bibnamefont
  {Attig}}, \bibinfo {author} {\bibfnamefont {J.}~\bibnamefont {Park}},
  \bibinfo {author} {\bibfnamefont {M.~M.}\ \bibnamefont {Scherer}}, \bibinfo
  {author} {\bibfnamefont {S.}~\bibnamefont {Trebst}}, \bibinfo {author}
  {\bibfnamefont {A.}~\bibnamefont {Altland}}, \ and\ \bibinfo {author}
  {\bibfnamefont {A.}~\bibnamefont {Rosch}},\ }\bibfield  {title} {\emph
  {\bibinfo {title} {{Universal principles of moir{\'{e}} band structures}}},\
  }\href {https://iopscience.iop.org/article/10.1088/2053-1583/ac1cf0
  https://iopscience.iop.org/article/10.1088/2053-1583/ac1cf0/meta} {\bibfield
  {journal} {\bibinfo  {journal} {2D Materials}\ }\textbf {\bibinfo {volume}
  {8}},\ \bibinfo {pages} {044007} (\bibinfo {year} {2021})}\BibitemShut
  {NoStop}%
\bibitem [{Sup()}]{Supple}%
  \BibitemOpen
  \href@noop {} {}\bibinfo {note} {See the Supplemental Material at DOI for
  details on (i) a symmetry analysis to obtain the form of the matrices
  $\hat{w}_{i i'}$ and $\hat{\lambda}_{ij}$ in Eq.~\eqref{eq:wireloctunneling},
  (ii) a simple effective Hamiltonian to fully reproduce the band structure in
  Fig.~\ref{fig2}c, (iii) the mean-field phase diagram, (iv) the $SU(4)$ spin
  wave theory, and (v) the semi-classical Monte Carlo simulations. It includes
  Refs.~\cite{Wietek2022, Colpa1978, gresista2023, muller1959, cardona2019,
  binninganalysis}}\BibitemShut {NoStop}%
\bibitem [{\citenamefont {Kugel'}\ and\ \citenamefont
  {Khomskiĭ}(1982)}]{Kugel1982}%
  \BibitemOpen
  \bibfield  {author} {\bibinfo {author} {\bibfnamefont {K.~I.}\ \bibnamefont
  {Kugel'}}\ and\ \bibinfo {author} {\bibfnamefont {D.~I.}\ \bibnamefont
  {Khomskiĭ}},\ }\bibfield  {title} {\emph {\bibinfo {title} {{The Jahn-Teller
  effect and magnetism: transition metal compounds}}},\ }\href {\doibase
  10.1070/PU1982v025n04ABEH004537} {\bibfield  {journal} {\bibinfo  {journal}
  {Soviet Physics Uspekhi}\ }\textbf {\bibinfo {volume} {25}},\ \bibinfo
  {pages} {231} (\bibinfo {year} {1982})}\BibitemShut {NoStop}%
\bibitem [{\citenamefont {Wietek}\ and\ \citenamefont
  {L{\"{a}}uchli}(2017)}]{Wietek2017}%
  \BibitemOpen
  \bibfield  {author} {\bibinfo {author} {\bibfnamefont {A.}~\bibnamefont
  {Wietek}}\ and\ \bibinfo {author} {\bibfnamefont {A.~M.}\ \bibnamefont
  {L{\"{a}}uchli}},\ }\bibfield  {title} {\emph {\bibinfo {title} {{Chiral spin
  liquid and quantum criticality in extended S= 12 Heisenberg models on the
  triangular lattice}}},\ }\href {\doibase
  10.1103/PHYSREVB.95.035141/FIGURES/5/MEDIUM} {\bibfield  {journal} {\bibinfo
  {journal} {Phys. Rev. B}\ }\textbf {\bibinfo {volume} {95}},\ \bibinfo
  {pages} {035141} (\bibinfo {year} {2017})}\BibitemShut {NoStop}%
\bibitem [{\citenamefont {Gong}\ \emph {et~al.}(2017)\citenamefont {Gong},
  \citenamefont {Zhu}, \citenamefont {Zhu}, \citenamefont {Sheng},\ and\
  \citenamefont {Yang}}]{Gong2017}%
  \BibitemOpen
  \bibfield  {author} {\bibinfo {author} {\bibfnamefont {S.~S.}\ \bibnamefont
  {Gong}}, \bibinfo {author} {\bibfnamefont {W.}~\bibnamefont {Zhu}}, \bibinfo
  {author} {\bibfnamefont {J.~X.}\ \bibnamefont {Zhu}}, \bibinfo {author}
  {\bibfnamefont {D.~N.}\ \bibnamefont {Sheng}}, \ and\ \bibinfo {author}
  {\bibfnamefont {K.}~\bibnamefont {Yang}},\ }\bibfield  {title} {\emph
  {\bibinfo {title} {{Global phase diagram and quantum spin liquids in a spin-
  12 triangular antiferromagnet}}},\ }\href {\doibase
  10.1103/PHYSREVB.96.075116/FIGURES/14/MEDIUM} {\bibfield  {journal} {\bibinfo
   {journal} {Phys. Rev. B}\ }\textbf {\bibinfo {volume} {96}},\ \bibinfo
  {pages} {075116} (\bibinfo {year} {2017})}\BibitemShut {NoStop}%
\bibitem [{\citenamefont {Szasz}\ \emph {et~al.}(2020)\citenamefont {Szasz},
  \citenamefont {Motruk}, \citenamefont {Zaletel},\ and\ \citenamefont
  {Moore}}]{Szasz2020}%
  \BibitemOpen
  \bibfield  {author} {\bibinfo {author} {\bibfnamefont {A.}~\bibnamefont
  {Szasz}}, \bibinfo {author} {\bibfnamefont {J.}~\bibnamefont {Motruk}},
  \bibinfo {author} {\bibfnamefont {M.~P.}\ \bibnamefont {Zaletel}}, \ and\
  \bibinfo {author} {\bibfnamefont {J.~E.}\ \bibnamefont {Moore}},\ }\bibfield
  {title} {\emph {\bibinfo {title} {{Chiral Spin Liquid Phase of the Triangular
  Lattice Hubbard Model: A Density Matrix Renormalization Group Study}}},\
  }\href {\doibase 10.1103/PHYSREVX.10.021042/FIGURES/13/MEDIUM} {\bibfield
  {journal} {\bibinfo  {journal} {Phys. Rev. X}\ }\textbf {\bibinfo {volume}
  {10}},\ \bibinfo {pages} {021042} (\bibinfo {year} {2020})}\BibitemShut
  {NoStop}%
\bibitem [{\citenamefont {Chen}\ \emph {et~al.}(2022)\citenamefont {Chen},
  \citenamefont {Chen}, \citenamefont {Gong}, \citenamefont {Sheng},
  \citenamefont {Li},\ and\ \citenamefont {Weichselbaum}}]{Chen2022}%
  \BibitemOpen
  \bibfield  {author} {\bibinfo {author} {\bibfnamefont {B.~B.}\ \bibnamefont
  {Chen}}, \bibinfo {author} {\bibfnamefont {Z.}~\bibnamefont {Chen}}, \bibinfo
  {author} {\bibfnamefont {S.~S.}\ \bibnamefont {Gong}}, \bibinfo {author}
  {\bibfnamefont {D.~N.}\ \bibnamefont {Sheng}}, \bibinfo {author}
  {\bibfnamefont {W.}~\bibnamefont {Li}}, \ and\ \bibinfo {author}
  {\bibfnamefont {A.}~\bibnamefont {Weichselbaum}},\ }\bibfield  {title} {\emph
  {\bibinfo {title} {{Quantum spin liquid with emergent chiral order in the
  triangular-lattice Hubbard model}}},\ }\href {\doibase
  10.1103/PHYSREVB.106.094420/FIGURES/15/MEDIUM} {\bibfield  {journal}
  {\bibinfo  {journal} {Phys. Rev. B}\ }\textbf {\bibinfo {volume} {106}},\
  \bibinfo {pages} {094420} (\bibinfo {year} {2022})}\BibitemShut {NoStop}%
\bibitem [{\citenamefont {Sur}\ \emph {et~al.}(2022)\citenamefont {Sur},
  \citenamefont {Udupa},\ and\ \citenamefont {Sen}}]{Sur2022}%
  \BibitemOpen
  \bibfield  {author} {\bibinfo {author} {\bibfnamefont {S.}~\bibnamefont
  {Sur}}, \bibinfo {author} {\bibfnamefont {A.}~\bibnamefont {Udupa}}, \ and\
  \bibinfo {author} {\bibfnamefont {D.}~\bibnamefont {Sen}},\ }\bibfield
  {title} {\emph {\bibinfo {title} {{Driven Hubbard model on a triangular
  lattice: Tunable Heisenberg antiferromagnet with a chiral three-spin
  term}}},\ }\href {\doibase 10.1103/PHYSREVB.105.054423/FIGURES/13/MEDIUM}
  {\bibfield  {journal} {\bibinfo  {journal} {Phys. Rev. B}\ }\textbf {\bibinfo
  {volume} {105}},\ \bibinfo {pages} {054423} (\bibinfo {year}
  {2022})}\BibitemShut {NoStop}%
\bibitem [{\citenamefont {Kuhlenkamp}\ \emph {et~al.}()\citenamefont
  {Kuhlenkamp}, \citenamefont {Kadow}, \citenamefont {Imamoglu},\ and\
  \citenamefont {Knap}}]{Kuhlenkamp2022}%
  \BibitemOpen
  \bibfield  {author} {\bibinfo {author} {\bibfnamefont {C.}~\bibnamefont
  {Kuhlenkamp}}, \bibinfo {author} {\bibfnamefont {W.}~\bibnamefont {Kadow}},
  \bibinfo {author} {\bibfnamefont {A.}~\bibnamefont {Imamoglu}}, \ and\
  \bibinfo {author} {\bibfnamefont {M.}~\bibnamefont {Knap}},\ }\bibfield
  {title} {\emph {\bibinfo {title} {Tunable topological order of pseudo spins
  in semiconductor heterostructures}},\ }\href@noop {} {\ }\Eprint
  {http://arxiv.org/abs/2209.05506} {arXiv:2209.05506} \BibitemShut {NoStop}%
\bibitem [{\citenamefont {Stoudenmire}\ \emph {et~al.}(2009)\citenamefont
  {Stoudenmire}, \citenamefont {Trebst},\ and\ \citenamefont
  {Balents}}]{stoudenmire2009}%
  \BibitemOpen
  \bibfield  {author} {\bibinfo {author} {\bibfnamefont {E.~M.}\ \bibnamefont
  {Stoudenmire}}, \bibinfo {author} {\bibfnamefont {S.}~\bibnamefont {Trebst}},
  \ and\ \bibinfo {author} {\bibfnamefont {L.}~\bibnamefont {Balents}},\
  }\bibfield  {title} {\emph {\bibinfo {title} {{Quadrupolar correlations and
  spin freezing in $S=1$ triangular lattice antiferromagnets}}},\ }\href
  {\doibase 10.1103/PhysRevB.79.214436} {\bibfield  {journal} {\bibinfo
  {journal} {Phys. Rev. B}\ }\textbf {\bibinfo {volume} {79}},\ \bibinfo
  {pages} {214436} (\bibinfo {year} {2009})}\BibitemShut {NoStop}%
\bibitem [{\citenamefont {Hickey}\ and\ \citenamefont
  {Paramekanti}(2014)}]{hickey2014}%
  \BibitemOpen
  \bibfield  {author} {\bibinfo {author} {\bibfnamefont {C.}~\bibnamefont
  {Hickey}}\ and\ \bibinfo {author} {\bibfnamefont {A.}~\bibnamefont
  {Paramekanti}},\ }\bibfield  {title} {\emph {\bibinfo {title} {{Thermal Phase
  Transitions of Strongly Correlated Bosons with Spin-Orbit Coupling}}},\
  }\href {\doibase 10.1103/PhysRevLett.113.265302} {\bibfield  {journal}
  {\bibinfo  {journal} {Phys. Rev. Lett.}\ }\textbf {\bibinfo {volume} {113}},\
  \bibinfo {pages} {265302} (\bibinfo {year} {2014})}\BibitemShut {NoStop}%
\bibitem [{\citenamefont {Landau}\ and\ \citenamefont
  {Binder}(2014)}]{LandauBinder}%
  \BibitemOpen
  \bibfield  {author} {\bibinfo {author} {\bibfnamefont {D.~P.}\ \bibnamefont
  {Landau}}\ and\ \bibinfo {author} {\bibfnamefont {K.}~\bibnamefont
  {Binder}},\ }\href {\doibase 10.1017/CBO9781139696463} {\emph {\bibinfo
  {title} {{A Guide to Monte Carlo Simulations in Statistical Physics}}}}\
  (\bibinfo  {publisher} {Cambridge University Press},\ \bibinfo {year}
  {2014})\BibitemShut {NoStop}%
\bibitem [{\citenamefont {Mermin}\ and\ \citenamefont
  {Wagner}(1966)}]{Mermin-1966}%
  \BibitemOpen
  \bibfield  {author} {\bibinfo {author} {\bibfnamefont {N.~D.}\ \bibnamefont
  {Mermin}}\ and\ \bibinfo {author} {\bibfnamefont {H.}~\bibnamefont
  {Wagner}},\ }\bibfield  {title} {\emph {\bibinfo {title} {{Absence of
  Ferromagnetism or Antiferromagnetism in One- or Two-Dimensional Isotropic
  Heisenberg Models}}},\ }\href {\doibase 10.1103/PhysRevLett.17.1133}
  {\bibfield  {journal} {\bibinfo  {journal} {Phys. Rev. Lett.}\ }\textbf
  {\bibinfo {volume} {17}},\ \bibinfo {pages} {1133} (\bibinfo {year}
  {1966})}\BibitemShut {NoStop}%
\bibitem [{\citenamefont {Villain}\ \emph {et~al.}(1980)\citenamefont
  {Villain}, \citenamefont {Bidaux}, \citenamefont {Carton},\ and\
  \citenamefont {Conte}}]{Villain1980}%
  \BibitemOpen
  \bibfield  {author} {\bibinfo {author} {\bibfnamefont {J.}~\bibnamefont
  {Villain}}, \bibinfo {author} {\bibfnamefont {R.}~\bibnamefont {Bidaux}},
  \bibinfo {author} {\bibfnamefont {J.-P.}\ \bibnamefont {Carton}}, \ and\
  \bibinfo {author} {\bibfnamefont {R.}~\bibnamefont {Conte}},\ }\bibfield
  {title} {\emph {\bibinfo {title} {Order as an effect of disorder}},\ }\href
  {\doibase 10.1051/jphys:0198000410110126300} {\bibfield  {journal} {\bibinfo
  {journal} {J. Phys.}\ }\textbf {\bibinfo {volume} {41}},\ \bibinfo {pages}
  {1263} (\bibinfo {year} {1980})}\BibitemShut {NoStop}%
\bibitem [{\citenamefont {Chalker}\ \emph {et~al.}(1992)\citenamefont
  {Chalker}, \citenamefont {Holdsworth},\ and\ \citenamefont
  {Shender}}]{Chalker-1992}%
  \BibitemOpen
  \bibfield  {author} {\bibinfo {author} {\bibfnamefont {J.~T.}\ \bibnamefont
  {Chalker}}, \bibinfo {author} {\bibfnamefont {P.~C.~W.}\ \bibnamefont
  {Holdsworth}}, \ and\ \bibinfo {author} {\bibfnamefont {E.~F.}\ \bibnamefont
  {Shender}},\ }\bibfield  {title} {\emph {\bibinfo {title} {{Hidden order in a
  frustrated system: Properties of the Heisenberg Kagom\'e antiferromagnet}}},\
  }\href {\doibase 10.1103/PhysRevLett.68.855} {\bibfield  {journal} {\bibinfo
  {journal} {Phys. Rev. Lett.}\ }\textbf {\bibinfo {volume} {68}},\ \bibinfo
  {pages} {855} (\bibinfo {year} {1992})}\BibitemShut {NoStop}%
\bibitem [{\citenamefont {Wittig}\ \emph {et~al.}(2023)\citenamefont {Wittig},
  \citenamefont {Dominguez}, \citenamefont {Beule},\ and\ \citenamefont
  {Recher}}]{wittig2023}%
  \BibitemOpen
  \bibfield  {author} {\bibinfo {author} {\bibfnamefont {P.}~\bibnamefont
  {Wittig}}, \bibinfo {author} {\bibfnamefont {F.}~\bibnamefont {Dominguez}},
  \bibinfo {author} {\bibfnamefont {C.~D.}\ \bibnamefont {Beule}}, \ and\
  \bibinfo {author} {\bibfnamefont {P.}~\bibnamefont {Recher}},\ }\href@noop {}
  {\emph {\bibinfo {title} {Localized states coupled to a network of chiral
  modes in minimally twisted bilayer graphene}}} (\bibinfo {year} {2023}),\
  \Eprint {http://arxiv.org/abs/2303.03901} {arXiv:2303.03901
  [cond-mat.mes-hall]} \BibitemShut {NoStop}%
\bibitem [{\citenamefont {Wietek}\ \emph {et~al.}(2022)\citenamefont {Wietek},
  \citenamefont {Wang}, \citenamefont {Zang}, \citenamefont {Cano},
  \citenamefont {Georges},\ and\ \citenamefont {Millis}}]{Wietek2022}%
  \BibitemOpen
  \bibfield  {author} {\bibinfo {author} {\bibfnamefont {A.}~\bibnamefont
  {Wietek}}, \bibinfo {author} {\bibfnamefont {J.}~\bibnamefont {Wang}},
  \bibinfo {author} {\bibfnamefont {J.}~\bibnamefont {Zang}}, \bibinfo {author}
  {\bibfnamefont {J.}~\bibnamefont {Cano}}, \bibinfo {author} {\bibfnamefont
  {A.}~\bibnamefont {Georges}}, \ and\ \bibinfo {author} {\bibfnamefont
  {A.}~\bibnamefont {Millis}},\ }\bibfield  {title} {\emph {\bibinfo {title}
  {{Tunable stripe order and weak superconductivity in the Moir\'e Hubbard
  model}}},\ }\href {\doibase 10.1103/PhysRevResearch.4.043048} {\bibfield
  {journal} {\bibinfo  {journal} {Phys. Rev. Res.}\ }\textbf {\bibinfo {volume}
  {4}},\ \bibinfo {pages} {043048} (\bibinfo {year} {2022})}\BibitemShut
  {NoStop}%
\bibitem [{\citenamefont {Colpa}(1978)}]{Colpa1978}%
  \BibitemOpen
  \bibfield  {author} {\bibinfo {author} {\bibfnamefont {J.}~\bibnamefont
  {Colpa}},\ }\bibfield  {title} {\emph {\bibinfo {title} {Diagonalization of
  the quadratic boson hamiltonian}},\ }\href
  {https://www.sciencedirect.com/science/article/pii/0378437178901607}
  {\bibfield  {journal} {\bibinfo  {journal} {Physica A: Statistical Mechanics
  and its Applications}\ }\textbf {\bibinfo {volume} {93}},\ \bibinfo {pages}
  {327} (\bibinfo {year} {1978})}\BibitemShut {NoStop}%
\bibitem [{\citenamefont {Gresista}\ \emph {et~al.}()\citenamefont {Gresista},
  \citenamefont {Kiese}, \citenamefont {Trebst},\ and\ \citenamefont
  {Scherer}}]{gresista2023}%
  \BibitemOpen
  \bibfield  {author} {\bibinfo {author} {\bibfnamefont {L.}~\bibnamefont
  {Gresista}}, \bibinfo {author} {\bibfnamefont {D.}~\bibnamefont {Kiese}},
  \bibinfo {author} {\bibfnamefont {S.}~\bibnamefont {Trebst}}, \ and\ \bibinfo
  {author} {\bibfnamefont {M.~M.}\ \bibnamefont {Scherer}},\ }\bibfield
  {title} {\emph {\bibinfo {title} {{Spin-valley magnetism on the triangular
  moiré lattice with SU(4) breaking interactions}}},\ }\href@noop {} {\
  }\Eprint {http://arxiv.org/abs/2210.080256} {arXiv:2210.080256} \BibitemShut
  {NoStop}%
\bibitem [{\citenamefont {Muller}(1959)}]{muller1959}%
  \BibitemOpen
  \bibfield  {author} {\bibinfo {author} {\bibfnamefont {M.~E.}\ \bibnamefont
  {Muller}},\ }\bibfield  {title} {\emph {\bibinfo {title} {{A Note on a Method
  for Generating Points Uniformly on N-Dimensional Spheres}}},\ }\href
  {\doibase 10.1145/377939.377946} {\bibfield  {journal} {\bibinfo  {journal}
  {Commun. ACM}\ }\textbf {\bibinfo {volume} {2}},\ \bibinfo {pages} {19–20}
  (\bibinfo {year} {1959})}\BibitemShut {NoStop}%
\bibitem [{\citenamefont {Alzate-Cardona}\ \emph {et~al.}(2019)\citenamefont
  {Alzate-Cardona}, \citenamefont {Sabogal-Suárez}, \citenamefont {Evans},\
  and\ \citenamefont {Restrepo-Parra}}]{cardona2019}%
  \BibitemOpen
  \bibfield  {author} {\bibinfo {author} {\bibfnamefont {J.~D.}\ \bibnamefont
  {Alzate-Cardona}}, \bibinfo {author} {\bibfnamefont {D.}~\bibnamefont
  {Sabogal-Suárez}}, \bibinfo {author} {\bibfnamefont {R.~F.~L.}\ \bibnamefont
  {Evans}}, \ and\ \bibinfo {author} {\bibfnamefont {E.}~\bibnamefont
  {Restrepo-Parra}},\ }\bibfield  {title} {\emph {\bibinfo {title} {{Optimal
  phase space sampling for Monte Carlo simulations of Heisenberg spin
  systems}}},\ }\href {\doibase 10.1088/1361-648X/aaf852} {\bibfield  {journal}
  {\bibinfo  {journal} {Journal of Physics: Condensed Matter}\ }\textbf
  {\bibinfo {volume} {31}},\ \bibinfo {pages} {095802} (\bibinfo {year}
  {2019})}\BibitemShut {NoStop}%
\bibitem [{\citenamefont {Bauer}\ and\ \citenamefont
  {Freyer}(2020)}]{binninganalysis}%
  \BibitemOpen
  \bibfield  {author} {\bibinfo {author} {\bibfnamefont {C.}~\bibnamefont
  {Bauer}}\ and\ \bibinfo {author} {\bibfnamefont {F.}~\bibnamefont {Freyer}},\
  }\href {\doibase 10.5281/zenodo.3603347} {\emph {\bibinfo {title}
  {{BinningAnalysis.jl: Standard error estimation tools}}}} (\bibinfo {year}
  {2020})\BibitemShut {NoStop}%
\end{thebibliography}%

\appendix


\onecolumngrid
%
%
%
%
%
%
%

\section{Symmetry analysis for the network model}
In this Appendix, we discuss the symmetry of the \moire system considered in the main text, and identify the form of the coupling $\hat{w}$ and $\hat{\lambda}$ (Eq.~\eqref{eq:wireloctunneling}) using a symmetry analysis.



\subsection{Symmetry}
The underlying lattice or discrete symmetries of our \moire model are the \moire translation, the 120$^\circ$ rotation ($C_3$), the mirror ($M_y$) with respect to the $y$ axis, the inversion ($\mathcal{I}$), and the time reversal symmetry ($\mathcal{T}$) and combinations thereof. Since the \moire potential varies smoothly so that the large-momentum transfer is highly suppressed, it is a good approximation to consider the two valley sectors separately. 
Therefore, we only consider the valley-conserving symmetries generated by $C_3$, $M_y$, and $\mathcal{T} \mathcal{I}$, in the following. 

The underlying lattice symmetry group is the dehedral group $D_3$ with $C_3$ and $M_y$. More precisely, in order to deal with the spinor wave function properly,
one has to take into account a minus sign under $2 \pi$ rotation, and therefore consider the dicyclic group $\text{Dic}_3$, which extends $D_3$. 
The character table of  $\text{Dic}_3$ is as follows:
$$ \begin{tabularx}{0.8\textwidth} { 
  | >{\centering\arraybackslash}X 
  | >{\centering\arraybackslash}X 
  | >{\centering\arraybackslash}X 
  | >{\centering\arraybackslash}X 
  | >{\centering\arraybackslash}X 
  | >{\centering\arraybackslash}X 
  | >{\centering\arraybackslash}X |}
 \hline
  $\mathbf{Dic}_{3}$ & $\mathbf{A}$ & $\mathbf{B}$ & $\mathbf{C}$ & $\mathbf{D}$ & $\mathbf{E}$ & $\mathbf{F}$ \\
 \hline
 $\rho_{1}$ & $1$ & $1$ & $1$ & $1$ & $1$ & $1$  \\
\hline
 $\rho_2$ & $1$ & $1$ & $1$ & $-1$ & $-1$ & $1$  \\
\hline
 $\rho_3$ & $1$ & $-1$ & $-1$ & $i$ & $-i$ & $1$  \\
\hline
 $\rho_4$ & $1$ & $-1$ & $-1$ & $-i$& $i$ & $1$  \\
\hline
 $\rho_5$ & $2$ & $2$ & $-1$ & $0$ & $0$ & $-1$  \\
\hline
 $\rho_6$ & $2$ & $-2$ & $1$ & $0$ & $0$ & $-1$  \\
\hline
\end{tabularx} $$
with the 6 irreducible representations $\rho_{j=1,\cdots, 6}$ and the equivalent classes $\mathbf{A}, \mathbf{B},\mathbf{C},\mathbf{D},\mathbf{E},\mathbf{F}$ which are given by
$$\mathbf{A} \equiv \{1\}, \ \mathbf{B} \equiv \{C^3_3\}, \ \mathbf{C} \equiv \{C_3, C^5_3\},\ \mathbf{D} \equiv \{M_y C_3, M_yC^3_3, M_yC^5_3\}, \ \mathbf{E}  \equiv \{M_y, M_yC^2_3, M_yC^4_3\}, \ \mathbf{F} \equiv \{C^2_3, C^4_3\}.$$
For states with one electron per mode, only three of the representations, $\rho_3$,  $\rho_4$, $\rho_6$, are relevant  since the condition $C^{3}_3=-1$ should be fulfilled for the spinor wave functions. 
While $\rho_3$ and $\rho_4$ are one-dimensional representations, $\rho_6$ is a two-dimensional representation. 

As shown in Fig.~\ref{fig3}, the \moire system is effectively described by the network model where localized states form at the junction of three 1D channels (for each valley) and are weakly coupled to the channels. Below, employing a symmetry analysis, we shall find the form of the inter-channel coupling $\hat{w}$ and the coupling of 1D channels to localized states $\hat{\lambda}$ (Eq.~\eqref{eq:wireloctunneling}).

\subsection{Localized modes} 

Localized states can be labeled by representation $\rho_{j=3,4,6}$ of the group $\text{Dic}_{3}$. 
From the character table, one can find relevant matrices for the transformation $C_{3}$ and $M_y$ in each of the representations:
$C_3^{\rho_3} = -1, C_3^{\rho_4} = -1, C_3^{\rho_6} = e^{i\pi \sigma_z/3}$ and $M_y^{\rho_3} = -i, \ M_y^{\rho_4} = i,M_y^{\rho_6}=e^{i\pi \sigma_y/2}$, respectively. A simple way to identify the symmetry of a given localized state from the band-structure calculation is to analyze the symmetry properties of the eigenfunction of flat bands at the $\Gamma$ point. For each of the three representations we find examples in our band structure calculations, see Fig. 2c of the main text.

\subsection{One-dimensional channels}

From scaling, one finds that the width of the 1D channels is given by $d_{1D} \sim (v L /u_s)^{1/2}$ with $d_{1D} \ll L$ for $u_s \gg v G$. $d_{1D} \ll L$ renders that the coupling of 1D channels being far apart is highly suppressed, and thus it is a good approximation to only consider the coupling of the neighboring channels at the junction. The symmetry properties are determined by the spinor structure of the eigenfunctions. The three channels at the junction are related to each other by a $120^\circ$ rotation matrix, $C_3^{\text{1D}}$, and the mirror transformation matrix, $M_y^{\text{1D}}$, e.g., for the $K$ valley, written by
\begin{align}
C_3^{\text{1D}}=\left(
\begin{array}{ccc} 0 & 0 & -1 \\
             1 & 0 & 0 \\
              0& 1 &  0\end{array}
 \right) , \  M_y^{\text{1D}}=\left(
\begin{array}{ccc} -i & 0 & 0 \\
             0 & 0 & i \\
              0& i &  0\end{array}
 \right).
\end{align}
For the $K'$ valley, the mirror transformation matrix has an extra overall minus sign as $K$ and $K'$ are related by time-reversal and thus by  complex conjugation.
The relative minus sign in one of the matrix elements of $C_3$ and $M_y$ reflects the fact that the $2 \pi$ rotation preinor wave function gets a minus sign. Also $(\mathcal{I} \mathcal{T})^{\text{1D}} = C \mathds{1}_3$, where $C$ is the complex conjugation and $\mathds{1}_3$ is the 3$\times$3 unit matrix. The inter-channel tunneling $H_{w}$ is written as 
\begin{align} \label{supeq:wirewiretunneling}
H_{w}=&\sum_{\text{crossing at } \vec{R}_m}  \hat{w}_{i i'}\Psi^\dagger_{n,i,\alpha,\sigma}(\rho_{n,i,m})\Psi_{n',i',\alpha,\sigma}(\rho_{n',i',m}). 
\end{align}
By imposing the symmetry constraints, $\hat{w} = (C_3^{\text{1D}})^{\dagger}\hat{w} C_3^{\text{1D}}$, $\hat{w} =(M_y^{\text{1D}})^{\dagger}\hat{w} M_y^{\text{1D}}$, and $\hat{w} =((\mathcal{I} \mathcal{T})^{\text{1D}})^{\dagger}\hat{w} (\mathcal{I} \mathcal{T})^{\text{1D}}$, one can obtain $\hat w$, parameterized by a single real parameter $w$ as 
\begin{align} \label{supeq:wirewirecouplingmatrix}
\hat w=&w L \left(\begin{array}{ccc} 0 & 1 &-1 \\ 1 & 0 &1\\ -1 & 1 & 0\end{array}\right). 
\end{align}
The size of moir\'e unit cell, $L$, is used such that $w$ has units of energy.

\subsection{Coupling between localized and propagating modes}

The coupling of 1D channels to localized modes, Eq.~\eqref{eq:wireloctunneling}, depends on to which representation of $\text{Dic}_3$ the localized mode belongs. Similarly to the $\hat{w}$ matrix, this coupling matrix $\hat{\lambda}$ can be obtained by the symmetry constraints, $\hat{\lambda} = (C_3^{\rho_j})^{\dagger} \hat{\lambda} C_3^{\text{1D}}$, $\hat{\lambda}  = (M_y^{\rho_j})^{\dagger} \hat{\lambda} M_y^{\text{1D}}$, and $\hat{\lambda}  = ((\mathcal{I} \mathcal{T})^{\rho_j})^{\dagger} \hat{\lambda} (\mathcal{I} \mathcal{T})^{\text{1D}}$, for each of the representation $j = 3,4,6$. For the $K$ valley, it is given by 
\begin{align} \label{supeq:wireloccoupling}
H_{\lambda}=& \sum_{\text{crossing at } \vec{R}_m} 
\hat{\lambda}_{ij}\, d^\dagger_{m,\alpha,\sigma,j} \Psi_{n,i,\alpha,\sigma}(\rho_{n,i,m}) +h.c., \nonumber \\
\hat{\lambda}=& \lambda  \frac{\sqrt{L}}{\sqrt{3}} \left\{ \begin{array}{ll} (1,-1,1) & \rho_3\  \text{representation} \\[1mm]
(0,0,0) & \rho_4  \ \text{representation} \\[1mm]
\left(
\begin{array}{ccc} 1 & e^{i \pi/3} & e^{2 i \pi/3} \\
             -i & -i e^{-i \pi/3} & -i e^{-2 i \pi/3} \end{array}
 \right) & \rho_6\ \text{representation}
 \end{array}\right.  
\end{align}
 Interestingly, the localized modes in the $\rho_4$ representation do not couple to the closest 1D channels at all. This vanishing coupling can be understood by the symmetry of the system. For example, let us consider the coupling of a localized mode in the $\rho_4$ representation to the neighboring channel propagating in the $y$ direction. While the localized mode has the eigenvalue $i$ under the mirror transformation $M_y$, the channel has the eigenvalue $-i$. It implies that the coupling matrix $\hat{\lambda}$ has to have an extra minus sign under $M_y$, and therefore has to vanish. The same argument can be applied to the two other neighboring channels with the symmetry transformation $M_y C_3$, and $M_y C_3^2$. Nevertheless, tunneling to channels further away are still possible but exponentially suppressed in the ratio of potential and $v G$ as $e^{-L / d_{1D}} \sim e^{- (u_s L/ v)^{1/2}}$.

The coupling matrices for the $K'$ valley can be obtained by time reversal, which implies that the time-reversed partner of $\rho_3$ and $\rho_4$ have an identical coupling matrix $\hat \lambda$. Formally, under time-reversal $\rho_3$ maps to $\rho_4$ but this effect is compensated because also in Eq.~\eqref{supeq:wireloccoupling} the entries for $\rho_3$ and $\rho_4$ are exchanged when one switches from $K$ to $K'$. The matrix $\hat \lambda$ for $\rho_6$ in the $K'$ valley is obtained by complex conjugation of $\hat \lambda$. 

\section{Effective Hamiltonian in momentum space} 

In this section, we construct an effective Hamiltonian for the network model in momentum space and show that the band structure obtained from the diagonalization of the \moire potential model, Eq.~\eqref{eq:Heff}, is nicely fitted by the band structure obtained form this effective model. 

\begin{figure}[t]
	\centering
	\includegraphics[width= .7\columnwidth]{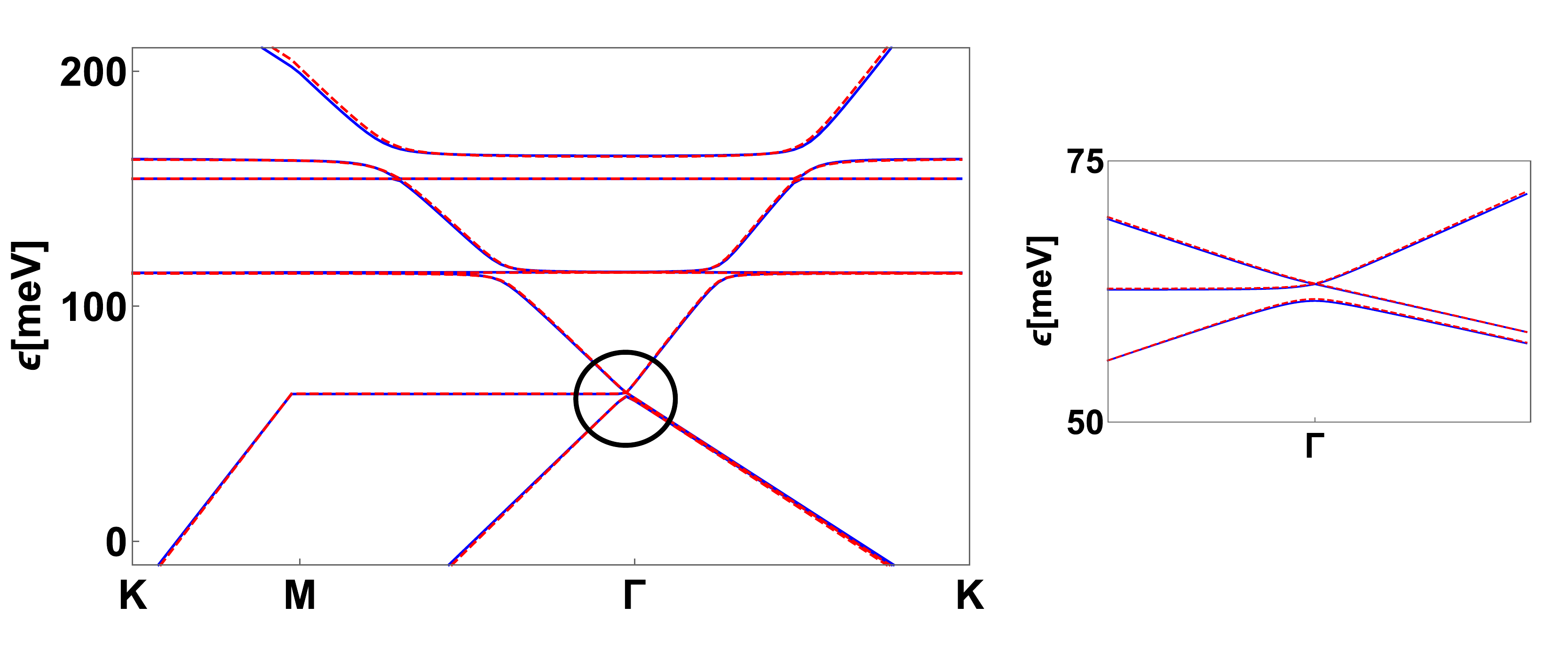}
	\caption{{\bf Comparison of two band structures.} Left: The band structure (blue solid curves) obtained from diagonalizing the original \moire potential model, Eq.~\eqref{eq:Heff}, and the band structure (red dashed curves) using the effective Hamiltonian, Eq.~\eqref{supeq:effectiveHam}. The parameters of Eq.~\eqref{supeq:effectiveHam} can be obtained by fitting the blue solid curves; $w = 0.2$meV, $\lambda_6 = 4.3$meV, $\lambda_3=11.3$meV, $\epsilon_{\rm{1D}}=63.4$meV, $\epsilon_6=114$meV, $\epsilon_4=154$meV, and $\epsilon_3=162$meV. Those band structures match remarkably, confirming the validity of our network model. For the blue curves, we have used the parameters,  $u_{AA} = u_{BB} = 0$ and $u_{AB} = u_{BA} = \frac{8 \pi v}{\sqrt{3} L }$. Right: a blowup of the band structures near the $\Gamma$ point (indicated by the black circle) where the three propagating bands meet. Weak level repulsion between the three bands is clearly shown, not visible in the left band structure. 
	}
	\label{Supfig:Fitting}
\end{figure}

As shown in Fig.~2(c) for the band structure in the main text (also the blue solid curves in Fig.~\ref{Supfig:Fitting}), the inter-channel coupling and the coupling of 1D channels to localized states are very weak. For example, a blowup of the band structure near the $\Gamma$ point (indicated by the black curcle) in Fig.~\ref{Supfig:Fitting} shows that a level-repulsion of the three 1D channels is an order of hundreds of $\mu$eV. 
Such weak couplings (i.e., $w, \lambda \ll v G$) allow us to construct an effective Hamiltonian from Eqs.~\eqref{eq:kineticenergy} and \eqref{eq:wireloctunneling}. Depending on the number of localized modes, $N_{\rm{loc}}$, within the energy window of $\epsilon \in [-v G/2, v G/2]$ near the chemical potential, this effective Hamiltonian $H_{\rm{eff}} (k_x, k_y)$ for a given valley can be described by a $(3+N_{\rm{loc}}) \times (3+N_{\rm{loc}})$ matrix for each momentum.  
For example, the band structure for the $K$ valley shown in Fig.~\ref{fig2}c can be reproduced with high precision by the $7 \times 7$ matrix, given by 
\begin{align}
H_{\rm{eff}} (k_x, k_y) = \left(
\begin{array}{ccccccc} - v k_y + \epsilon_{\rm{1D}} & w & -w & \frac{\lambda_6}{\sqrt{3}} & i \frac{\lambda_6}{\sqrt{3}} & \frac{\lambda_3}{\sqrt{3}} & 0  \\
             w & - \frac{v}{2} \left ( \sqrt{3} k_x - k_y \right) + \epsilon_{\rm{1D}} & w & \frac{\lambda_6}{\sqrt{3}} e^{-i\frac{\pi}{3}} & i\frac{\lambda_6}{\sqrt{3}} e^{i\frac{\pi}{3}} & -\frac{\lambda_3}{\sqrt{3}} &0\\
              -w & w &  \frac{v}{2} \left ( \sqrt{3} k_x + k_y \right) + \epsilon_{\rm{1D}} &\frac{\lambda_6}{\sqrt{3}} e^{-2i\frac{\pi}{3}} & i\frac{\lambda_6}{\sqrt{3}} e^{2i\frac{\pi}{3}} & \frac{\lambda_3}{\sqrt{3}} &0 \\
              \frac{\lambda_6}{\sqrt{3}} & \frac{\lambda_6}{\sqrt{3}} e^{i \frac{\pi}{3}} &\frac{\lambda_6}{\sqrt{3}} e^{2i \frac{\pi}{3}} & \epsilon_6 & 0&0&0
              \\ -i \frac{\lambda_6}{\sqrt{3}} & -i\frac{\lambda_6}{\sqrt{3}} e^{i \frac{\pi}{3}} & -i\frac{\lambda_6}{\sqrt{3}} e^{2i \frac{\pi}{3}} & 0 &\epsilon_6  &0&0\\
              \frac{\lambda_3}{\sqrt{3}} & -\frac{\lambda_3}{\sqrt{3}} &\frac{\lambda_3}{\sqrt{3}} & 0 &0 & \epsilon_3&0 \\ 
              0&0&0&0&0&0& \epsilon_4
              \end{array}
 \right).  \label{supeq:effectiveHam}
\end{align}
The first three rows and columns of this effective Hamiltonian describe the 1D channels propagating along the three directions, the 4- and 5th rows and columns describe degenerate localized states that belong to the $\rho_6$ representation with energy $\epsilon_6$, and the 6 and 7th row and column correspond to localized states that belongs to the $\rho_3$ and $\rho_4$ representation with energy $\epsilon_3$ and $\epsilon_4$, respectively. As shown in Eqs.~\eqref{supeq:wirewiretunneling} and \eqref{supeq:wireloccoupling}, the form of the coupling terms is determined by the symmetry of the system and the representation of the localized states. 

Figure \ref{Supfig:Fitting} compares band structures obtained from two different Hamiltonians.
The band structure plotted in blue dashed curves is numerically obtained from the diagonalization of the \moire model, Eq.~\eqref{eq:Heff}. On the other hand, the band structure plotted in blue solid line is obtained from the $7 \times 7$ effective Hamiltonian in Eq.~\eqref{supeq:effectiveHam} with the parameters of $w = 0.2$meV, $\lambda_6 = 4.3$meV, $\lambda_3=11.3$meV, $\epsilon_{\rm{1D}}=63.4$meV, $\epsilon_6=114$meV, $\epsilon_4=154$meV, and $\epsilon_3=162$meV, which can be found by fitting to the blue dashed curves. Those band structures remarkably match well, including silent features. It confirms the validity of our network model.

\section{Mean-field phase diagram}

In this section, we discuss the mean-field phase diagram of the Hamiltonian, Eq.~\eqref{eq:simspinvalleyHam}, given by 
\begin{align} \label{eq:spinvalleyHam_supple}
    H_{\rm{sv}} &= J_2 \sum_{\langle m_1 \to m_2 \rangle_c}(1 + \vec{\sigma}_{m_1} \cdot \vec{\sigma}_{m_2} )  (
    e^{i \varphi} \tau^{+}_{m_1} \tau^{-}_{m_2} + h.c.)
 + J_3 \!\!\!\! \sum_{p = \triangleright/\triangleleft,(m_1, m_2, m_3)_p} \!\!\!\! (-1)^p  \vec{\sigma}_{m_1}\cdot(\vec{\sigma}_{m_2} \vec{\times} \vec{\sigma}_{m_3}) \Big(\prod_{i=1}^3 P_{m_i}^{+} - \prod_{i=1}^3 P_{m_i}^{-} \Big) \nonumber \\ 
 &+ J_2' \sum_{\langle m_1, m_2 \rangle_c}(1 + \vec{\sigma}_{m_1} \cdot \vec{\sigma}_{m_2} )  (1 + \tau^{z}_{m_1} \tau^{z}_{m_2})\,\,\,\,.
\end{align}
Compared with the Hamiltonians (Eqs.~\eqref{eq:twospininteraction}-\eqref{eq:twospintermJprime}) obtained from a perturbation theory in $J$ and $w$, the Hamiltonian contains only the nearest neighbor interactions. 
The continuous symmetries of $H_{\rm{sv}}$ are $U(1) \times SU(2) \times SU(2)$ generated by $\tau^z$, $P^+ \vec \sigma$ and $P^- \vec \sigma$. 
Assuming that the filling of the localized states is unity, the states on the $SU(4)$ space are spanned by a 4-component complex vector. In this basis, we solve the self-consistent mean-field equations iteratively with the fixed unit cell. We use 4 different unit cells with one-, two-, three-, four-sublattices. We find that either a one-sublattice or a three-sublattice solution has the lowest energy. Therefore we consider only these cases in the following.

The model is parametrized by 3 dimensionless parameters $J_3/J_2$, $J_2'/J_2$ and $\varphi$. Experimentally, we expect $|J_3/J_2|, |J_2'/J_2| \ll 1$ while $\varphi$ can take arbitarily large values. The parameter $\varphi$ can be viewed as an Aharonov-Bohm phase acquired by a particle with charge $\tau^z$ moving along a side of the triangular loop.
As $3 \varphi$ is the total phase accumulated along the triangular loop, it is possible to use a transformation to change the total phase by $2 \pi$ which is equivalent to a change of $\varphi$ by $n 2\pi/3$, $n \in \mathbb Z$.
This is achieved by doing $\tau^z$ rotations of the spins on the A, B, C (see the inset of Fig.~\ref{fig4:phasediagram}) by $0$, $2\pi/3$, $4 \pi/3$
\begin{align} \label{supeq:tauzrotation}
\varphi\rightarrow \varphi+\frac{2\pi}{3},\,\,\,\vec{\tau}_{\rm A} \rightarrow \vec{\tau}_{\rm A},\,\,\, \vec{\tau}_{\rm B} \rightarrow e^{i \tau^z_{\rm B} \frac{2\pi}{3}}\vec{\tau}_{\rm B}  e^{-i \tau^z_{\rm B} \frac{2\pi}{3}},\,\,\, \vec{\tau}_{\rm C} \rightarrow e^{i \tau^z_{\rm C} \frac{4\pi}{3}}\vec{\tau}_{\rm C}  e^{-i \tau^z_{\rm C} \frac{4\pi}{3}}.
\end{align}
If one uses the Aharonov-Bohm analogy described above, this would be a gauge transformation. In our system, however, a rotation by $\tau^z$ changes the physical state. The transformation maps, for example, a ferromagnetic state obtained for $\varphi=\pi$ to a state with $120^\circ$ order for $\varphi=\pm \pi/3$, see below.

\begin{figure}[t]
	\centering
    \includegraphics[width=.7\columnwidth]{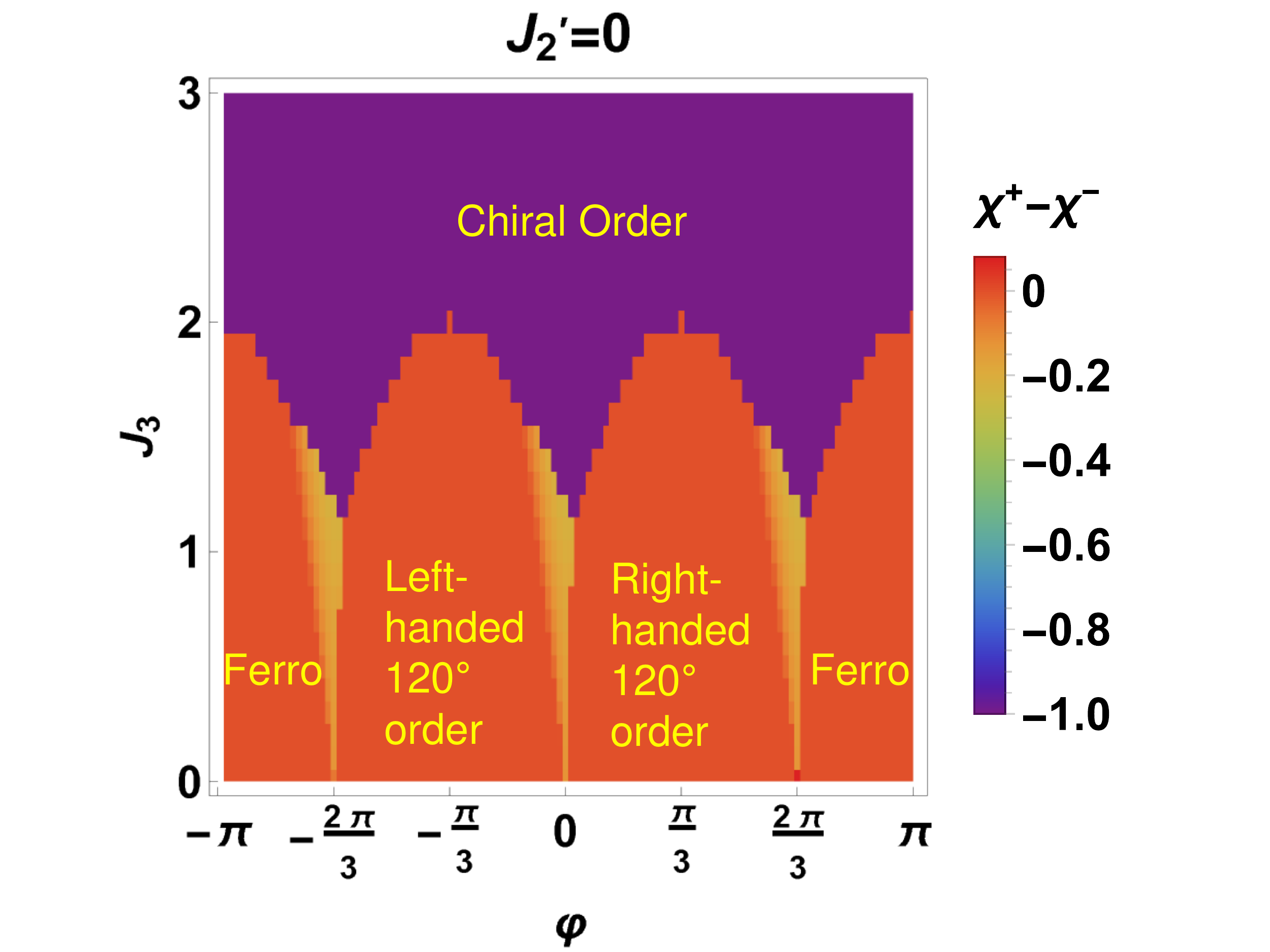}
	\caption{
	{\bf Mean-field phase diagram} with the order parameter $(\chi^{+} - \chi^{-})$ in $(\varphi, J_3)$ parameter space. $J_2' = 0$ and $J_2 = 1$. This diagram extends Fig.~\ref{fig4:phasediagram}a, which only covers small $\varphi$ (defined modulo $2 \pi/3$) and weak $J_3$ near the highly singular points. While three different types of coplanar phase (two $120^\circ$ orders and a ferromagnetic order) are realized in weak $J_3$, the chiral order achieves the maximal values  of chirality $\chi^{+}- \chi^- = -1$ for sufficiently strong $J_3>0$ (purple region). 
   }\label{fig5}
\end{figure}

As $J_2 >0$ is the largest term, we first analyze the case $J_3= J'_2=0$. 
For $\varphi=0$, the Hamiltonian, Eq.~\eqref{eq:spinvalleyHam_supple}, becomes a variant of the Kugel-Khomskii model \cite{Kugel1982}
\begin{align} \label{sup:H0}
H_0&=H_{\rm{sv}} =2J_2 \sum_{\langle m_1 \to m_2 \rangle_c}(1 + \vec{\sigma}_{m_1} \cdot \vec{\sigma}_{m_2} )  
    (\tau^{x}_{m_1} \tau^{x}_{m_2} + \tau^{y}_{m_1} \tau^{y}_{m_2} )\nonumber \\
    &= 2J_2\sum_{\langle m_1 \to m_2 \rangle_c} \vec{\mu}^1_{m_1}\cdot \vec{\mu}^1_{m_2}+ \vec{\mu}^2_{m_1} \cdot \vec{\mu}^2_{m_2},
\end{align}
 with the 4-component vectors given by $\vec \mu^1_m=(\tau^x_m, \tau^y_m \sigma^x_m, \tau^y_m \sigma^y_m, \tau^y_m \sigma^z_m)$ and  $\vec \mu^2_m=(\tau^y_m, \tau^x_m \sigma^x_m, \tau^x_m \sigma^y_m, \tau^x_m \sigma^z_m)$.
 Numerically, we find from our mean-field analysis that the ground state has the properties that 
 the vectors $\langle \vec \mu^n_m\rangle$, $n=1,2$, have the norm $1$, and show 120$^\circ$ order, such that  $\langle \vec \mu^n_{m_1}\rangle\cdot  \langle \vec \mu^n_{m_2}\rangle=\cos(2 \pi/3)=-\frac{1}{2}$ for neighboring sites. 
 Thus, this specific type of 120$^\circ$ order is realized with 4-component vectors. More precisely, two different types of the 120$^\circ$ order are realized in the ground-state manifold of $H_0$: a right-handed and left-handed 120$^\circ$ order. The states of the right (left)-handed 120$^\circ$ order rotate in anti-clockwise (clockwise) order around the $\tau^z$ orientation along the triangular loop ($A \to B \to C$) as 
 \begin{align} 
 \psi_B = e^{\mp i\tau^z \frac{2\pi}{3}}\psi_A,\,\,\,\,\,\, \psi_C = e^{\mp i\tau^z \frac{4\pi}{3}}\psi_A. 
 \end{align}
 Those 120$^\circ$ orders are staggered such that the states in the neighboring triangles circulate in the opposite direction.

Surprisingly, the above described 120$^\circ$ orders have an extra degree of freedom. 
This degree of freedom can be revealed by analyzing the valley-projected magnetization vectors, $\langle \vec{\sigma} P^\pm \rangle$, which are
length $1/2$ in the ground-state manifold. In one of the two valley, the magnetization is always ferromagnetic, but  in the other valley a non-coplanar spin configuration is possible. This non-coplanar spin configuration can be fully characterized by the opening angle $0 \leq \theta \leq \pi$ of the magnetization vectors of the three neighboring sites. While the opening angle $0$ and $\pi$ correspond to two distinct spin ferromagnetic states, see below, $\pi/2$ corresponds to the coplanar $120^\circ$ states. 

It is possible to write down analytically the spinor wavefunctions in the A, B and C sublattices within the ground-state manifold.
Up to rotations using the continuous symmetries $U(1) \times SU(2) \times SU(2)$, they take the form
\begin{align} \label{supple:parameterizationofgroundstate}
\psi^{+}_A &=\frac{1}{\sqrt{2}} \left ( \begin{array}{c} 1  \\ 1  \\ 0  \\ 0 \end{array} \right ),\,\,\,\,\,
\psi^{+}_B =\frac{1}{\sqrt{2}} \left ( \begin{array}{c} 1  \\ -\frac{1}{2} + i \frac{\sqrt{3}}{2} \cos \theta \\ 0  \\ i \frac{\sqrt{3}}{2} \sin \theta \end{array} \right ),\,\,\,\,\,
\psi^{+}_C =\frac{1}{\sqrt{2}} \left ( \begin{array}{c} 1  \\  -\frac{1}{2} - i \frac{\sqrt{3}}{2} \cos \theta  \\ 0  \\ -i \frac{\sqrt{3}}{2} \sin \theta \end{array} \right )
\end{align}
if the spins in the $+$ valley order ferromagnetically. If the spins in the $-$ valley order ferromagnetically, one finds instead (again up to transformations by the continuous symmetries $U(1) \times SU(2) \times SU(2)$)
\begin{align} \label{supple:parameterizationofgroundstate2}
\psi^{-}_A &=\frac{1}{\sqrt{2}} \left ( \begin{array}{c} 1  \\ 1  \\ 0  \\ 0 \end{array} \right ),\,\,\,\,\,
\psi^{-}_B =\frac{1}{\sqrt{2}} \left ( \begin{array}{c}  -\frac{1}{2} - i \frac{\sqrt{3}}{2} \cos \theta \\ 1  \\ -i \frac{\sqrt{3}}{2} \sin \theta   \\ 0 \end{array} \right ),\,\,\,\,\,
\psi^{-}_C =\frac{1}{\sqrt{2}} \left ( \begin{array}{c}  -\frac{1}{2} + i \frac{\sqrt{3}}{2} \cos \theta \\ 1  \\ i \frac{\sqrt{3}}{2} \sin \theta   \\ 0 \end{array} \right ).
\end{align}
In both cases, $\theta$ is the opening angle characterizing the chiral spin order.
An opening angle $\varphi \neq 0, \pi, \pi/2$ leads to a finite valley-projected chirality 
$\chi^{\pm} = \langle \vec{\sigma} P^\pm \rangle_A \cdot ( \langle \vec{\sigma} P^\pm \rangle_B \vec{\times}\langle \vec{\sigma} P^\pm \rangle_C)$. Using the $\Psi^+$ solutions above, we obtain
\begin{align}
\chi^{+}=0, \qquad \chi^{-} = \frac{3}{16} \sqrt{3} \cos \theta \sin^2 \theta,  \label{eq:chiplus}
\end{align}
while for the $\Psi^-$ solution we find
\begin{align}
\chi^{+} = - \frac{3}{16} \sqrt{3} \cos \theta \sin^2 \theta, \qquad \chi^-=0. \label{eq:chiminus}
\end{align}
Ground states have an arbitrary $\theta$, and thus an arbitrary  $-\frac{1}{8} \le \chi^{-} \le \frac{1}{8}$ for the $\psi^+$ states
or 
$-\frac{1}{8} \le \chi^{+} \le \frac{1}{8}$ for the $\psi^-$ states.  For $\theta =0$ or $\theta=\pi$, the $\psi^+$ and $\psi^-$ solutions coincide (up to trivial phases).
The states with $\theta = 0$ or $\pi$ in Eq.~\eqref{supple:parameterizationofgroundstate} are spin-ferromagnet in both of the valleys, but have the distinct 120$^\circ$ order with the left or right handedness, respectively. 
 
Expanding $H_0$, at a energy minimum, in terms of three 4-component complex states, $|\psi_m^{0} \rangle \rightarrow |\psi_m^{0} \rangle + |\delta \psi_m \rangle$ with $m=A,B,C$, up to the second order, we obtain seven zero-modes among $3 \times (8-2) = 18$ degrees of freedom. Those seven zero-modes show explicitly that the ground-state manifold is 7-dimensional. Six of this seven modes arise from the spontaneously broken continuous symmetries, the 7th mode, in contrast, is related to a change of $\theta$, thus links states which are not related by symmetry. An exception of this counting argument are the ferromagnetic states at $\theta=0$ or $\theta=\pi$, which have a higher symmetry. They have nine zero modes, from which five arise from spontaneously broken symmetries while two each describe changes of the opening angle $\theta$ either in the valley $+$ or the $-$ sector. Note that there are two such modes per sector as magnetization vectors can tilt in two different directions starting from the ferromagnetic configuration.

In the main text, we discuss the mean-field phase diagram arising from small perturbations around the $\varphi=0$ point, see Fig.~3.
 In Fig.~\ref{fig5} we show the analog of Fig. 3a but for an extended parameter range where $\varphi$ varies from $-\pi$ to $\pi$ and we also allow for large values of $J_3$. We find three types of non-chiral phases (red) which show either ferromagnetic or a $120^\circ$ order in the vectors $\langle \vec{\mu}_{n}\rangle$ with $n = 1,2$.

 Importantly, the phase diagram shows singular points not only at $\varphi=0$ but also at $\varphi = \pm \frac{2 \pi}{3}$. Those singular points can be understood by the enhanced symmetry of the $\varphi=0$ state discussed above and the transformation of Eq.~\eqref{supeq:tauzrotation} which can be used to map the states at $\varphi = \pm \frac{2 \pi}{3}$ to $\varphi=0$. 
 Due to the $\tau^z$ rotation, Eq.~\eqref{supeq:tauzrotation}, as adding $\varphi$ by $2\pi/3$ successively, the phase changes from a left-handed $120^\circ$ ordered phase $\rightarrow$ a right-handed $120^\circ$ phase $\rightarrow$ a ferromagnetic phase (more precisely, in the vectors  $\langle \vec{\mu}_{n}\rangle$ with $n = 1,2$), and back to the left-handed $120^\circ$ phase again. 
 Such a transition between different types of $120^\circ$ order was recently studied in the \moire Hubbard model~\cite{Wietek2022}.

  Since the $\varphi=0$ and $\varphi = \pm 2\pi/3$ points with $J_3 = J_2' =0$ have a large, degenerate ground-state manifold, even small perturbations which lift this degeneracy can lead to a giant effect close to all three points as shown in Fig.~\ref{fig5}.
  Note that the state with finite chiral order in either the $+$ or $-$ sector, also breaks the discrete valley symmetry leading to a  finite $\langle \tau^z \rangle \neq 0$. For sufficiently large $J_3$ (cf. Fig.~\ref{fig5}), the $\tau^z$ symmetry is maximally broken with $|\langle \tau^z \rangle | = 1$ and  also the chirality takes its maximal value, $\chi^+ - \chi^- = -1$ (purple region). 

\section{Spin-wave theory}

In this section, we perform a spin wave calculation in $J_2' = J_3 = 0$ and $\varphi = 2 \pi n /3$, $n \in \mathbb{Z}$, where ground states are highly degenerate. 
The motivation of this spin wave calculation is to investigate the effect of thermal and quantum fluctuation in such a highly degenerate ground-state manifold. We show that by the thermal order-by-disorder mechanism, the system selects spin ferromagnetic states in both of the valley sectors from the ground-state manifold. This is contrasted with that the quantum order-by-disorder mechanism favors 120$^\circ$ spin order in one valley and ferromagnetic order in the other valley. The thermal order-by-disorder mechanism leads to a mass gap linear in temperature for the soft modes related with the opening angle $\theta$. Expanding around the classical ground states (i.e., spin-ferromagnet in both of the valleys), we show that a finite chirality $\chi^\pm$ is induced at finite temperatures by the $\theta$-related soft modes. The non-analytic temperature dependence of the chirality arises from the linear mass gap of the soft modes in temperatures. 
These results remarkably match with results of the classical Monte-Carlo simulation. 

We first start by developing a spin wave theory for $SU(4)$ operators. Spin wave theories become exact in certain large $M$ limits, where $M$ parametrize  representations of the group. 
In the $SU(2)$ case one uses the size of the spin $s$ (with $M=2 s$) and performs a $1/s$ expansion. 
In the $SU(4)$ case, we choose a totally symmetric representation of the $SU(4)$ operators by (i) writing the  operators
with {\em bosonic} creation and annihilation operators, $a^\dagger_{i}, a_{i}$ ($i=1,\dots,4$ and we suppress an extra site index here), and (ii) fix the number of bosons (per site) to be $M$ using
\begin{align}
\hat{\Gamma}^{\ell} &= \sum_{i,i'=1,\dots,4} \hat{a}_i^{\dagger} \gamma_{i i'}^{\ell} \hat{a}_{i'}, \qquad 
M = \sum_{i=1}^4 \hat{a}_{i}^{\dagger} \hat{a}_{i}.
\end{align}
Furthermore, we add an extra factor $1/M$ in front of the Hamiltonian, $H \to H/M$, to make the large $M$ limit well defined, see below. The $M = 1$ case corresponds to the fundamental representation of $SU(4)$, realized if a single electron is localized on each site. While we are interested in this limit, the large $M$ theory to useful to derive a spin-wave theory in a controlled way.

Within the functional integral formalism, the partition function of our system is expressed as 
\begin{align}
 \mathcal{Z} = \int D[\bar{a}, a] D\lambda e^{- \int d\tau \left (  \sum_{i, s, \vec{n}} \bar{a}_{i, s, \vec{n}} \partial_{\tau} a_{i, s, \vec{n}} + \frac{1}{M} H (\bar{a}, a) + i \sum_{s, \vec{n}} \lambda_{s, \vec{n}} (M-\sum_{i = 1}^{4} \bar{a}_{i, s, \vec{n}} a_{i, s, \vec{n}}  
 ) \right)}. 
\end{align}
where $a_{i, s, \vec{n}}$ are {\em complex} fields and the Lagrange multipliers $\lambda_{s, \vec{n}}$ are used to implement the constraint on each site.
Rescaling the boson fields as $a_{i, s, \vec{n}} = \sqrt{M} \tilde{a}_{i, s, \vec{n}}$ and using that $H$ is quartic in these operators, we arrive 
\begin{align} \label{supeq:action}
 \mathcal{Z} = \int D[\bar{\tilde a}, \tilde a] D\lambda e^{-M S_{\textrm{eff}}},\qquad S_{\textrm{eff}} = 
 \int_0^\beta d\tau  \left(\sum_{i, s, \vec{n}}\bar{\tilde{a}}_{i, s, \vec{n}} \partial_{\tau} \tilde{a}_{i, s, \vec{n}} +  H(\bar{\tilde{a}}, \tilde{a}) - i \sum_{s, \vec{n}} \lambda_{s, \vec{n}} \left(\sum_{i = 1}^{4} \bar{\tilde{a}}_{i, s, \vec{n}} \tilde{a}_{i, s, \vec{n}} -1 \right) \right)
\end{align}
Due to the factor $M$ in front of  $S_{\textrm{eff}}$, the functional integral in the large $M$ limit is dominated by its saddle point and fluctuations around the saddle point, which are controlled by $1/M$.
Saddle point solutions can be obtained by solving $\frac{\partial S_{\textrm{eff}}}{\partial \tilde{a}_{i,s, \vec{n}}}\vert_{\tilde{a}_{\textrm{sp}}, \lambda_{\textrm{sp}}}
=\frac{\partial S_{\textrm{eff}}}{\partial \bar{\tilde{a}}_{i,s, \vec{n}}}\vert_{\tilde{a}_{\textrm{sp}}, \lambda_{\textrm{sp}}}=\frac{\partial S_{\textrm{eff}}}{\partial \lambda_{s, \vec{n}}}\vert_{\tilde{a}_{\textrm{sp}}, \lambda_{\textrm{sp}}}=0$. 
Static saddle point solutions exactly correspond to the zero-temperature mean-field solutions with $\lambda_{s, \vec{n}} =  \lambda_{\textrm{sp}}$ being the mean-field energy per site. Expanding $S_{\textrm{eff}}$ up to second order around the saddle point solutions, the resulting action captures physics of spin wave excitations. Although this saddle point approximation becomes more accurate with large $M$, it also provides a good approximation even to the $M = 1$ case. Hereafter $M$ is set to 1, unless otherwise stated.  

To understand the effect of quantum or thermal fluctuation on the degenerate ground-state manifold of $H_0$ (Eq.~\eqref{sup:H0}), we use the approach explained above for $J_2' = J_3 = 0$ and $\varphi = 2 \pi n /3$, $n \in \mathbb{Z}$ where the mean-field solution is highly degenerate. As a reference state around which the action (Eq.~\eqref{supeq:action}) is expanded, 
we take the spinor wavefunctions $\psi^0_{s} (\alpha = \pm, \theta) = \psi_{s}^{\alpha} (\theta)$ (Eqs.~\eqref{supple:parameterizationofgroundstate} and \eqref{supple:parameterizationofgroundstate2}) that depend on the opening angle $\theta$ and $\alpha$. 
$\alpha$ represents the valley sector in which the spin has the ferromagnetic order. 
Then, low-energy states associated with the spin wave excitation can be generally written, up to normalization, as 
\begin{align} \label{eq:lowenergystates}
\psi_{s} (\alpha, \theta) \sim \psi_{s}^{0} (\alpha, \theta)+ \frac{1}{\sqrt{M}}\sum_{i = 1}^{3} \hat{a}_{s, i} \delta \psi_{s, i} (\alpha, \theta).
\end{align}
Here $\delta \psi_{s, i}$ are three 4-component unit vectors perpendicular to the reference state $\psi_{s}^{0} (\pm, \theta)$.
At this point, we find it useful to switch back from the functional integral formalism to the operator formalism, where it is more easy to keep track of commutation relation.
 Expanding to the second order in $\hat{a}_{s, i}$ and performing the fourier transform to momentum space, the Hamiltonian has a Bogoliubov-de Gennes (BdG) form  
\begin{align} \label{eq:BDGham}
    H_{\textrm{BdG}}  =\frac{1}{2} \sum_{\vec{k}} \begin{pmatrix} \hat{a}_{\vec{k}}^{\dagger} & \hat{a}_{-\vec{k}} \end{pmatrix} \begin{pmatrix}
A(\vec{k}) & B(\vec{k})\\
B^*(-\vec{k}) & A^*(-\vec{k})
\end{pmatrix} 
\begin{pmatrix} \hat{a}_{\vec{k}} \\ \hat{a}_{-\vec{k}}^{\dagger} \end{pmatrix} - \frac{1}{2} \sum_{\vec{k}} \textrm{Tr} A(\vec{k}). 
\end{align}
Here $A (\vec{k})$ and $B(\vec{k})$ are $9 \times 9$ matrices in the basis of sublattice $s = A, B, C$ 
and $i = 1,2,3$ representing three directions perpendicular to $\psi^0_{s} (\alpha, \theta)$, and fulfill the condition $A (\vec{k}) = A^{\dagger} (\vec{k})$
and $B(\vec{k}) = B^T (-\vec{k})$, respectively. The diagonalization of the bosonic BdG Hamiltonian, Eq.~\eqref{eq:BDGham}, should be taken with special care. To fulfill the bosonic commutation relation for the eigenmodes, the transformation matrix $T(\vec{k})$ for the diagonalization has to satisfy the paraunitarity condition $T^{\dagger} (\vec{k}) \Sigma^z T (\vec{k})= T(\vec{k})  \Sigma^z T^{\dagger} (\vec{k})  = \Sigma^z$ with the third Pauli matrix $\Sigma^z$ acting on the Nambu space. As a consequence, $T(\vec{k})$ and the corresponding eigen energies are obtained from diagonalizing the matrix $\Sigma^z H(\vec{k})$ instead~\cite{Colpa1978}. Using this diagonalization scheme, one obtains 
\begin{align} \label{eq:BdGdiag}
    H_{\textrm{BdG}}  = \frac{1}{2}\sum_{\vec{k}}  \begin{pmatrix} \hat{\gamma}_{\vec{k}}^{\dagger} & \hat{\gamma}_{-\vec{k}} \end{pmatrix} \begin{pmatrix}
E(\vec{k}) & 0 \\
0 & E (-\vec{k})
\end{pmatrix} 
\begin{pmatrix} \hat{\gamma}_{\vec{k}} \\ \hat{\gamma}_{-\vec{k}}^{\dagger} \end{pmatrix} - \frac{1}{2} \sum_{\vec{k}} \textrm{Tr} A(\vec{k}) = 
\sum_{\vec{k}} \sum_{n=1}^9 E_{\vec{k}, n} (\alpha, \theta) \left (\hat{\gamma}^{\dagger}_{\vec{k},n} \hat{\gamma}_{\vec{k},n} +\frac{1}{2} \right)
- \frac{1}{2} \sum_{\vec{k}} \textrm{Tr} A(\vec{k}).
\end{align}
Here $E (\vec{k})$ is a diagonal matrix with positive elements $E_{\vec{k},n} > 0$. 
In the ground state the occupation of the finite-energy states vanishes and therefore the ground-state energy is given by
\begin{align} 
E_0\approx E_{\text{MF}}+\frac{1}{2}\sum_{\vec{k}, n} E_{\vec{k}, n} (\alpha, \theta)- \frac{1}{2} \sum_{\vec{k}} \textrm{Tr} A(\vec{k})
\label{eq:E0}
\end{align}
where $E_{\text{MF}}$ is the mean-field energy. It turns out that $\textrm{Tr} A(\vec{k})$ is independent on the expansion point $(\alpha,\theta)$. $E_0$ also includes corrections arising from quantum fluctuations, computed to leading order in $1/M$.
Similarly, the free energy at low temperatures can be approximated by 
\begin{align}
F\approx E_{\text{MF}}+\frac{1}{2}\sum_{\vec{k}, n} E_{\vec{k}, n} (\alpha, \theta)- \frac{1}{2} \sum_{\vec{k}} \textrm{Tr} A(\vec{k})+\frac{1}{\beta} \sum_{\vec{k}, n}  \log (1-e^{-\beta E_{\vec{k},n} (\alpha, \theta)}). \label{eq:F}
\end{align}

\begin{figure}[h!]
	\centering
	\includegraphics[width= \columnwidth]{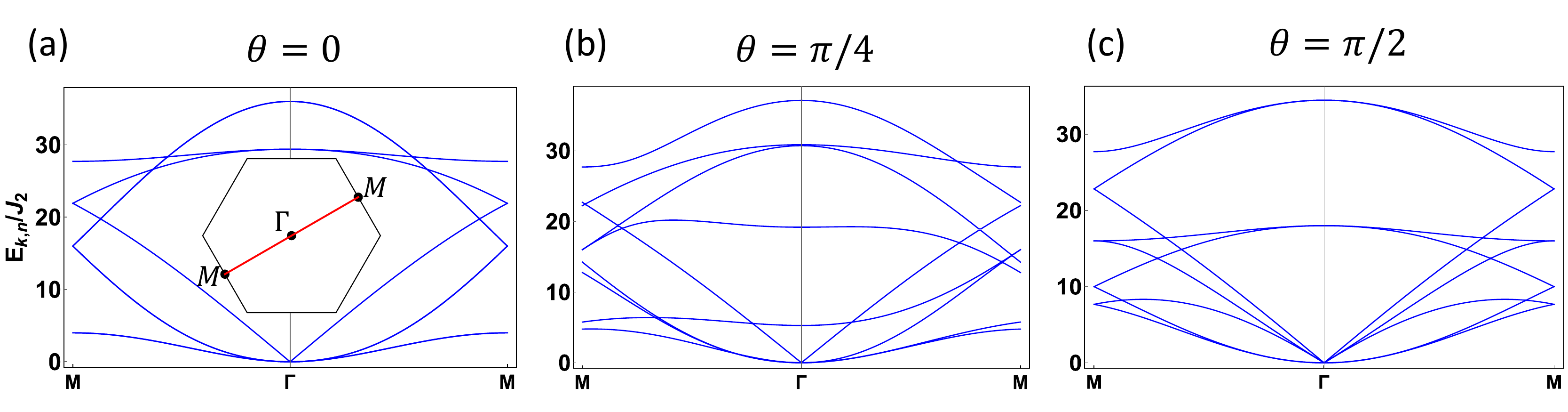}
	\caption{ {\bf Spin-valley excitation spectra} for an expansion around (a) a state with ferromagnetic spin order ($\theta = 0$), (b) a state with chiral spin chirality ($\theta = \pi/4$) and (c) a coplanar spin-state  ($\theta = \pi/2$). The spectra are drawn along the red line ($M - \Gamma - M$) in the Brillouin zone shown in the inset of panel (a). The excitation spectra differ in the number of Goldstone modes. 
   Panel (a): four quadratic modes, (more precisely, for each valley, a Goldstone mode corresponding to the ferromagnetic order and a $\theta$-related soft mode) and a linear mode, Panel (b): three quadratic modes and a linear mode, Panel (c): two quadratic and three linear modes.}
	\label{Figsup_spinwaveex}
\end{figure}

The spectrum of excitation $E_{\vec{k}, n} (\alpha, \theta)$ depends on the chosen mean-field state $\psi^0_{s} (\alpha, \theta)$
even in cases where the mean-field energy is exactly the same. This is shown in Fig.~\ref{Figsup_spinwaveex} where the excitation spectrum is shown for an expansion around (i) a state with ferromagnetic spin-order ($\theta = 0$), (ii) a state with chiral spin chirality ($\theta = \pi/4$) and (iii) a coplanar spin-state  ($\theta = \pi/2$). The excitation spectrum differs in the number of Goldstone modes and also in its high-energy spectrum. 
Thus both $E_0$ and $F$ will depend on the chosen ground state within the mean-field ground-state manifold. 
Nature will select the state with the lowest (free-) energy. This is an example of the ``order by disorder'' mechanism, where quantum or thermal fluctuations select one specific ordered state out of a larger manifold.

To be able to compare with the result of classical Monte Carlo calculations, it is useful to evaluate the free energy \eqref{eq:F}
in the classical limit, $E_{\vec{k},n} \ll T$, where we obtain
\begin{align}
F^{\text{cl}} \approx E_{\text{MF}}+ T \sum_{\vec{k}, n}  \log (\beta E_{\vec{k},n} (\alpha, \theta)). \label{eq:Fclassical}
\end{align}

\begin{figure}[t]
	\centering
	\includegraphics[width= \linewidth]{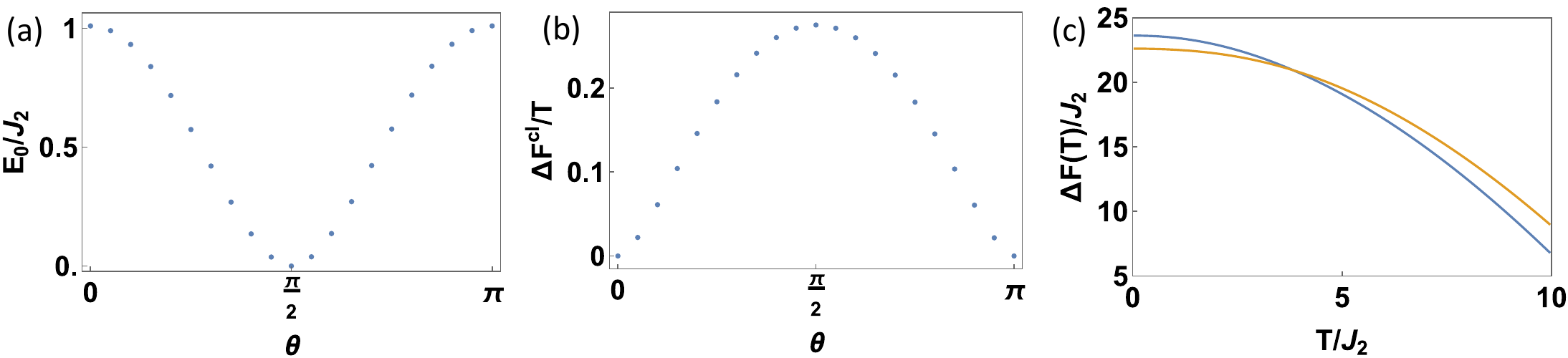}
	\caption{{\bf Order-by-disorder mechanisms.} 
 (a) The ground-state energy per site (Eq.~\eqref{eq:E0}) as function of $\theta$, parameterizing the opening angle of valley-projected spins in the three sublattices. By the quantum order-by-disorder mechanism, the state with the opening angle $\theta = \pi/2$ (i.e., the spin-coplanar 120 degree order) is selected from the ground-state manifold. (b) 
 In the classical model, the free energy $F^{\text{cl}}$, Eq.~\eqref{eq:Fclassical}, obtains at low $T$ a correction linear in $T$ from thermal fluctuations. In contrast with the quantum order-by-disorder mechanism, the thermal order-by-disorder leads to the selection of the spin ferromagnetic order in both of the valley sectors ($\theta = 0$ or $\pi$).
 (c) Free energy as function of temperature. As the thermal order-by-disorder mechanism selects the distinct ground states from the quantum order-by-diorder mechanism, a first-order phase transition occurs at an intermediate temperature, $k_B T \sim 3.8 J_2$.}
	\label{Supfig:freeenergy}
\end{figure}

Figure \ref{Supfig:freeenergy}(a-b) shows the free energy with different ground states, characterized by the opening angle $\theta$. 
As shown in Fig~\ref{Supfig:freeenergy}(a), the free energy at zero temperature has a minimum at the state with the opening angle $\theta = \pi/2$, i.e., the spin-coplanar 120 degree order in one valley (keeping the ferromagnetic order in the other valley). 
The selection of the states are achieved by the quantum order-by-disorder mechanism. 
In contrast, the thermal fluctuations select distinct states from the quantum fluctuations as shown from the classical free energy in \ref{Supfig:freeenergy}(b). The thermal order-by-disorder mechanism leads to a selection of the spin ferromagnetic order in both of the valley sectors ($\theta = 0$ or $\pi$). Thus, our system is one of the rare cases where quantum and classical fluctuations select very different types of ground states. Technically this arises, because the classical fluctuations select the state where the {\em geometric} average (sum of logarithms) of the energies $E_{\vec{k}, n}$ is lowest, while quantum fluctuations select the state with the lowest {\em arithmetic} average. 

In Fig.~\ref{Supfig:freeenergy}(c), we show the free energy, Eq.~\eqref{eq:F}, as function of temperature for the states with $\theta=\pi/2$ and $\theta=0,\pi$. Formally, the calculation predicts a first order transition from the spin-coplanar state to the spin-ferromagnetic state upon increasing $T$. The transition temperature, $T\approx 3.8\,J_2$, is, however, so high that the expansion around the $T=0$ mean-field, which underlies  Eq.~\eqref{eq:F}, is not expected to be valid any more. Our classical Monte-Carlo simulations (see main text) show that there is no long-ranged order at this temperature.

The discussion of the free energy given above explains that the classical Monte Carlo calculations reported in the main text
obtain a spin-ferromagnetic ground state ($\theta=0,\pi$) in the limit $T\to 0$. For the classical model, the spin-wave theory should become exact for low $T$, as it captures Gaussian fluctuations around the classical ground state. We can therefore use it to explain two numerical results shown in Fig.~4(b) and (c). The spin-chirality, $\chi^+-\chi^-$, obtains a finite expectation value at $T>0$, which rises in a singular way as function of temperature. Furthermore, the ferromagnetic order parameter obtains a correction linear in $T$ with a prefactor which increases for increasing system size. Remarkably, such a system-size dependence is largely absent for the spin-chirality and only visible at the lowest temperatures, see inset of Fig.~4(b).

The suppression of the order parameter is a well-known consequence of the Mermin-Wagner theorem: the thermal occupation of the Goldstone modes gives rise to a correction of order $-T\log{1/N}$ to the order parameter, where $N$ is the linear system size. This effect is clearly visible in the numerics. In an infinite system, $N\to \infty$, long-range order is expected to be absent at any finite $T$ with a  correlation length  which is exponentially large in $1/T$. 

More surprising is the finite chirality and its unusual temperature dependence. It is straighforward to expand the valley-projected chirality operators
\begin{align}
 \chi^{\pm} = \langle \vec{\sigma}_{m_1} P^\pm_{m_1} \cdot ( \vec{\sigma}_{m_2} P^\pm_{m_2} \vec{\times} \vec{\sigma}_{m_3} P^\pm_{m_3})\rangle \approx \chi^{\pm}_{\text{cl}} +\frac{1}{V} \sum_{\vec{k},n}
\chi^{\pm}_{\vec{k},n} (\,\langle \hat{\gamma}^{\dagger}_{\vec{k}, n} \hat{\gamma}_{\vec{k}, n}\rangle +1/2)\label{eq:chipm}
 \end{align}
where $V=N^2$ is the number of sites in the system, $\chi^{\pm}_{\text{cl}}$ is the chirality of the mean-field ground state
 and $\chi^{\pm}_{\vec{k},n}$ is a numerically determined weight factor which encodes how much chirality an excitation with quantum numbers $n$ and $\vec k$ carries.
 
To compare to our classical Monte Carlo simulation, we use this formula expanding around the  spin-ferromagnetic state, $\theta=0,\pi$, where $\chi^{\pm}_{\text{cl}}=0$. In the classical limit, $T \gg E_{\vec{k}, n}$, we have to replace $\langle \hat{\gamma}^{\dagger}_{\vec{k}, n} \hat{\gamma}_{\vec{k}, n}\rangle+\frac{1}{2}$ by $T/E_{\vec{k}, n}$ and we obtain
\begin{align} 
    \langle \hat{\chi}^\pm \rangle = \frac{1}{V} \sum_{\vec{k}, n} \chi^{\pm}_{\vec{k},n} 
  \frac{ T}{E_{\vec{k}, n} (\alpha, \theta)}. \label{eq:chipmClassical00}
\end{align}
  Importantly, $\chi^{\pm}_{\vec{k},n}$ turns out to be finite for $\vec k \to 0$ for one of the modes, which we label by $n_{\pm}$, with $E_{\vec{k}, n_\pm}\approx c_\theta k^2$ for $\vec k \to 0$. Numerically, we obtain $\chi^{\pm}_{0,n_\pm}=\mp 2.60$ (when expanding around $\theta=0$, signs are opposite when expanding around $\theta=\pi$) and $c_\theta \approx 3.1$. The $n_\pm$ modes  describe fluctuations of $\theta$ which naturally give rise to a finite spin chirality (note that  $\chi^{\pm} \approx \mp \frac{3\sqrt{3}}{16}  \theta^2$ according to Eqs.~\eqref{eq:chiplus} and \eqref{eq:chiminus}).

Thus, Eq.~\eqref{eq:chipmClassical} predicts for the $\theta=0$ state a nominally {\em divergent} contribution to the chirality of the form
\begin{align} 
    \langle \hat{\chi}^\pm \rangle \approx \frac{1}{V} \sum_{\vec{k}} \chi^{\pm}_{0,n_\pm} 
  \frac{ T}{J_2 c_\theta k^2} \approx  \mp 0.11\, \frac{T}{J_2} \ln\left[\frac{k_0}{k_\text{min}}\right]. \label{eq:chipmClassical0}
\end{align}
 where $k_0$ denotes a UV cutoff to the $k$ sum and we introduced ad hoc an minimal momentum $k_{\text{min}}$ 
as an infrared cutoff. In three dimensions, the analog calculation would give $T (c_1  - c_2 k_\text{min})$.

Thus the question arises, what sets the value of the infrared cutoff $k_{\text{min}}$. Importantly, it is {\em not} set by the system size but by the fact that the $\theta$-modes $n_\pm$ are {\em not}  true Goldstone modes. While within mean-field changes of $\theta$ in the two valley-sectors do not cost any energy. This is, however, not the correct result. As shown in Fig.~\ref{Supfig:freeenergy}(b), the free energy near the classical minima  ($\theta_0 = 0$, $\pi$) is approximately described by  
\begin{align} \label{supeq:freeenergyaroundminima}
F_{\alpha, \theta}^{\textrm{cl}} \approx F_{\alpha, \theta_0}^{\textrm{cl}} + \frac{1}{2} c_1 T (\theta -\theta_0)^2.
\end{align}
This term induces a finite mass $\sim T$ to the $n_\pm$ modes 
resulting in an effective IR cutoff $k_\text{min} \sim \sqrt{T}$, 
\begin{align} 
    \langle \hat{\chi}^+-\hat{\chi}^- \rangle \approx  \pm 0.22\, \frac{T}{J_2} \ln\left[\sqrt{T_0/T}\right]. \label{eq:chipmClassical}
\end{align}
  where the sign depends on whether we expand around $\theta=0$ or $\theta=\pi$ and $T_0$ is some UV cutoff energy.
Eq.~\eqref{eq:chipmClassical} explains the singular temperature dependence observed in the Monte Carlo numerics, Fig.~\ref{Fig5:NumericsH0} and also the approximate absence of finite-size effects in this quantity as long as $k_\text{min} N \gtrsim 1$. In three dimensions, the analog calculation would give a correction of the form $c_1 T  - c_2 T^{3/2}$.

\begin{figure}
	\centering
	\includegraphics[width= .6\columnwidth]{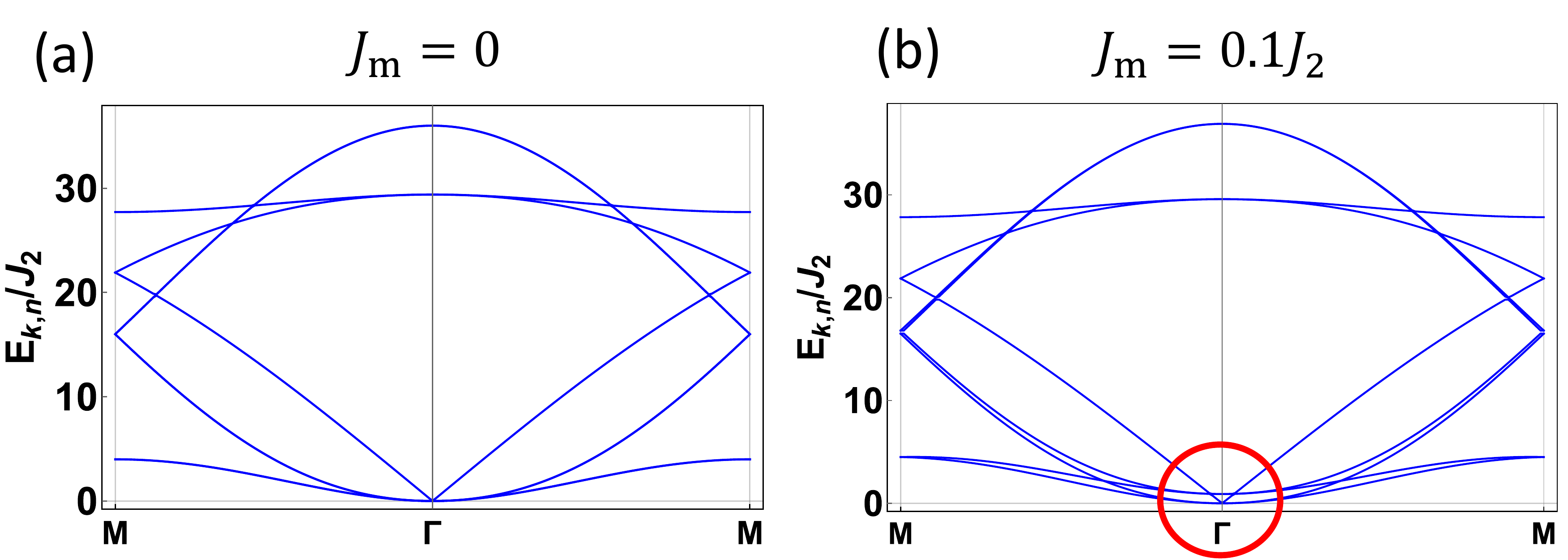}
	\caption{{\bf Mass gap of $\theta$-modes}, $n_\pm$ modes (see the text), induced by the Hamiltonian Eq.~\eqref{supeq:massgapHam} to stabilize the ferromagnetic spin order.
 The excitation spectra are obtained from an expansion of a ferromagnetic state ($\theta= 0$), which is an energy minimum of the mean-field theory. The two $n^\pm$ modes from two valleys acquire the mass gap as shown in near $\Gamma$ point (the red circle).}
	\label{Figsup_stabilizer}
\end{figure}

To see how the mass gap enters the dispersion of the $n_{\pm}$ modes, we add a Hamiltonian, given by 
\begin{align} \label{supeq:massgapHam}
    H_{\text{mass}} = J_{\text{m}} \sum_{m, m'} ( (\vec{\sigma} P^+)_m \cdot  (\vec{\sigma} P^+)_{m'} +  (\vec{\sigma} P^-)_m \cdot  (\vec{\sigma} P^-)_{m'}). 
\end{align}
Note that this Hamiltonian stabilizes the spin-ferromagnetic state ($\theta = 0, \pi$), but opens up a mass gap in the $n^{\pm}$ modes. 
Expanding around the $\theta = 0$ state, one of the classical minima with $H_0 + H_{\text{mass}}$, we obtain the excitation spectra shown in Fig.~\ref{Figsup_stabilizer}. Each of the $n_{\pm}$ modes acquires a mass gap with dispersion $E_{\vec{k}, n_{\pm} }\sim \vec{k}^2 + J_{\text{m}}^2$. Therefore, the mass gap nicely provides an effective infrared cutoff $k_{\text{min}} \sim J_{\text{m}}$, as discussed above.   

The discussion given above applies to the classical model. 
In the quantum case, when we expand around a ferromagnetic solution (stabilized, e.g., by $J_2'<0$), $|\langle \hat{\chi}^+-\hat{\chi}^- \rangle|$ is  finite even for $T=0$. 
In the spin-planar phase, $\theta=\pi/2$ (stabilized by quantum fluctuations in the pure $J_2$ model and also obtained for $J_2'>0$), in contrast, $\chi^\pm$ vanishes by symmetry, see below.

Two symmetries are most important for the discussion of the spin chirality. First, a 180$^\circ$ rotation of both spin and space 
around, e.g., the $\hat y$ axis  maps $\chi^\pm$ to $-\chi^\pm$. Second, the inversion symmetry maps $\chi^\pm$ to $-\chi^\mp$.

The spin-ferromagnetic state (with a 120$^\circ$ valley order) breaks 180$^\circ$ rotation symmetry but is inversion symmetric. 
Thus $\chi^++\chi^-=0$ while $\chi^+-\chi^-$ is finite. The state with coplanar 120$^\circ$ spin-order (realized in the quantum $J_2$ model), in contrast, has both symmetries and thus $\chi^\pm=0$. This is also reflected in Fig.~\ref{Fig6:2dplotsphasediagram} of the main text which shows that $|\chi^+ - \chi^-|$ is only finite in the spin-ferromagnetic phase.


\section{Semi-classical Monte Carlo}
\label{suppl:mc}
In this section, we provide a more detailed description of the our semi-classical Monte Carlo implementation. We note that a very similar description (by some of us) for a filling of two instead of one electron per site can be found in Ref.~\cite{gresista2023}. We then conclude this manuscript by presenting additional numerical data elucidating the type of phase transitions separating the disordered and the two ordered states found in the $J_2-J_2^\prime$ model (see Fig.~\ref{Fig6:2dplotsphasediagram}).

\subsection{Implementation}
To calculate finite-temperature observables, we perform semi-classical Monte Carlo calculations using the Metropolis algorithm \cite{LandauBinder} with local updates. Instead of a classical spin configuration, however, we need to update the product-state wavefunction $|\Psi \rangle=\prod_{m} |\Psi_{m}\rangle$, where $|\Psi_{m}\rangle$ is a single-site, 4-component, normalized wave function. To this end, we parameterize the single-site wave-function as
\begin{equation}
\label{eq:local-state}
    |\psi_m\rangle = \sum_{j=1}^4 b_m^j|\gamma^j\rangle\,,
\end{equation}
with normalized, 4-dimensional, complex-valued vectors $|\mathbf{b_m}| = 1$. The states $|\gamma^j\rangle$ constitute a basis of the local Hilbert space, for which we simply choose 
\begin{equation}
    |\gamma^j\rangle \in \left\{ |\uparrow +\rangle, |\downarrow +\rangle, |\uparrow -\rangle, |\downarrow -\rangle \right\}
\end{equation}
where $\sigma = (\uparrow, \downarrow)$ is the spin and $\alpha = (+, -)$ the valley quantum number, labeling the eigenvalues of $\sigma^z$ and $\tau^z$, respectively. Subtracting the normalization and a local arbitrary phase, a state can therefore be parametrized by $N^2\cdot(8-2)$ real numbers. In the Monte Carlo calculation, however, it turns out beneficial to simply include the redundancy of the phase, which does not affect any of our observables, and work with all components of $\mathbf{b}_m$. To perform a local Metropolis update, we consequently need to be able to uniformly sample the space of normalized, complex valued, 4-dimensional vectors. Such vectors can be understood to live on a 7-dimensional hypersphere (7-sphere), parameterized by the real- and imaginary part of each component. To uniformly sample on a 7-sphere, one can simply draw $7+1$ normally distributed numbers and then normalize the resulting vector \cite{muller1959}. Sampling on the full sphere, however, leads to very low acceptance rates for low temperatures, which in turn results in a slow convergence of the results. Instead, we adapt Ref.~\cite{cardona2019} and utilize the \emph{Gaussian trial move}, which generates a new local state in the `vicinity' of the original as
\begin{equation}
    \mathbf{b}_m^\prime = \frac{\mathbf{b}_m + \sigma_g \mathbf{\Gamma}}{|\mathbf{b}_m + \sigma_g \mathbf{\Gamma}|},
\end{equation}
where $\mathbf{\Gamma}$ is a 4-dimensional complex vector, with the real and imaginary part of each component sampled from a normal distribution. The value of $\sigma_g$ controls the `step-size' of the update. Staring with a large $\sigma_g = 60$ and then adjusting $\sigma_g$ every ten Monte Carlo sweeps according to 
\begin{equation}
\label{eq:update-cone-width}
    \sigma_g \to \frac{0.5}{1-R} \sigma_g,
\end{equation}
where $R$ is the acceptance rate during the last ten sweeps, this very quickly tunes the overall acceptance rate to approximately 50~\% leading to significant speedup in  convergence at lower temperatures. 

We begin each Monte Carlo run with a thermalization phase, typically lasting for $N_t = 1 \cdot 10^6$ sweeps, in which the temperature is continuously lowered from a large initial value of $T_I = 3 |J_2|$ to the desired temperature $T$. More precisely, for the first $3/4 N_t$ sweeps the temperature is lowered by a multiplication with the factor $ (T/T_I)^\frac{4}{3 N_t}$ after each sweep. For the remaining $\frac{1}{4} N_t$ sweeps the temperature is kept constant. During the thermalization phase $\sigma_g$ is adjusted using the procedure described above.
After thermalization, we start the measurement phase, typically for $N_m = 10 \cdot 10^6$ sweeps, where we keep $T$ and $\sigma_g$ constant and perform measurements every tenth sweep. The statistical evaluation of the measurements is done using the \texttt{BinningAnalysis} Julia Package~\cite{binninganalysis}.

\subsection{Phase transitions}
\begin{figure*}
    \centering
    \includegraphics{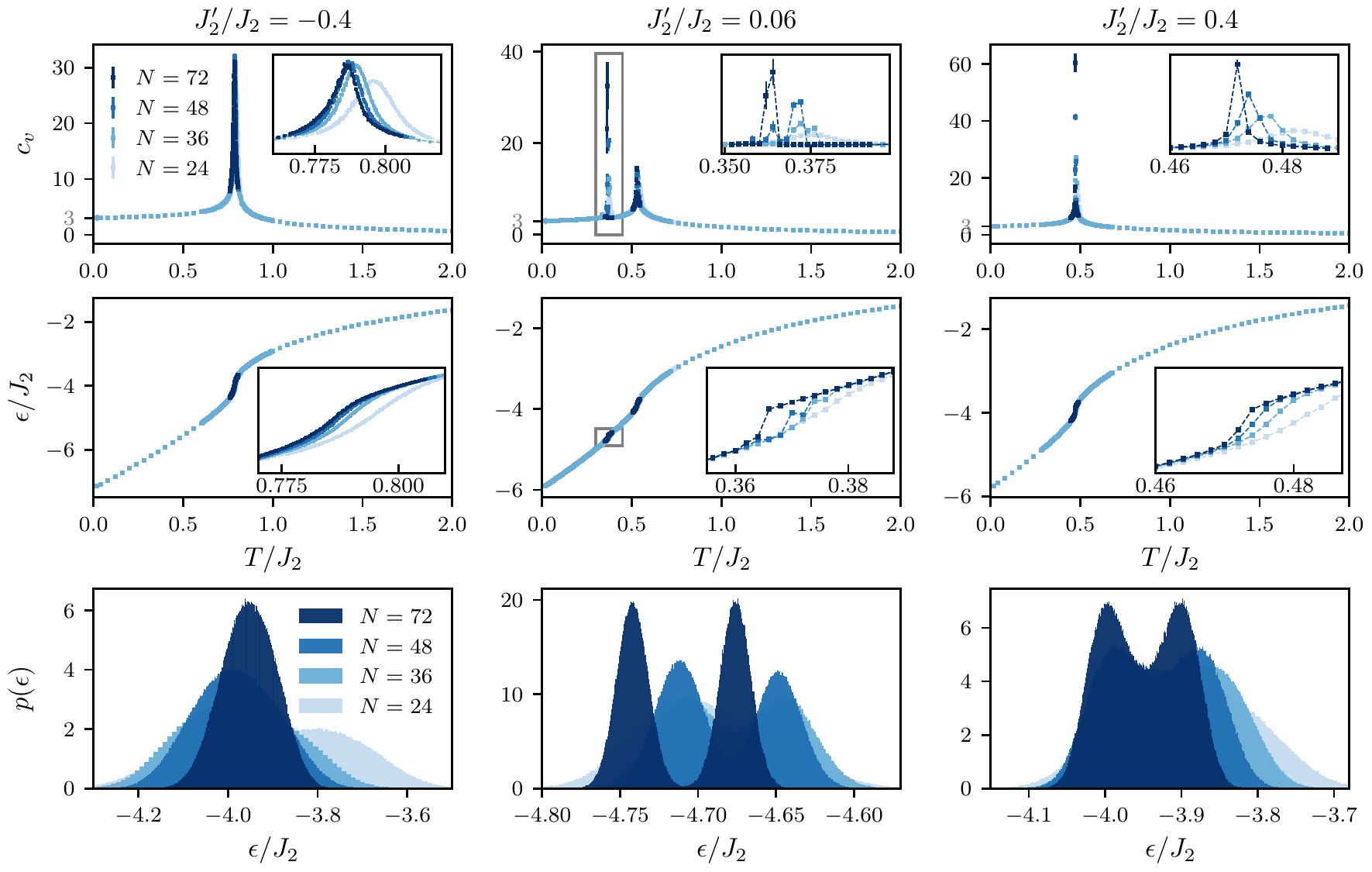}
    \caption{{\bf Thermal phase transitions in the $\mathbf{J_2-J_2^\prime}$ model.} Monte Carlo data showing the specific heat $c_v$, energy per site $\epsilon$ and energy distribution $p(\epsilon)$ in the three principal phase transitions we observe, obtained for linear system sizes $N = 24, 36, 48, 72$. At $J_2'/J_2 = -0.4$ (left column) a transition from the disordered state to the ordered state with $120^\circ$ order in $\mu^{1, 2}$ and ferromagnetic order in $\boldsymbol{\sigma}P^{\pm}$ occurs. The continuous decrease in energy and the single peak structure of the energy distribution at the transition temperature  suggest a continuous or a weak first-order phase transition. At $J_2'/J_2 = 0.06$ (middle column) the peak in $c_v$ at higher temperature corresponds to the same transition as for $J_2'/J_2 = -0.4$. The second peak, which is shown in the insets, corresponds to the transition between the two ordered states, where the spin changes from ferromagnetic to $120^\circ$ order in one valley. The kink in the energy and the bimodal structure of the energy distribution, which become more pronounced with increasing $N$, imply a first-order transition. For  $J_2'/J_2 = 0.4$ (right column) the system directly transitions from the disordered state into the ordered state with $120^\circ$ spin order, also showing signatures of a first-order transition.}
    \label{fig:phase-transitions}
\end{figure*}
\begin{figure}
    \centering
    \includegraphics{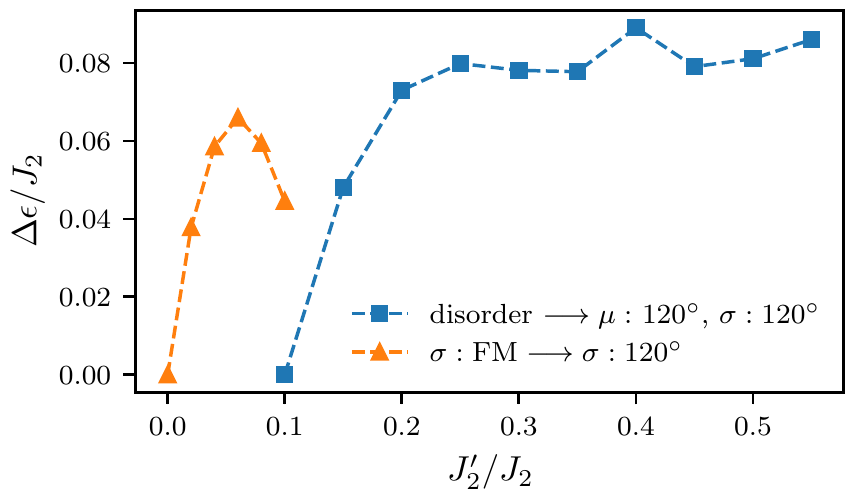}
    \caption{{\bf Latent heat  of the first-order phase transitions.} The latent heat $\mathbf{\Delta \epsilon}$ is obtained from the Monte Carlo data at lattice size $N = 72$ by fitting double Gaussians to the energy distribution at the critical scale and calculating the distance between the two peaks. For $0 < J_2^\prime/J_2 < 0.1$ a first-order phase transition between the two ordered states of Fig.~\ref{Fig6:2dplotsphasediagram} occurs, at which in one valley the spin transitions from ferromagnetic to coplanar $120^\circ$ order. At $J_2^\prime > 0.1 J_2$ the disordered state directly transitions into  the state with $120^\circ$ spin order. Both transitions show a sizable latent heat, indicating a strong first-order transition.}
    \label{fig:latent-heat}
\end{figure}
The finite-temperature phase diagram of the $J_2-J_2^\prime$ model shown in Fig.~\ref{Fig6:2dplotsphasediagram} features three distinct phases: A disordered phase at high temperature, a state with $120^\circ$ order in $\boldsymbol{\mu}^{1, 2}$ and ferromagnetic order in $\boldsymbol{\sigma}$, and a similar state where the spin instead shows $120^\circ$ order in one valley. Fig.~\ref{fig:phase-transitions} shows Monte Carlo data for all three of the corresponding phase transitions separating the different phases. 

The transition from the disordered state into the state with ferromagnetic spin order ($J_2^\prime/J_2<0.1)$ features a seemingly continuous energy as a function of temperature and the energy distribution at the transitions shows only one Gaussian peak, indicating a continuous phase transition, a thermal crossover or a weak first-order transition. As discussed in the main text and shown in Fig.~\ref{fig5}, the sharp rise of the chirality $|\langle \chi^+-\chi^-\rangle|$, accompanied by the breaking of a discrete $\mathbb Z_2$ symmetry, is mostly independent from $N$, strongly suggesting a phase transition instead of a crossover. Very close to $T_c$, however, $|\langle \chi^+-\chi^-\rangle|$ strongly fluctuates between different Monte Carlo runs, even when repeating runs at fixed $T$, leading to large statistical errors which prohibit us from determining the precise nature of the phase transition. 

In contrast, the transition separating the two ordered states ($0 < J_2^\prime/J_2 < 0.1$), as well as the transition between the disordered phase and the phase with spin $120^\circ$ order ($J_2^\prime/J_2 > 0.1$) show a discontinuity in the energy and a bimodal energy distribution at the critical temperature, both becoming more pronounced for larger lattice sizes $N$. This suggests that both phase transitions are of first-order. To measure the strength of the first-order transitions, we obtain the associated latent heat by fitting double Gaussians to the energy distribution at the critical temperature and calculating the distance between the two peaks. The resulting latent heat for both first-order transitions is shown in Fig.~\ref{fig:latent-heat}, which exhibit a sizable latent heat of up to $\Delta \epsilon/J_2 \approx 0.07$ and $\Delta \epsilon/J_2 \approx 0.08$, respectively, indicating strong first-order transitions. When approaching the transition into the ferromagnetic spin order, where the thermal phase transition appears continuous, the latent heat smoothly vanishes.
\end{document}